\documentclass[a4paper, 9pt]{elsarticle}

\usepackage[english]{babel}
\usepackage{fontenc}
\usepackage{titlesec}
\usepackage{float}
\usepackage[hyphens]{url}
\usepackage{physics}
\usepackage{graphicx}
\usepackage{amsmath,amsthm,amssymb}
\usepackage[colorlinks=true,citecolor=blue]{hyperref}
\usepackage{subfigure}
\usepackage{color}
\usepackage{multirow}
\usepackage{enumerate}
\usepackage{epstopdf}
\usepackage{soul}
\usepackage{textcomp}
\newcommand{\ba}{\begin{array} }
	\newcommand{\ea}{\end{array} }
\newcommand{\bae}{\begin{eqnarray}}
\newcommand{\eae}{\end{eqnarray}}
\newcommand{\bea}{\begin{eqnarray*}}
	\newcommand{\eea}{\end{eqnarray*}}
\newcommand{\be}{\begin{equation}}
\newcommand{\ee}{\end{equation}}

\newcommand{\pr}{{\bf Proof}~~}

\def\to{{\rightarrow}}

\def\to{{\rightarrow}}

\usepackage{mathtools}

\usepackage{amsthm}
\usepackage{amsmath}

\usepackage{titlesec}
\titleformat{\paragraph}
{\normalfont\normalsize\bfseries}{\theparagraph}{1em}{}
\titlespacing*{\paragraph}{0pt}{3.25ex plus 1ex minus .2ex}{1.5ex plus .2ex}

\numberwithin{equation}{section}

\usepackage{multicol}
\usepackage[utf8]{inputenc}
\usepackage{csquotes}
\usepackage{subfiles}

\begin{document}

\markboth{Extreme events}{Review}

\title{Extreme events in dynamical systems and random walkers: A review}

\author{Sayantan Nag Chowdhury $^{1,*}$, Arnob Ray $^{1,*}$, Syamal K. Dana $^{2,3,4}$ and Dibakar Ghosh $^{1,\dagger}$}
\address{$^1$ Physics and Applied Mathematics Unit, Indian Statistical Institute, 203, B. T. Road, Kolkata 700108, India\\
	$^2$ Division of Dynamics, Lodz University of Technology, 90-924 Lodz, Poland\\
	$^3$ Department of Mathematics, Jadavpur University, Kolkata 700032, India\\
    $^4$ Department of Mathematics, National Institute of Technology, Durgapur 713209, India}
\date{Received: date / Accepted: date}

\begin{abstract}
	Extreme events gain the attention of researchers due to their utmost importance in various contexts ranging from finance to climatology. 
	An observable that deviates significantly from its long-time average will have adverse consequences for the system. This brings such recurrent events to the limelight of attention in interdisciplinary research.
	There is a need for research efforts in many systems in the real world to find solutions that can predict and mitigate the unfavorable effects of these recurring events. A comprehensive review of recent progress is provided to capture recent improvements in analyzing such very high-amplitude events from the point of view of dynamical systems and random walkers.
	We emphasize, in detail, the mechanisms responsible for the emergence of such events in complex systems. Several mechanisms that contribute to the occurrence of extreme events have been elaborated that investigate the sources of instabilities leading to them. In addition, we discuss the prediction of extreme events from two different
	contexts, using dynamical instabilities and data-based machine learning algorithms. Tracking of instabilities in the phase space is not always feasible and precise knowledge of the dynamics of extreme events
	does not necessarily help in forecasting extreme events. Moreover, in most studies on high-dimensional systems, only a few degrees of freedom participate in extreme events’ formation. Thus, the notable inclusion of prediction through machine learning is of enormous significance, particularly for those cases where the
	governing equations of the model are explicitly unavailable.
	Besides, random walks on complex networks can represent several transport processes, and exceedances of the flux of walkers above a prescribed threshold may describe extreme events. We unveil the theoretical studies on random walkers with their enormous
	potential for applications in reducing extreme events. We cover the possible controlling strategies, which may be helpful to mitigate extreme events in physical situations like traffic jams, heavy load of web requests,
	competition for shared resources, floods in the network of rivers, and many more. This review presents an overview of the current trend of research on extreme events in dynamical systems and networks, including
	random walkers, and discusses future possibilities. We conclude this review with an extended outlook and compelling perspective, along with the non-trivial challenges for further investigation.

\vskip 2truecm
* These Authors equally contributed to the manuscript\\
$\dagger$ Corresponding author (diba.ghosh@gmail.com)

\end{abstract}


\date{\today}
\maketitle

\tableofcontents

\section{Introduction}\label{introduction}

\par Extreme events give rise to massive challenges among different scientific communities and become one of the active topics in interdisciplinary researches due to the disastrous impact and irregular occurrences of these low-probability events. Examples of extreme events are numerous, and their emergent behavior is visible in several processes ranging from environmental disasters to call drops in cellular networks, economic drawdowns, epileptic seizures, global pandemics, to name a few. The studies on a broad spectrum of extreme events deserve special attention to get rid of its erratic behavior. This review aims to present a repertoire of the recent trend of research on this important interdisciplinary topic using dynamical systems and various random walk models that recreate some realistic situations in daily life.

\subsection{Extreme events in real-life situations}
\par Extreme events, one of the essential unifying paradigms, gain significant recognition among many scientific disciplines for their severe detrimental consequences and potential applications. Various aspects of this interdisciplinary topic have customary involvement with our society rather say, in the progress of our civilization. In our recent experiences, the super cyclone Amphan is one of the worst storms recorded over the Bay of Bengal, packing winds reaching 185 kmph and triggering torrential rains across a wide swath of West Bengal (India) from deltaic regions to urban neighborhoods of Kolkata on May 20, 2020. Another example may be considered as the spreading of the pandemic COVID-19 worldwide, which does not require any further introduction. These are the few intriguing real-life instances that belong to the genre of extreme events. One or two of these examples attest to the necessity of understanding extreme events from the fundamental zero ground level. 

\par Concerning the widespread impact, extreme events have been studied from different perspectives in various fields like oceanography \cite{pelinovsky2008extreme}, climatology \cite{web_5}, sociology \cite{webb2002sociology}, finance \cite{embrechts2013modelling,sabir2014record}, ecology \cite{gutschick2003extreme} and so on from several decades. Despite being statistically improbable, examples of such events have been documented in the form of diverse natural disasters like earthquakes \cite{pisarenko2003characterization}, epidemic spreading \cite{mcmichael2015extreme}, cyclones \cite{emanuel2005increasing}, floods \cite{sachs2012black,buchele2006flood}, droughts \cite{hoerling2013anatomy}, harmful algal blooms \cite{anderson2012progress}, regime shifts in ecosystems \cite{scheffer2003catastrophic,folke2004regime}, hurricanes \cite{barlow2011influence}, global warming-related changes in climate and weather \cite{sura2011general,michel2008extreme,ren2018research}, asteroid impacts \cite{huppert2006extreme}, solar flares \cite{buzulukova2017extreme}, tsunamis \cite{mascarenhas2006extreme,bird2011human} to name but a few. Natural hazards \cite{alexander2018natural}, significantly considered as extreme events from the beginning of our civilization, hamper the progress of human evolution with its adverse effect. 

\par As much as our society has progressed and technology has improved, we have faced different threats of extreme events in the form of share market crash \cite{johansen2002endogenous,krause2015econophysics}, power blackouts \cite{dobson2007complex,kinney2005modeling}, industrial accidents \cite{zio2013industrial,salzano2013public}, acts of terrorism \cite{kunreuther2004dealing}, mass panics \cite{helbing2001traffic} and many more. Researches regarding extreme events are also desirable due to their substantial negative impact on society, either in the form of economic downfall or with respect to human causalities or even in some cases both \cite{sapsis2018new, albeverio2006extreme}. Generally, extreme events have a catastrophic side, which has an impact on our society as well as on nature. In some instances, extreme events are originated by human activities, irrespective of unintentional or intentional motivation. For example, wars or revolutions are one of the staggering specimens of such ruinous events that lead to the change of economics and politics of a community \cite{hobsbawm2020age}. 

\subsection{Extreme events: Characteristics and challenges}
\par There is no strict definition \cite{albeverio2006extreme,mcphillips2018defining,akhmediev2010editorial} available for extreme events in the literature due to its extensive acceptance in diverse fields, especially in the context of natural events since a relatively smaller event can even make a huge damage. When talking about the characteristics of such large impact events, the first word that comes into anyone's mind is unpredictability \cite{albeverio2006extreme}. These devastating events appear from nowhere without any clue, increasing the overall peril in terms of its havoc. In some cases, early warning signals of upcoming extreme events may be possible to find, thanks to the advancement of science and technology \cite{boers2014prediction,carrara1999use,laptev2017time,denny2009prediction,thompson2002stratospheric,scheffer2009early,karnatak2017early}. But in most of the cases, researchers are trying to reduce the damage and mitigate the harmful impact on society using several approaches \cite{chapman1996project,nott2006extreme}. Besides, {\it extreme value theory} \cite{coles2001,de2007extreme}, a field of statistics, is found to be helpful for understanding the probabilities associated with extreme events. The application of extreme value theory from this perspective is acknowledged in various fields such as hydrology \cite{bruun1998comparison} and finance \cite{gilli2006application}. 
Besides, a statistical approach can analyze extreme events in
fluid dynamics \cite{sapsis2021statistics}. Interested readers may consult Ref.\ \cite{majumdar2020extreme} for a quick review of the extreme value theory.

\par However, the study of extreme value theory includes the limiting distribution (if exists) of maxima or minima of an observable. Obviously, in terms of magnitude and intensity, extreme events are the extrema of the time evolution of an observable generating a statistical transition from symmetric
near-Gaussian statistics to a highly skewed probability density function \cite{mohamad2016probabilistic,mohamad2015probabilistic,mohamad2016probabilistic1}. Here, an observable \cite{sapsis2018new} is defined as a function of state variables related to the differential equations, which can be measured. The characterization of extreme events can also be done from the time domain instead of the spatial domain. The point process technique is one of the premiers
specimens, which gives us an insight into the time occurrence of the extreme events. Recurrence of extreme events is reflected in studies of return intervals \cite{santhanam2012extreme,santhanam2017record} in time between extreme events. The statistically uncorrelated events are distributed according to the exponential distribution \cite{ansmann2013extreme,santhanam2008return}, or Poisson distribution \cite{von2001statistical}. Poisson point process \cite{kingman2005p} is one of the effective tools in the existing literature to study the statistics of return intervals between the extreme events for such uncorrelated events. On the other hand, when extreme events form clusters ({\it i.e.,}\ concentrated in time) \cite{eichner2007statistics,santhanam2005long}, the statistically correlated events \cite{singh2020inferring} lead to a different dynamical process generating stretched exponential distribution \cite{eichner2007statistics,altmann2005recurrence} of the return time intervals. More recent studies \cite{santhanam2008return,blender2008nonlinear} have shown that Weibull distribution is a good representation for the return interval distribution of long-range correlated data.
But, the name `extreme' reflects another attribute, which is the infrequentness of extreme events. There are several examples, where extreme events are treated as rare events \cite{dematteis2018rogue,cousins2015unsteady, kyul2018heavy}. But, there is a clear thin dissimilarity among the rare events and extreme events  \cite{sapsis2018new}. Already, vast researches have been made on extreme events, where they found the frequent occurrence of extreme events in space and time  \cite{majda2014conceptual,grooms2014stochastic,majda2015intermittency,cai2001dispersive}.

\par Rare events are events with a low frequency associated with a random mechanism \cite{cameron2013regression,ross2006first}. Examples of such events might include the chances of being dying on your birthday or being born on a leap day. The chances that the person you're dating is a millionaire are on the slim side. Although humans being born with teeth or being born with an extra finger, may be considered as rare events, but they definitely do not belong to the category of extreme events. In cases of extreme events, several instabilities are incorporated due to various non-trivial mechanisms \cite{sapsis2018new}. One of the examples of the extreme event in the human body is epileptic seizures in the brain \cite{lehnertz2006epilepsy,pisarchik2018extreme,frolov2019statistical}. This is an example of extreme events, which occurs frequently \cite{albeverio2006extreme}. This type of example motivates researchers from different scientific communities to inspect this topic beyond the statistical properties of the systems.        

\subsection{Dynamical system: A comprehensive tool of study}
\par Up to now, the interest has been focused on the need to find new tools for characterizing extreme events based on several folds, viz. (i) causes, (ii) characteristics, (iii) early warning detections and predictions, and (iv) mitigations. In oceanography, research of extreme events is explored around {\it ocean rogue waves} significantly and several theories, quantifications, and predictions are flourished \cite{ryszard1996ocean,zanna2008oceanic,slunyaev2009rogue,cousins2019predicting,birkholz2015predictability}, which also lead to a new direction of scientific research. Later, an analogy similar to ocean rogue wave has been drawn with nonlinear optical systems, and thereby a new field is opened, known as {\it optical rogue wave} \cite{solli2007optical,bonatto2011deterministic,won2011surely,akhmediev2013recent,akhmediev2016roadmap,jin2017generation,web_15}. Indeed, researchers become interested in thinking about the extreme events or similar types of phenomena from a nonlinear dynamical system perspective \cite{ghil2011extreme,lucarini2016extremes,farazmand2019extreme,grigoriu2019discussion}. The important characteristic of extreme events is their irregular occurrence, and hence this signature may be observed in the chaotic evolution of trajectories of nonlinear dynamical systems \cite{strogatz2016nonlinear,meiss2007differential}. 

\par From this perspective, some new motivations have emerged in the dynamical systems. From the point of view of extreme events, the trajectory of a dynamical system evolves within its bounded attractor most of the time but occasionally visits the outside of the bounded region. This excursion is reflected as a large amplitude deflection (as a form of spikes or bursts) in a dynamical variable of the system due to the appearance of the region of instability \cite{babaee2016variational,farazmand2016dynamical} in the state space. The amplitudes of the variable deviate significantly from the central tendency (regular behavior) of the observable. These infrequent, but recurrent occurrences of large amplitude events reveal qualitative similarities, in the sense of dynamical systems, with the existing data sets of real-life calamities\ \cite{ray2019intermittent}.

\par Recently, a trend of research has started with isolated dynamical systems to observe extreme events-like scenarios and then understand the underlying mechanisms of origin of such events and suggest a possible method of prediction and control. In many dynamical systems, intermittent large deviation in the amplitude of a state variable is seen in their temporal dynamics. It is found that regions of instability \cite{sapsis2018new,babaee2016variational,ashwin2005instability,blanchard2019analytical} are always associated in phase space, and this fact is responsible for producing extreme events in a dynamical system, which are described using a set of first-order ordinary differential equations \cite{simmons2016differential}. Parameters play a crucial role to control the intrinsic dynamics of the nonlinear systems, and by tuning a parameter's value, the system bifurcates its qualitative behavior and can exhibit rich dynamics \cite{kuznetsov2013elements}.

\subsection{Extreme events in isolated and coupled dynamical systems: Causes}
\par  The extreme events may be observed near a bifurcation point when the transition between two states is occurred, such as switching from the periodic dynamics to chaotic or switching from chaotic dynamics of one feature to another with different features. Scientists are utterly interested in finding the systems which exhibit extreme events and in which route these are generated. One of the most important route for originating extreme events in chaotic systems is {\it intermittency} \cite{ott2002chaos,alligood1996chaos}. Two types of intermittency such as the interior crisis-induced intermittency \cite{grebogi1983crises,grebogi1987critical} and Pomeau Manneville (PM) intermittency \cite{pomeau1980intermittent} have been reported as leading to the originate extreme events \cite{zamora2013rogue,kingston2017extreme}. Another phenomenon is noise-induced intermittency that helps the switching between the coexisting states \cite{horsthemke1984noise,van1994noise,forgoston2018primer}. Extreme events \cite{pisarchik2011rogue} may also emerge due to noise-induced intermittency \cite{gwinn1985intermittent} in multistable systems \cite{feudel2008complex}. Besides, if any system possesses singularity \cite{jeffrey2009two}, then it may be capable of generating extreme events through sliding bifurcation \cite{teixeira1993generic} due to the presence of a discontinuous basin boundary \cite{kumarasamy2018extreme}.


\par But, the study of extreme events is not limited within the confined regime of the isolated dynamical system. Extreme events are noticed in coupled systems connected via different coupling functions. Researchers focus on collective behaviors \cite{boccaletti2002synchronization,pikovsky2003synchronization,rakshit2018synchronization} in coupled chaotic systems within the last three decades. Investigation of extreme events leads to a new path of exploration in coupled dynamical systems. Here, extreme events emerge due to instability of synchronization manifold \cite{mendoza2004convective}. The trajectories being repelled by the unstable objects such as saddle point or saddle orbit of synchronization manifold, ultimately comes back to the invariant manifold \cite{josic1998invariant}, creating finite-size,  short-lived, intermittent excursion away from that invariant manifold. This phenomenon is known as {\it attractor bubbling} \cite{ashwin1994bubbling,heagy1994characterization,heagy1995desynchronization}. The factors responsible for generating instability of synchronization manifold are systems' heterogeneity or presence of noise or in some cases both. It causes on-off intermittent bursts \cite{ding1997stability} along the transverse direction of the synchronization manifold, and these bursts may be reported as extreme events in the coupled system  \cite{cavalcante2013predictability}. Imperfect phase synchronization \cite{zaks1999alternating,park1999phase} and instability of antiphase spike and burst synchronization \cite{wang2008bursting} are also found to be responsible for generating extreme events \cite{ansmann2013extreme,karnatak2014route,mishra2018dragon}. 

\subsection{Extreme events in static and time-varying dynamical networks}
\par Recently in the $21^{st}$ century, the attention of researchers has been shifted to a new discipline named network science \cite{barabasi2016network,newman2018networks,albert2002statistical,newman2003structure,watts1998collective}. This research discipline offers fresh new insights into complex systems. Several systems, like brain networks \cite{frolov2019statistical,amor2015extreme}, ecological networks \cite{moitra2019emergence,kundu2021persistence,chaurasia2020advent,kundu2017survivability}, social networks \cite{adger2005social,stieglitz2018sense}, owe their functionality to a complex network as their backbone. Several emergent dynamical processes, including percolation \cite{parshani2010interdependent,buldyrev2010catastrophic,gao2012networks}, diffusion \cite{gomez2013diffusion,alves2016characterization}, epidemic spreading \cite{saumell2012epidemic,taylor2015topological,saha2020infection}, chimera states \cite{abrams2004chimera,majhi2016chimera,kundu2021amplitude,bera2017chimera,panaggio2015chimera,majhi2019chimera,khaleghi2019chimera,PARASTESH2020,majhi2017chimera,maksimenko2016excitation,kundu2018chimera}, suppression of oscillations \cite{saxena2012amplitude,dixit2021dynamic,kundu2018resumption,dixit2021emergent,resmi2011general,ray2020aging,kundu2018augmentation}, synchronization \cite{arenas2008synchronization,hramov2004approach,chowdhury2019convergence,majhi2019emergence,sitnikova2014time,chowdhury2019synchronization, rakshit2021relay} and cooperation among unrelated individuals \cite{gomez2012evolution,nag2020cooperation,szolnoki2010reward,perc2017statistical,chowdhury2021eco,perc2010heterogeneous,perc2008social,chowdhury2021complex}, are noticeable in this multidisciplinary field. Recently, some studies on extreme events in coupled dynamical networks have been done under different context \cite{marwan2015complex}.

\par This is important to understand the origin of extreme events like epileptic seizures \cite{o2013spreading}. It is important to develop a deeper understanding of the complexities for addressing the queries like when a large population bloom may occur \cite{yao2011mathematical}, or how does the spreading of epidemics via social networks \cite{wu2008community,liu2014information} take place. Unfortunately, in most of the cases, precise knowledge of the physical model does not necessarily help to understand the mechanism behind extreme events. This problem particularly seems to be more pronounced in the case of high-dimensional complex systems, where only a low-dimensional subset of the many interacting variables participates in the formation of extreme events. In such systems, it is really a challenging task to propose an efficient strategy for the prediction and mitigation of extreme events. The complex interactions among all state variables create difficulty to isolate the ones that underpin extreme events.

\par Fortunately, these challenges attract the attention of an increasing number of scientists nowadays, and few initial investments are proposed in this direction to recognize the role of interplay between system’s intrinsic dynamics and network’s topology in the causation of extreme events \cite{ansmann2013extreme,ray2020extreme}. Actually, dynamical systems rarely remain isolated, and the static network formalism maybe, in some cases, representing over-simplified scenarios ignoring the possible time-varying interactions \cite{holme2012temporal,ghosh2022synchronized} of physical and social networks. This leads to another fundamental puzzle is how the mobility of agents affects the dynamics of a time-varying network and thereby may lead to extreme events. Earlier, few attempts are made to scrutinize the consequence of mobile agents \cite{frasca2008synchronization, fujiwara2011synchronization, porfiri2006random} and the effect of attractive-repulsive interactions in networks of oscillators \cite{majhi2020perspective,hong2011kuramoto,chowdhury2021antiphase,wang2011synchronous,chowdhury2020effect, sar2022swarmalators}  from different points of view. However, a lack of studies related to the phenomenon of extreme events in time-varying dynamical networks create a dramatic void for characterizing such complex infrastructures. These crucial questions in the context of extreme events have been brought to the spotlight recently \cite{chowdhury2019extreme,9170822} by considering the temporal networks with co-existing attractive-repulsive coupling. On-off intermittency among the coupled chaotic oscillators is found to be responsible for the genesis of extreme event. We provide a concise yet richer and more detailed outlook of the spectacular progress of research on this topic in the context of the dynamical networks.

\subsection{Extreme events in random walkers}
\par Transportation network \cite{helbing2001traffic} is an important realization of a spatial network \cite{barthelemy2011spatial} from the
perspectives of both theory and application. This type of network carries a huge amount of load in the form of either vehicular movement or the flow of some commodities. Examples include but are not limited to road networks, railways, air routes, the Internet, and power grids. In the modern information era, the welfare and security of modern societies increasingly
depend on the correct functioning of such communication networks.
The study of the information flow through communication networks has been introduced with several goals \cite{wang2009abrupt,de2009congestion,echenique2005dynamics,germano2006traffic,zhao2005onset,ashton2005effect,tadic2007transport}. For example, transmission of information packet in discrete unit via the Internet is an important relevant scenario \cite{kim2009jamming}. In these packet based communication networks, data are created at certain nodes in the Internet, and travel to their destination networks along an optimal path, sharing a common line (buffer) in the network. The advancement of technology in day-to-day life leads to the continuous growth of most communication networks. When the number of packets in a network is high, misfunction in the form of congestion phenomena is observed due to limited capacity of processing and storage of each node and link of the networks. The efficient performance of these systems is affected by slowing down the traffic, clogging large regions of the network.

\par The Google search engine received approximately 2.9 million search requests per minute by the end of $2009$ \cite{web_1}. As per Wikipedia, the popular social networking site Facebook had $500$ million users in July $2010$, and it crossed the $2$ billion user mark in June $2017$. According to the company's data at the July $2010$ announcement, half of the site's membership used Facebook daily, for an average of $34$ minutes, while $150$ million users accessed the site by mobile. In October $2012$, Facebook's monthly active users passed one billion, with $600$ million mobile users, $219$ billion photo uploads, and $140$ billion friend connections \cite{web_2}. Twitter handled about $600$ tweets per second in early $2010$ \cite{web_3}. Most of these websites are unprepared to handle such a large number of congestion in the form of HTTP requests, resulting in an increment of the transit time. On the road networks, examples of such congestion are not less. The China National Highway $110$ traffic jam, dated August $13$, $2010$, is one of the premier specimens of such traffic jams that lasted for nine days, slowing down thousands of vehicles for more than 100 kilometers \cite{web_4}. These numbers represent extreme events and could potentially disrupt the services, and hence the lifestyle of human beings.

\par Motivated by these facts, a section is devoted to understand the emerging extreme events in networks of random walkers. As we have already discussed, information flowing through an edge and a node is a critical task, as each node and edge has its limited capacity. This limited handling capacity of the ingredients of a network and the inherent fluctuations in the flux passing through them may constitute extreme events. Extreme events in networks of random walkers are often observed in the form of the heavy load of HTTP requests, power blackouts, traffic jams, gridlock on highways, etc. Therefore, systematic studies on extreme events in networks of random walkers deserve special attention. The articulation and development of effective control strategies are in demand to avoid the catastrophic consequences of extreme events. A few available efficient and physically implementable methods, along with their analytical theories to understand its working, are reviewed in this report.

\subsection{Prediction and Suppression: Fundamental necessitate and difficulty}
\par Prediction of extreme events to trigger early warning signals is very relevant issue for every field of studies like finance \cite{sornette2002predictability,johansen1999predicting}, climatology \cite{wehner2004predicted}, oceanography \cite{islas2005predicting} and so on. The devastating effect of extreme events can be avoided if a prediction can be made well in advance. We discuss two different approaches for this purpose. One is the dynamical system approach using the instability regions \cite{farazmand2016dynamical,cousins2014quantification}, and another is the machine learning approach \cite{guth2019machine,blanchard2019learning,qi2020using,saha2020predicting,narhi2018machine, meiyazhagan2022prediction}. In the dynamical system approach, people are interested in identifying the instability region in the phase space. When the trajectory passes through this region, a large excursion of trajectory occurs. This instability region of phase space may appear as a form of channel-like structure \cite{zamora2013rogue}, or due to the presence of saddle point \cite{cavalcante2013predictability}, or the singularity \cite{kumarasamy2018extreme}. Such approaches have recently been investigated. Besides the dynamical system approach, recently, machine learning techniques \cite{jordan2015machine} have been used to predict extreme events \cite{pyragas2020using,lellep2020using} in the dynamical system from the time series. In this context, reservoir computing framework \cite{jaeger2004harnessing,pathak2018model} or other neural network frameworks \cite{schmidhuber2015deep,lecun2015deep,goodfellow2016deep,hassoun1995fundamentals} are used for model-free forecasting of such disastrous events.

\par Control or suppression of extreme events \cite{galuzio2014control,bialonski2015data,joo2017extreme,farazmand2019closed,bree2016controlling,perrone2014controlling, sudharsan2021constant, sudharsan2021emergence} to minimize the damage from the devastating effect of extreme events is one of the most challenging issues till now. This is, of course, in principle not possible to control any natural disasters like Tsunami, floods, cyclones, droughts etc. But one can attempt to design a controller to avoid huge losses in human-made systems like power grid \cite{li2017networked,bie2017battling}, financial crisis \cite{herrera2014statistics}, traffic jamming \cite{to2001jamming} and many more. So, one can plan to design some control policies in dynamical systems after knowing the dynamical instability of the manifold that causes extreme events. Few methods such as the feedback control \cite{desoer1980foundations,ikhlef2008time,yamashita2015continuous}, corrective resetting \cite{mayol2012anticipated}, threshold-activated coupling scheme \cite{sinha1990adaptive,sinha1993adaptive} are investigated for control purposes in the dynamical systems \cite{ray2019intermittent,cavalcante2013predictability,zamora2014suppression,suresh2018influence,varshneysuppression}. 

\subsection{Brief outline of the report}
\par Our motive is to review the recent development of extreme events associated with dynamical
systems and related to random walk. We enrich this review with an extensive introductory preamble on relevant concepts. The Chronicle of this review article is separated by some sections, which are organized as follows: 
%
In Section \ref{Dynamical origin of extreme events}, 
we summarize different mechanisms that are capable of triggering extreme events. We provide the discussions on different models generating extreme events. Section\ \ref{Extreme events due to random walk in complex networks} contains a wide variety of models on random walks taking place on top of complex networks. Possible controlling strategies of extreme events in random walker related problems, an utmost important issue, are included. Section \ref{prediction} includes a rather complete overview of available prediction schemes for extreme events using dynamical instabilities and machine learning approaches. Depending on the nature of dynamical instabilities, some powerful prediction algorithms have been proposed, which are reviewed. We also encourage the model-free prediction using machine learning, which is beneficial and useful, particularly for the cases where the governing models are unavailable.  Then, we also analyze reservoir computer, one of the promising approaches of model-free prediction. Section \ref{control} emphasizes a discussion on existing controlling strategies of dynamical systems. We consider some of the promising aspects of several methods along with various coupling configurations, which are helpful for mitigating extreme events in particular cases. Section \ref{Experimental observation of extreme events} provides a concise summary of the well-designed experimental studies on extreme events. Section \ref{Conclusion and Future directions} sketches our conclusive remarks, which will be helpful for the readers. We summarize the main features of extreme events, observed so far in dynamical systems and in networks of random walkers, with perspective unsolved problems. Some open questions regarding the extreme events, which are not explored yet, are accumulated for the future progress of this field.

\section{Formation of extreme events in dynamical systems}\label{Dynamical origin of extreme events}  
\par A trend of research has started, in the last two decades, on extreme events in dynamical models and laboratory experiments in laser systems \cite{akhmediev2016roadmap,zamora2013rogue,gomel2019extreme}, electronic circuits \cite{cavalcante2013predictability,singhee2012method} and others. In the current literature, several mechanisms are found that trigger occasional large events in dynamical systems. Extreme events have been recognized as occasional large deviation in amplitude of the temporal evolution of a state variable or a suitably chosen observable. In other words, the trajectory of a dynamical system evolves within a bounded region in state space, most of the time, but occasionally travels to a distance far away from that region in response to a  parameter beyond a critical value. This largely deviated value of the trajectory is reflected as occasional large amplitude events in an observable's temporal dynamics. The events are called extreme events if these are larger than a predefined threshold height. In this section, we revisit the existing processes that lead to the formation of extreme events in isolated, two coupled dynamical systems, and networks of dynamical systems.
But before that, we discuss here a general algorithm that may be followed in the exploration of extreme events in the dynamical systems.

\begin{enumerate}

	\item In a dynamical system, one may notice a sudden large expansion of an attractor with the variation of a system parameter. This observation leads to the following systematic steps for further exploration of extreme events in the system.
	
	
	\item Collect a long-term time series $x_i~(i=1, 2, ..., n)$ of a state variable for a system parameter, if it reveals occasional large amplitude events. Otherwise, define an appropriate observable since large events might not always be visible from a temporal dynamics of a state variable.
	
	\item Extract the extreme events 
	using either the block maxima method \cite{ghil2011extreme,lucarini2016extremes, rocco2014extreme, buishand1989statistics} or the peak over threshold approach \cite{gilli2006application, leadbetter1991basis, solari2012unified}. In the block maxima method, one can divide the time series $x_i$ into $b$ bins each containing $k$ data, {\it i.e.,} $n=bk$ and extract the $b$ number of extreme events. Then one can try to fit the collected extreme events with the generalized extreme value (GEV) distribution \cite{majumdar2020extreme}. If the best fit is not compatible with the  GEV distribution, then we have to conclude that the block maxima method is not suitable for the extraction of extreme events. One of the reasons may be the short time series, {\it i.e.,} $n$ is too small. If the block maxima method fails to fit the distribution, or a cluster of events exists in the observed time series, then we will use peak over threshold approach. Most popular practice is to define a threshold $T=m \pm d \sigma$ ($d \in \mathbb{R}^+$), where $m$ is the sample mean of the gathered data, and $\sigma$ is the corresponding standard deviation. Choose an appropriate value of $d$ so that a sufficient number of extreme events are available for statistical modeling purposes.
	
	\item After collecting the time series by simulating the system numerically for a sufficiently long interval, one can use that the collected data for prediction and predictability of extreme events. One can draw probability distribution of extreme events too for estimating their return time. A major advantage of dynamical system related studies is that we can originate an enormously large number of events using numerical simulations. These 
	large numbers of collected data may give us clues on how to extract information from simulated data in absence of real data, and this may also be helpful for more accurate characterization of the statistical properties.
		
	\item We can repeat the study with several system parameters and locate the parameter space of a system, where such extreme events may appear.  
	
	\item After a confirmation of the existence of extreme events for a particular set of parameters in a system, the sources of instabilities (saddle point, saddle orbit, and other sources of singularity if present in the system) in the system can be investigated  for understanding the generation of extreme events.
	
\end{enumerate}
	
\par Over the last few decades, researchers have considered enormous numerical frameworks as well as experiments to interpret the underlying processes that can trigger extreme events in dynamical systems. Several mechanisms are found responsible for such an emerging phenomenon. However, due to the lack of a general underlying process, scientific communities are still motivated to explore it from different perspectives. Till now, systematic studies have been made using various models (excitable systems, neuronal models, electronic circuit models, optical systems) to see the response of such models due to parameter variation, external forcing, and even for induced noise. So far, a few nonlinear processes have been identified that originate extreme events  such as interior crisis-induced intermittency \cite{grebogi1982chaotic,ditto1989experimental}, Pomeau-Manneville intermittency \cite{jeffries1982observation}, attractor hopping in multistable system \cite{kraut2002multistability} due to external or internal noise of a system, and many more. This list has recently been extended to coupled oscillators. On-off intermittency \cite{xie1995off}, in-out intermittency \cite{ashwin1999transverse}, imperfect phase synchronization \cite{zaks1999alternating} are such examples that may take responsibility for the origination of extreme events in coupled systems. Even, few examples are found that are system-specific and motivate us to continue these research works for further investigations. The list is incomplete, and many other processes may exist that are yet to be explored. 

\par In the following subsections, we elaborately discuss various processes that can yield such extreme events in the dynamical systems. Initially, we discuss the results on isolated dynamical systems and then focus on two coupled dynamical systems. Lastly, we explore a few recent works on static and time-varying networks. 
	
\subsection{Isolated (uncoupled) dynamical systems}\label{single}	
\par One of the vital characteristics of extreme events is its irregular occurrences. This striking feature of extreme events makes the chaotic dynamical systems\ \cite{dudkowski2016hidden,sayeed2020behavioral,jafari2015recent,ray2020another,nag2020hidden,vaidyanathan2015analysis} a prominent candidate for the studies of such a non-equilibrium phenomenon. Here, we review the nonlinear processes that lead to extreme events in isolated, {\it i.e.}, uncoupled dynamical systems. 	

\subsubsection{Crisis-induced Intermittency}\label{interior}
\par Crisis-induced intermittency or {\it crisis} is a commonly observed mechanism through which a sudden transition occurs from one state to another in dynamical systems. Due to crisis, attractors generally vanish or enlarge suddenly in the phase space. Three possible types of crisis are found in the literature \cite{grebogi1983crises,grebogi1987critical, grebogi1986critical}. Attractor annihilation due to collision between a chaotic attractor and its basin boundary or any unstable equilibrium point is known as \textit{exterior or boundary crisis} \cite{grebogi1982chaotic}. Another kind of crisis is the \textit{attractor merging crisis} \cite{chossat1988symmetry} that is manifested as a merging of several chaotic attractors to form a new chaotic attractor. The third one, \textit{interior crisis} \cite{grebogi1982chaotic, ditto1989experimental}, is crucial for the origin of extreme events, as this type of crisis seems to be responsible for a route to extreme events in many numerical and experimental studies. Interior crisis is manifested when a trajectory of chaotic attractor meets the stable manifold of a saddle or an unstable periodic orbit and, as a result, the size of the chaotic attractor immediately enlarges. This sudden expansion of the chaotic attractor happens at a critical parameter value that may trigger extreme events. 
\par
\begin{figure}
	\centerline{
		\includegraphics[scale=0.08]{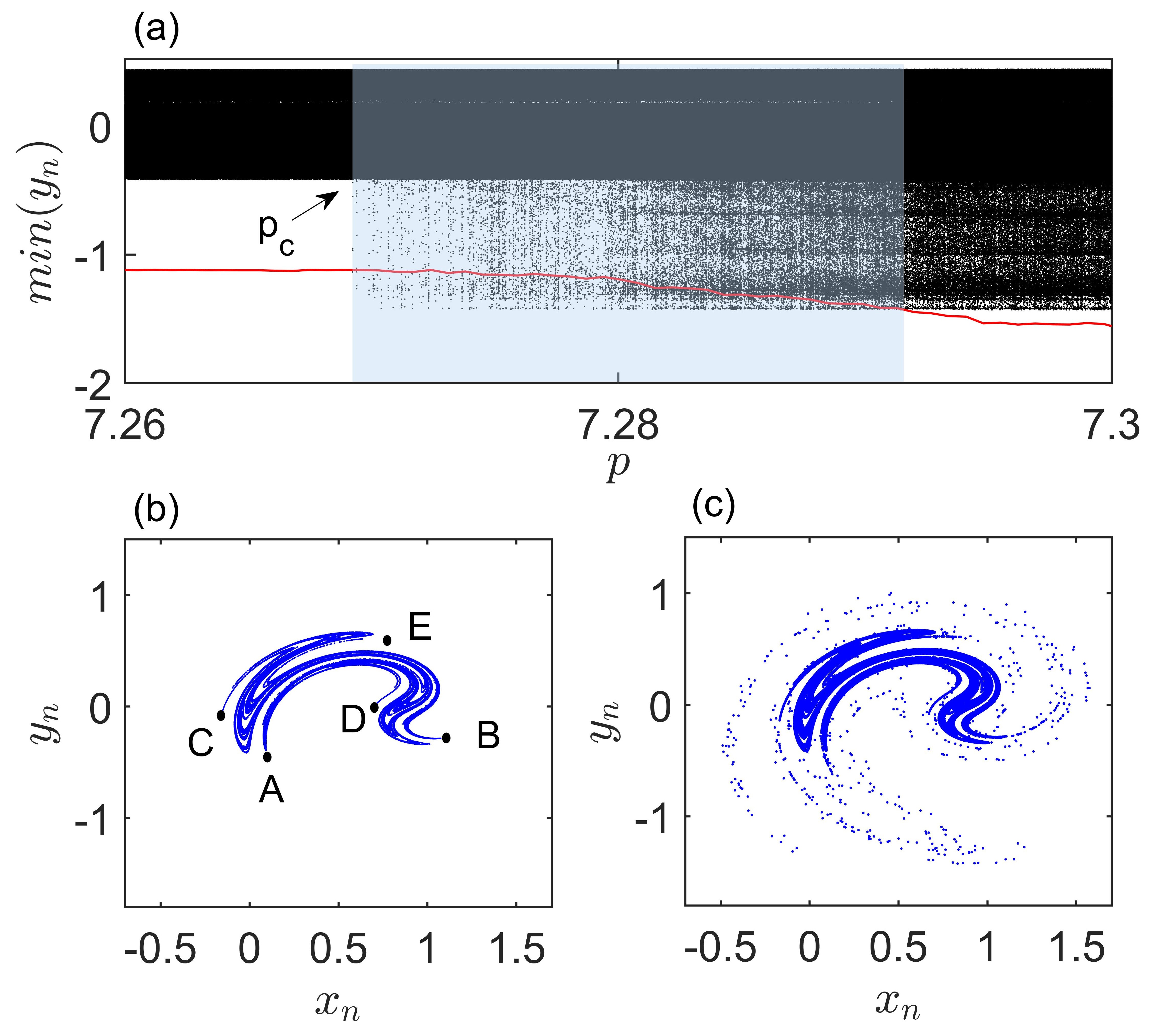}}
	\caption{(a) {\bf Bifurcation diagram of the Ikeda map:}  
		Local minima of $y_{n}$ are plotted against the bifurcation parameter $p \in [7.26, 7.30]$. A sudden expansion of the size of the attractor is displayed at a crisis point $p_c(\approx7.2689)$ and when $p$ crosses $p_c$, then a widened attractor persists. The red line indicates extreme events indicating threshold $T$.  Existence of extreme events is visible in a range of $p$ values marked by a gray-shaded region. (b)-(c) {\bf Phase portraits of the map:} (b) Chaotic attractor at pre-crisis point, $p=7.268$ where one unstable period-5 orbit (black circles) is shown as $A \rightarrow B \rightarrow C \rightarrow D \rightarrow E$ \cite{grebogi1987critical} that coexists with a chaotic attractor (blue). (c) Post-crisis scenario at $p=7.278$, the trajectory occasionally travels locations (sparse blue points) far away from the pre-crisis attractor (dense blue). Model description: $z_{n+1} =a+b~z_{n} exp \Big[ik-\dfrac{ip}{(1+|z_{n}|^2)}\Big]$, where $z_{n}=x_{n}+iy_{n}$, and $a$, $b$, $k$, and $p$ are the amplitude of the external input to the laser, dissipation parameter, laser empty-cavity detuning parameter, and a parameter related to linear phase across the resonator, respectively. Other parameters: $a=0.85$, $b=0.9$, and $k=0.4$.}
	\label{fig_1}
\end{figure}

 For example, such an interior crisis has been reported earlier \cite{grebogi1987critical} in the Ikeda map describing the evolution of laser across a nonlinear optical resonator. In the bifurcation diagram as shown in Fig.\ \ref{fig_1}(a), a sudden expansion of chaotic attractor is observed at a critical value, $p=p_c$ (marked by an arrow). Ray et al.\ \cite{ray2019intermittent} assigned a threshold as $T=m-5\sigma$, where $m$ is the mean, and $\sigma$ is the standard deviation of all local minima in the iteration of the state $y_n$. If any event ($min(y_{n})$) falls below $T$, then the event is recognized as an extreme event. The variation of threshold $T$ (red line) is drawn to identify the range of $p\in(7.269, 7.29)$ (shaded region in Fig.\ \ref{fig_1}(a)) where the system exhibits extreme events. Beyond this range  of $p$, $T$ lies below the events ($min(y_{n})$), when no more extreme events appear. In the pre-crisis scenario ($p<p_c$), the chaotic trajectory (blue attractor) coexists with an unstable period-$5$ orbit (black circles) in Fig.\ \ref{fig_1}(b). At the crisis point $p_c$, the chaotic trajectory collides with the stable manifold of the unstable period-$5$ orbit. As a result, a sporadically spread chaotic attractor is observed beyond the crisis point, as shown in Fig.\ \ref{fig_1}(c).

\par Few more works along the same line have also documented that shows the interior crisis as a responsible route for the emergence of extreme events. The Li{\'e}nard-type system \cite{chandrasekar2005unusual} with an external sinusoidal forcing can generate extreme events for a suitable choice of parameters, as reported in Ref.\ \cite{kingston2017extreme}. The interior crisis also originates extreme events in
a memristor-based driven Li{\'e}nard system\ \cite{kingston2020extreme} and  the parametrically excited Li\'enard system\ \cite{suresh2020parametric}. Extreme events  also emerge in the forced anharmonic oscillator in the presence of nonlinear damping and linear damping \cite{kaviya2020influence}, and damped and driven velocity-
dependent mechanical system\ \cite{sudharsan2021symmetrical}. 
Extreme events also emerge in a fractional dynamical system derived from a Li\'enard-type oscillator\ \cite{ouannas2021chaos}.

\begin{figure}
	\centerline{
		\includegraphics[scale=0.35]{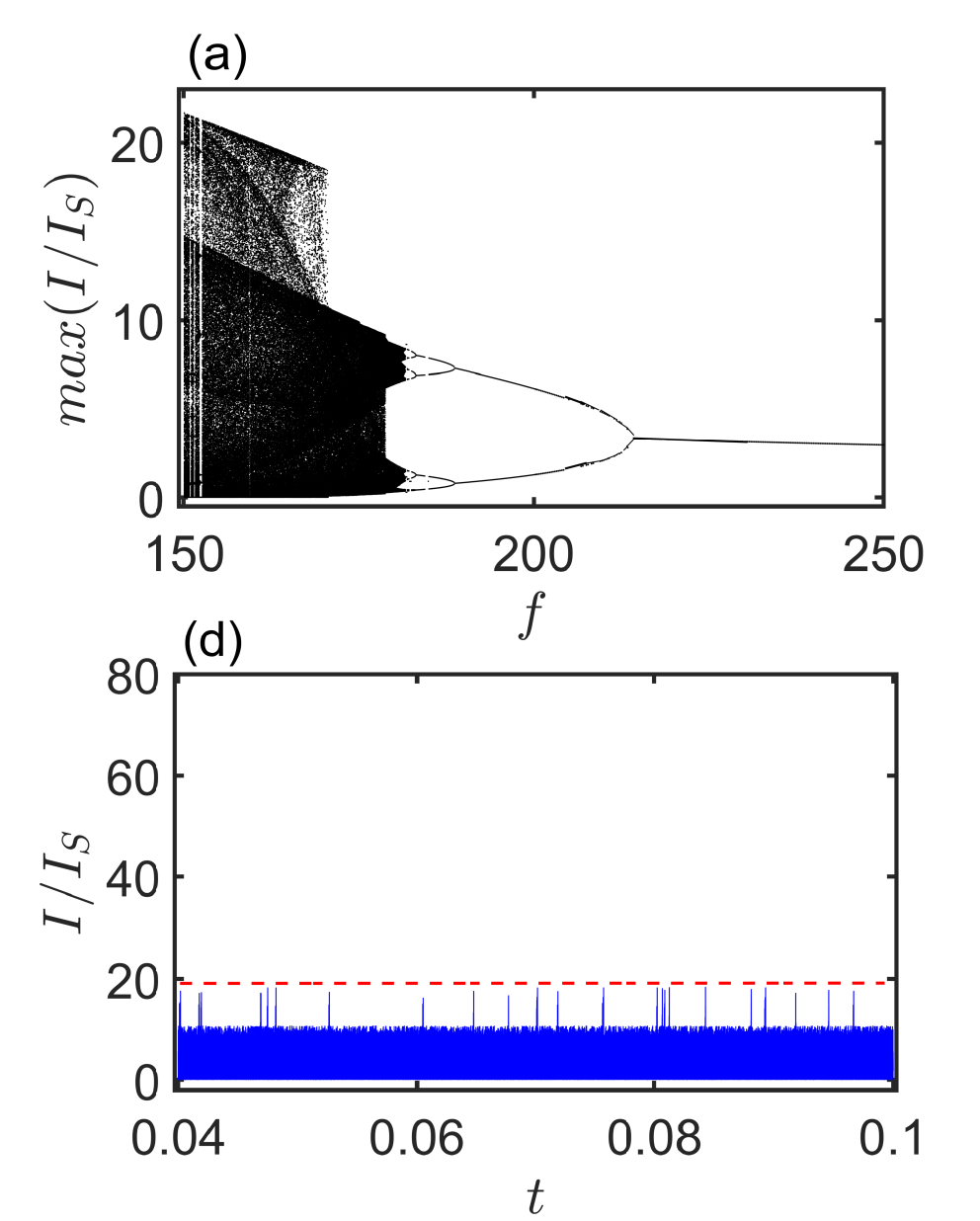}
		\includegraphics[scale=0.35]{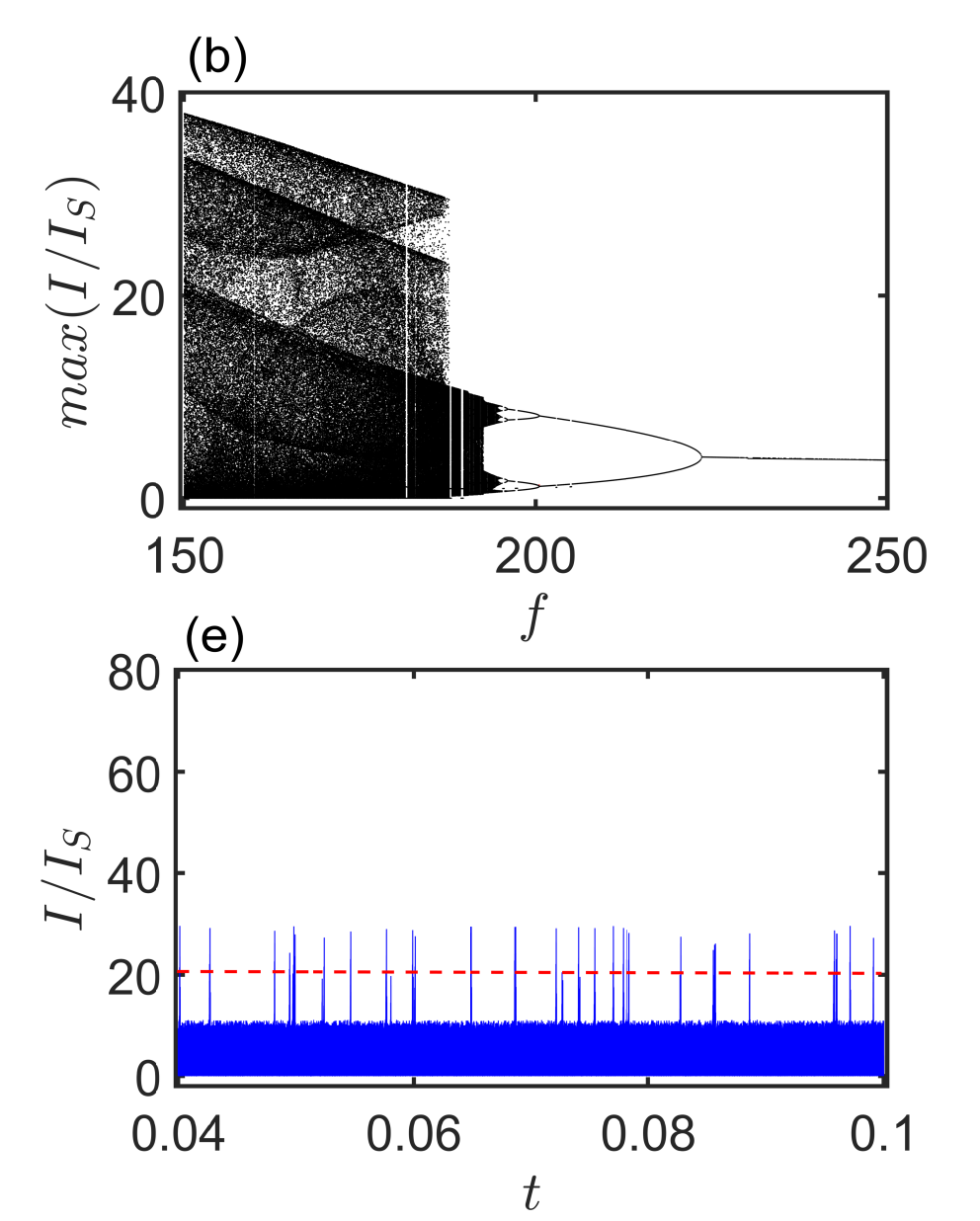}
		\includegraphics[scale=0.35]{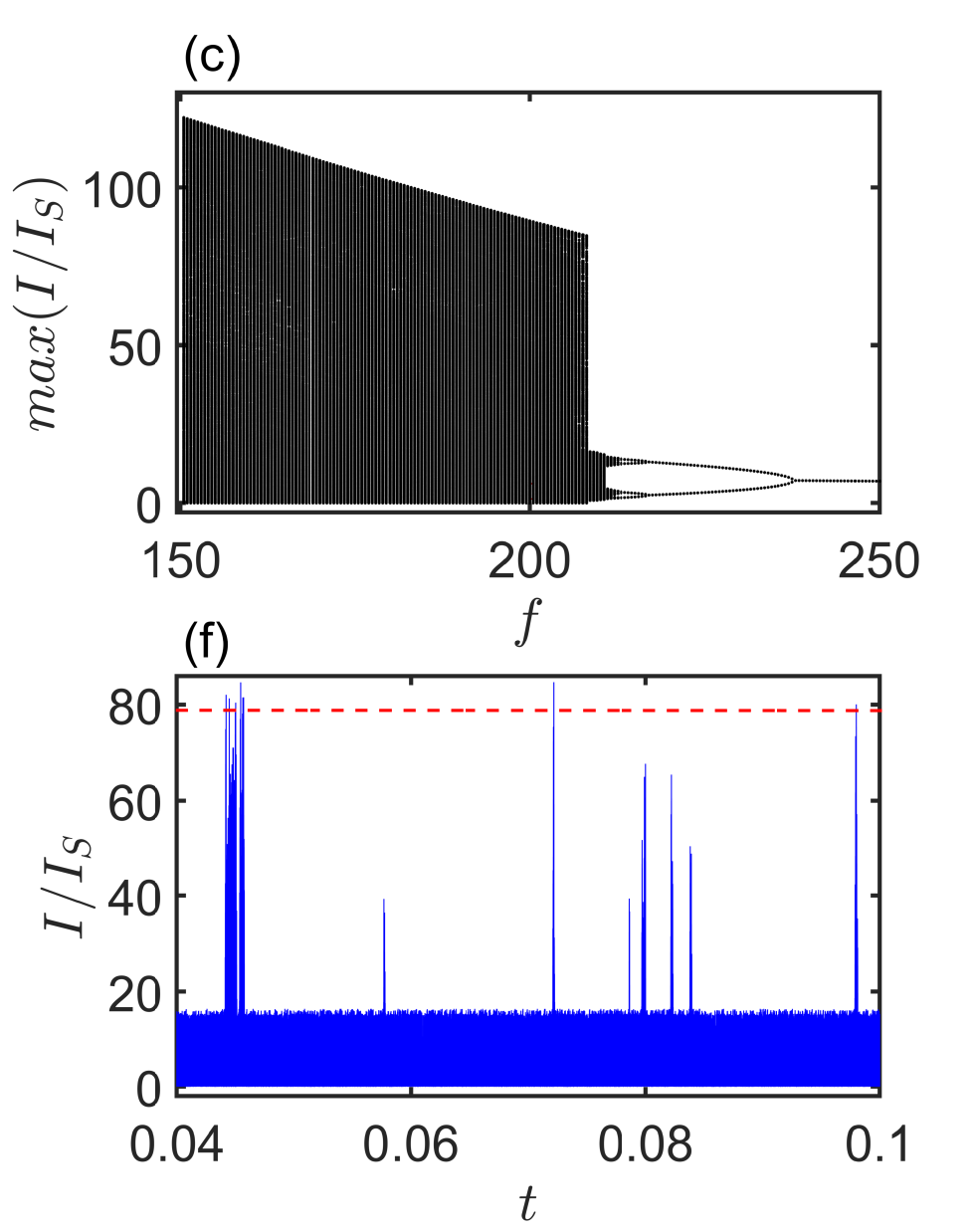}}
	\caption{(a)-(c) {\bf Bifurcation diagrams of laser system\ \eqref{eq.4} for different values of system parameter:}
		Local maxima of laser intensity $(I/I_S)$	
		with respect to modulation frequency $f$ are plotted for three different values of $a$: (a) $a = 0.05$, (b) $a = 0.075$, and (c) $a = 0.19$. Sudden expansion of each chaotic attractor is clearly shown for each figure depicting interior crisis. But it is also distinguishable that the size of expansion of the attractor is gradually increasing by increment the value of $a$. (d)-(f) {\bf Time series of the laser intensity:} Three temporal evolution of laser intensity for three different values of $f$ are portrayed at (d) $f=170.4,$ (e) $f=187.5$, and (f) $f=208.15$ for respective values of (a) $a = 0.05$, (b) $a = 0.075$, and (c) $a = 0.19$. Also extreme event qualifying threshold $T=m+d\sigma$ (red dashed line) is plotted for each case. The values of $d$ to calculate $T$ are $d=4$ for (d), $d=4$ for (e), and $d=10$ for (f). (e) and (f) depicts extreme events and super extreme events, respectively, whereas extreme events do not appear in subfigure (a). 
		Parameters: $\gamma = 1.978 \times 10^5$ $\rm{s}^{-1}$, $\tau =3.5 \times 10^{-9}$ s, $k_0=0.17$ and $N_0=0.175$.} 
	\label{fig_3}
\end{figure}

A question may arise whether the interior crisis or a sudden expansion of a chaotic attractor always triggers extreme events near the critical value of transition in any system. Of course, the answer is no, because of the choice of predefined threshold $T$. $T$ can be chosen in such a way that infrequent events do not appear as extreme events. Again, one can assign $T$ so that such events become super extreme events. Bonatto et al.\ \cite{bonatto2017extreme} focused those issues by investigating from the perspective of the formation of extreme events in a $CO_2$ laser model \cite{arecchi1982experimental, chizhevsky1997attractor, chizhevsky2001multistability}. The mathematical representation of the system  is 
\begin{equation}
\begin{array}{l}\label{eq.4}	
\dot{I} =\dfrac{I(N-k(t))}{\tau},~~~ \dot{N} =(N_0-N) \gamma -IN,
\end{array}
\end{equation}
where $k(t)=k_0(1+a \cos(2\pi ft))$ is an external periodically modulating signal of amplitude $a$ and frequency $f$, $k_0$ is a scaling factor. $I$ is proportional to the radiation density, $N$ and $N_0$ are the gain and the unsaturated gain in the active medium, $\tau$ is the transit time of light, and $\gamma$ is the gain decay rate.

\par This system undergoes a period-doubling cascade leading to chaos with decreasing values of forcing frequency $f$ \cite{arecchi1982experimental} as shown in Figs.\ \ref{fig_3}(a)-(c). For all the three different values of $a$, a sudden expansion of chaotic attractor occurs at three critical values of $f$. The sudden expansion, for all three cases, occurs due to the interior crisis that happens due to a collision of a chaotic attractor of this system with an unstable period-$3$ orbit. Now, the time evolution of laser intensity $\frac {I} {I_S}$ for all the cases near their corresponding critical value of crisis, are plotted in Figs.\ \ref{fig_3}(d)-(f), respectively. For the first case, we find that events (local maxima of the temporal dynamics) are larger than usual dynamics but not yet cross a predefined threshold $T=m+4 \sigma$ (red dashed line), and so none of the events qualify as extreme events. This confirms  the fact that interior crisis may not lead to extreme events although the signature of an expansion of a chaotic attractor exists at a critical parameter. Meanwhile, large intensity pulses cross the threshold $T=m+4 \sigma$, confirming the appearance of extreme events in the second case. But, in the last case, the amplitudes of the large intensity events are highly enhanced at the crisis point, as shown in Fig.\ \ref{fig_3}(c). Clearly it is seen that the amplitudes of the large intensity events cross a higher threshold $T=m+10\sigma$ as shown in Fig.\ \ref{fig_3}(f) at $f=208.15$. 
To calculate the extreme event indicating threshold $T=m+d \sigma$, $d$ is chosen either $4$ or $8$ in Ref.\ \cite{bonatto2017extreme}. An event deviated for more than $10\sigma-12\sigma$ has a larger impact than a $4\sigma$ event. 
Not only the consequence but also these $10\sigma-12\sigma$ deviated extreme events lead to different order solutions. Considering the impact of such large value of extreme events, they are defined as \textit{super-extreme events} in analogy with the \textit{super-rogue waves} observed in a higher-order rational solution of the nonlinear Schr\"odinger equation \cite{chabchoub2012super} to distinguish such giant rare laser pulses from the conventionally known extreme events.

\par Now, we focus on other particular examples describing two processes that are different from the earlier mentioned examples for the emergence of extreme events through the crisis. 

\paragraph{Crisis due to crossing of two bifurcation processes}\label{papd}
\begin{figure}[H]
	\centerline{
		\includegraphics[scale=0.08]{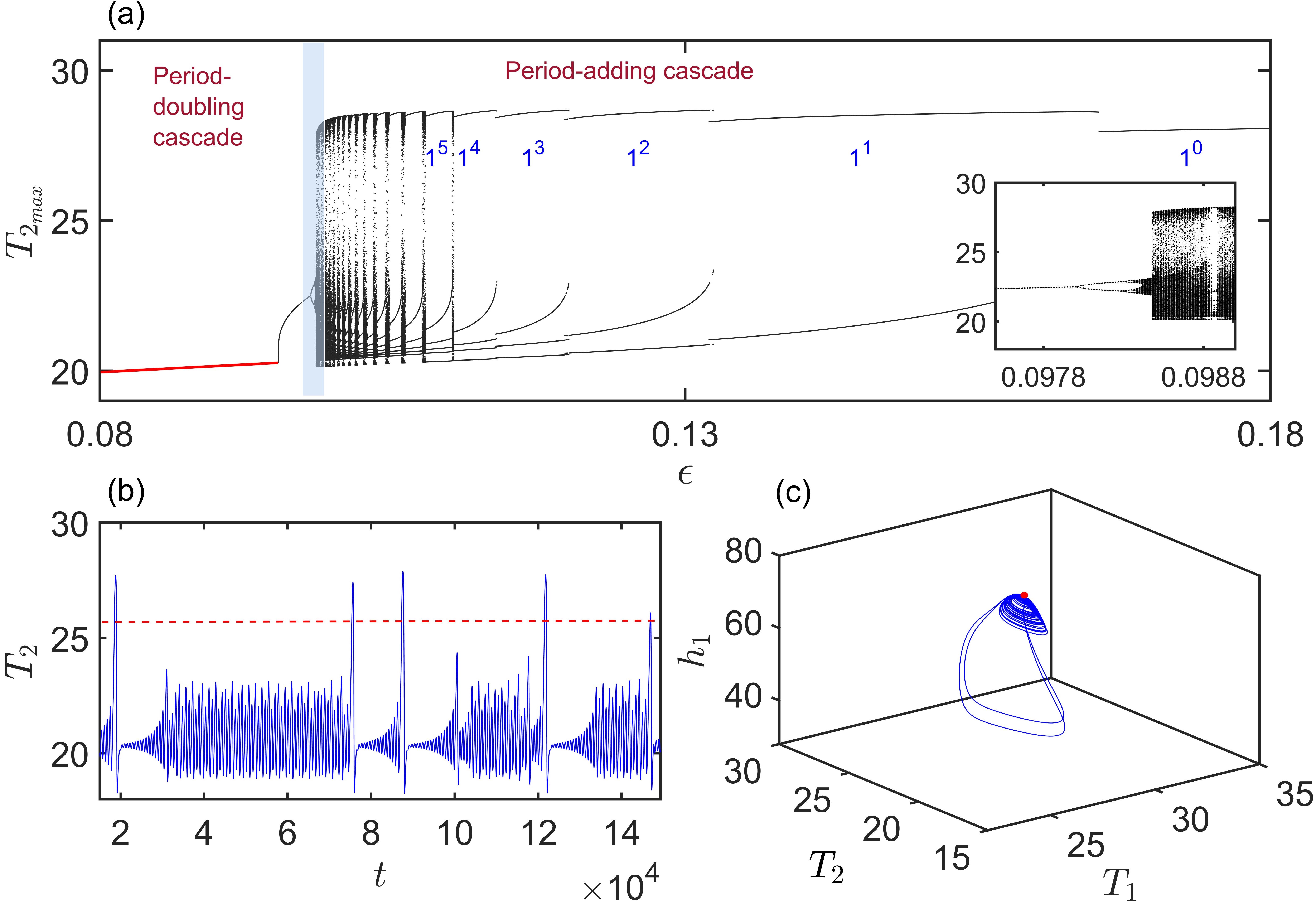}}
	\caption{(a) {\bf Collision between two opposite bifurcation processes:} The variation of local maxima of $T_2$ ($T_{2_{max}}$) with respect to $\epsilon$ is drawn and a stable focus (red line) persists for a range of $\epsilon \in (0.0,0.098522)$. A period-$1$ stable limit cycle emerges in the system, and the chaotic attractor appears in the system by varying the parameter $\epsilon$ via period doubling cascades of bifurcation.  As seen in the opposite side of the bifurcation diagram, when the value of $\epsilon$ is decreased (say, from $\epsilon=0.18$), a large amplitude period-1 (denoted by $1^0$) limit cycle emerges. The $1^0$ orbit transits to period-2  (denoted $1^1$) oscillation. A Farey sequence of $1^1-1^2-1^3 \cdots 1^{\infty}$  oscillations is observed for decreasing the value of $\epsilon$, and it is known as a period-adding cascade of bifurcation \cite{desroches2012mixed}. 
		These two different bifurcation processes from two opposite directions of $\epsilon$ meet at a point (located by a box region). An enhanced version of that box region in the inset clearly exhibits the sudden expansion of the chaotic attractor.
		(b) {\bf Time evolution of chaotic attractor from post-crisis regime:} Time evolution of $T_2$ is plotted for $\epsilon=0.0985$ along with the threshold line $T$ (red dashed line), where $T=25.5$. Irregular occurrence of large spikes in this time series is observed. (c) {\bf Three dimensional chaotic attractor after the crisis:} The chaotic attractor generating extreme events is depicted in this figure corresponding to (b). When the trajectory occasionally leaves the bounded region for the large excursions, extreme events are observed in time evolution of $T_2$ (in b). The position of saddle focus is located at $(76.43, 27.28, 20.33)$ (red circle). Parameter values: $r=\frac{1}{400}\rm{day}^{-1}$, $\alpha=\frac{1}{180}\rm{day}^{-1}$, $\mu=0.0026~ \rm{K}^{-1} \rm{day}^{-1}$, $\frac{\mu b L}{\beta}=22 \rm{m} \rm{K}^{-1}$, $T_{r}=29.5^0$ C, $T_{r0}=16^0$ C, $H=100$, $L=15\times10^6$ m, $\zeta=1.3$, $z_{0}=75$ m, $h_{*}=62$ m.} 
	\label{fig_4}
\end{figure}

\par Another type of interior crisis occurs, generally in slow-fast systems, where two bifurcation processes meet each other. A cascade of period-doubling bifurcation from one direction may cross a period-adding cascade of bifurcation at a critical parameter value from the opposite direction. Due to this crossing of two opposing bifurcation cascades, an enlargement of the chaotic attractor is observed, and such a transition is also termed as interior crisis\ \cite{fan1995crisis}. This process causes extreme events in a slow-fast dynamical system\ \cite{timmermann2003}, which describes the onset of El Ni{\~n}o-Southern Oscillation (ENSO) and this system is given by, 
\begin{equation}
\begin{array}{l}\label{eq.6}	
\dfrac{dh_1}{dt}=r\bigg(-h_{1}-\dfrac{bL\mu(T_{2}-T_{1})}{2\beta}\bigg),~~~
\dfrac{dT_1}{dt}=-\alpha(T_{1}-T_{r})-\epsilon\mu(T_{2}-T_{1})^2,\\

\dfrac{dT_2}{dt}=-\alpha(T_{2}-T_{r})+\zeta\mu(T_{2}-T_{1})\Bigg(T_{2}-T_{r}+\dfrac{1}{2}(T_{r}-T_{r0})\bigg[1-\tanh\frac{\Big(H+h_{1}+\frac{bL\mu(T_{2}-T_{1})}{\beta}-z_{0}\Big)}{h_*}\bigg]\Bigg),
\end{array}
\end{equation} 
where $h_{1}$, $T_{1}$, and $T_{2}$ represent the thermocline depth of the western Pacific, equatorial sea surface temperatures of the western and eastern Pacific, respectively. 
Interpretations of all parameters are given in Refs.\ \cite{timmermann2003,ray2020understanding}.

\par Figure\ \ref{fig_4}(a) locates a collision point (in a shaded box) of two advancing bifurcations (period-doubling and period-adding) from opposing directions of variation in the strength of zonal advection, $\epsilon$. A zoomed version of the shaded box depicts a sudden change in the size of the chaotic attractor at $\epsilon\approx0.09845$ in the inset of Fig.\ \ref{fig_4}(a). Near this critical value of $\epsilon$, occasional large amplitude oscillations are observed along with small amplitude oscillations, and the large amplitude events exceeding a predefined threshold are identified as extreme events (See Fig.\ \ref{fig_4}(b)). The threshold is determined using a mean-excess plot \cite{coles2001, ghosh2010discussion}. When the trajectory moves to a close vicinity of the saddle focus, it is repelled, spiraling away on the unstable manifold in a plane. Occasionally, the trajectory goes for a long excursion when it passes through a channel-like structure \cite{ansmann2016self}. The trajectory is reinjected into the saddle focus along its stable manifold, and the process is repeated (See Fig.\ \ref{fig_4}(c)). Due to this mechanism, extreme events emerge in this system\ \eqref{eq.6} as described in Ref.\ \cite{ray2020understanding}.

\par Such a type of interior crisis due to the collision of advancing period-doubling and period-adding cascades against a variation of a system parameter is observed in another diffusively coupled heterogeneous FitzHugh–Nagumo system (Bonhoeffer–van der Pol model) \cite{bonhoeffer1948activation, fitzhugh1961impulses, nagumo1962active} as reported by Ansmann et al.\ \cite{ansmann2013extreme}. 

\paragraph{External crisis-like process}
\par The crisis process initiates rogue waves in the form of extreme intensity pulses in an optically injected laser \ \cite{zamora2013rogue}. The mechanism of extreme events in the laser model follows external crisis-like process that is different compared to previously discussed mechanisms.

\par
\begin{figure}[H]
	\centerline{
		\includegraphics[scale=0.8]{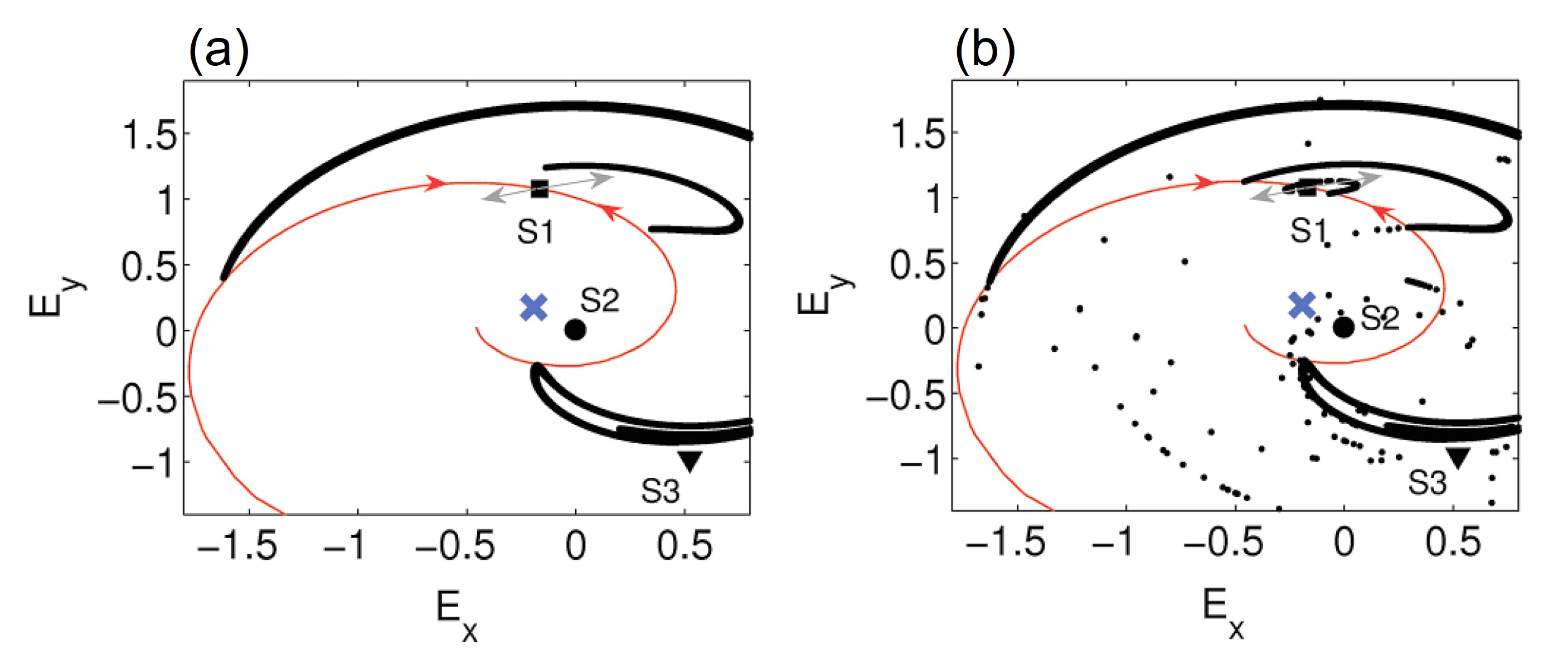}}
	\caption{\textbf{Poincar{\'e} surface of section of the attractor:} Poincar{\'e} surface of section (the plane at N = 1.0036) of the attractor is displayed here. (a) Pre-crisis state is described where the attractor (bold black dots) reaches close to the stable manifold (red curve) of saddle point, S1 (solid black square), but never goes beyond this. On the other hand, at (b) post-crisis scenario, the attractor collides with the stable manifold of saddle point (S1). After this collision, the trajectory enters the region of phase space of the stable manifold (blue cross) of unstable focus S2 (solid black dot). The trajectory moves along the stable manifold of S2 and moving away to form the high amplitude events (indicated by sporadic dots). S3 is an unstable focus due to which a chaotic attractor appears in the system. Parameters are $D=0.0$, $\alpha=3.0$, $P_{inj}=60$ n$s^{-2}$, $\nu =2.20$, $\gamma_n=1$ n$s^{-1}$, $\kappa=300$ n$s^{-1}$. Here, $\Delta \omega= 2 \pi \times 0.49$ GHz. Figure is adapted with permission from Ref.\ \cite{zamora2013rogue}.}
	\label{fig_2}
\end{figure}

 Zamora-Munt et al.\ \cite{zamora2013rogue} showed that external crisis-like process occurs in a continuous-wave optically injected laser, and initiates optical rogue waves in the form of extreme intensity pulses. They have considered the model representing the evolution of the slow envelope of the complex electric field $E (=E_x+i E_y)$ and carrier density $N$ \cite{wieczorek2005dynamical, ohtsubo2012semiconductor} as given by
\begin{equation}
\begin{array}{l}\label{eq.2}	
\dot{E}=\kappa(1+i\alpha)(N-1)E+i \Delta \omega E +\sqrt{P_{inj}}+\sqrt{D}\xi,
\\
\dot{N}=\gamma_{n}(\nu-N-N|E|^2),
\end{array}
\end{equation}
where $\kappa$ is the field decay rate, $\gamma_{n}$ denotes the carrier decay rate, $\alpha$ represents the line width enhancement factor, $\nu$ indicates the injection current, and $P_{inj}$ is the injection strength, $D$ is the noise strength, $\Delta \omega$ is the detuning between the lasers, and $\xi$ is the complex Gaussian white noise representing spontaneous emission. 
 Here, the observable $|E|$ is taken as $|E|=\sqrt{E_x^2+E_y^2}$. For this case study, the predefined threshold is chosen as $T=\langle |E|_{max} \rangle+8\sigma$ for defining extreme events. Here, the mechanism of crisis is different from the other described cases above. In this case, after a specific value of the critical bifurcation parameter, the trajectory of the chaotic attractor collides with the stable manifold of an unstable equilibrium point (say, S1). Thereby it reaches the vicinity of another equilibrium point (say, S2) along its stable manifold. Then the trajectory finally gets repulsion and traverses for a long excursion along the unstable manifold of that equilibrium point (S2). The mechanism of extreme events is delineated by using the Poincar{\'e} surface of section at a plane drawn in two-dimensional plane for describing the pre-crisis and post-crisis scenarios in Figs.\ \ref{fig_2}(a) and \ref{fig_2}(b), respectively.
\subsubsection{Pomeau-Manneville intermittency}\label{PM}
\par
\begin{figure}[H]
	\centerline{
		\includegraphics[scale=0.425]{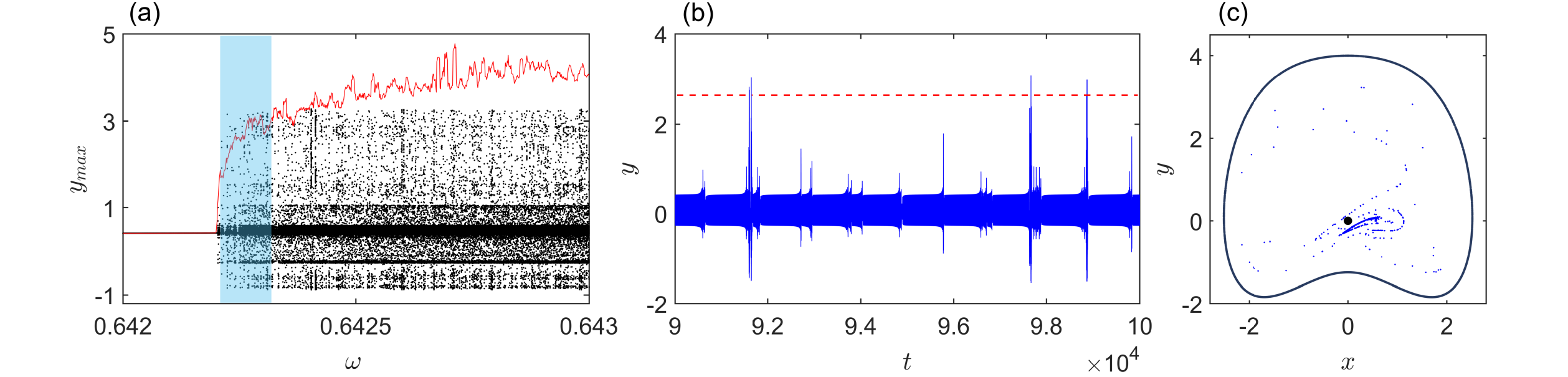}}
	\caption{(a) \textbf{PM intermittency and extreme events:} Bifurcation diagram of $y_{max}$ against $\omega$ displays a sudden transition from period-1 to chaos that occurs at $\omega \approx 0.6423$ via PM intermittency. A predefined threshold $T=\langle y_{max} \rangle + 8\sigma$ (red line) is plotted against $\omega$. It captures a range of $\omega$ (shaded region) for which extreme events emerge in the system. (b) \textbf{Temporal evolution of $y$:} A time series shows a laminar phase of almost periodic oscillation intercepted intermittently by large amplitude chaotic bursts for $\omega=0.6423$ as chosen from the shaded region. $T$ is denoted by a  horizontal line (dashed red line). (c) \textbf{Poincaré surface of section:} Poincaré surface of section of the system dynamics is plotted for $\omega=0.6423$ in the $xy$-plane. The trajectory reaches a close vicinity of the saddle orbit (black solid circle) around the $(0,0)$ and infrequently collides with the saddle orbit and repelled to a faraway distance (sparsely distributed scattered blue dots) that corresponds to the intermittency bursting episodes. 
		The sparsely distributed points (blue dots) are confined within the boundary of the quasiperiodic orbit (black cycle).  
		Other parameters are fixed at $\alpha=0.45$, $\beta=0.50$ and $\gamma=-0.50$.} 
	\label{fig_5}
\end{figure}

 Pomeau and Manneville first reported the intermittency \cite{pomeau1980intermittent} that causes a transition from a periodic state to a chaotic state of dynamical systems via saddle-node bifurcation at a critical value of system parameter. 
The time evolution of the system shows almost periodic oscillation (laminar phase) intercepted irregularly by chaotic bursts (turbulent phase). The chaotic bursts in many systems appear to trigger very large amplitude events that have been recognized as extreme events when quite a few of the large events are seen really larger than the significant threshold height $T$. 
Figure\ \ref{fig_5}(a) shows a plot of local maxima $y_{max}$ in a range of  forcing frequency $\omega\in [0.642,0.643]$, where sudden transition from a period-1 limit cycle to large amplitude oscillation occurs at a critical value of $\omega$. Beyond the parameter's transition value, the system's time evolution shows almost periodic oscillation (laminar phase) is intercepted irregularly by chaotic bursts (turbulent phase). The chaotic bursts in many systems appear to trigger occasional large amplitude oscillations as in Fig.\ \ref{fig_5}(b).
Such a route to extreme events has been shown by Leo Kingston et al.\ \cite{kingston2017extreme} in the forced Li{\'e}nard-type system \cite{chandrasekar2005unusual}. This system is described as
\begin{equation}
\begin{array}{l}\label{eq.3}	
\dot{x} =y,\\
\dot{y} =-\alpha xy-\gamma x-\beta x^3+F sin(\omega t),\\
\end{array}
\end{equation}
where $\alpha, \beta$ and $\gamma$ represent nonlinear damping,  strength of nonlinearity and intrinsic frequency, respectively. $F$ and $\omega$, respectively, are the amplitude and frequency of the external forcing. $\omega$ is varied to observe extreme events in this system. 
Note that, the origin is a saddle equilibrium point of the autonomous Li{\'e}nard system ($F=0$). But, a saddle orbit arises 
around the origin in the forced Li{\'e}nard system\ (\ref{eq.3}) \cite{Guckenheimer1983}. Near the critical transition, extreme events occur, and the range of extreme events (shaded region) is identified using a predefined threshold $T$ (marked by a red line). 
Figure\ \ref{fig_5}(c) displays a Poincar{\'e} surface of section of the attractor in $xy$-plane. Sparsely distributed points (blue dots) are observed in state space within a closed cycle (blue circle). Extreme events emerge in the system\ (\ref{eq.3}) because the chaotic attractor infrequently collides with the saddle orbit around the origin (black dot). The scattered points represent the large intermittent events as shown in Fig.\ \ref{fig_5}(b). 

\par Extreme events via this intermittency route have been reported in many other systems. It may appear from periodic or quasiperiodic motion \cite{kingston2021}. However, we should be aware that intermittency may not always result in extreme events. The extreme events are generated in a semiconductor laser with an external feedback \cite{lang1980external} due to intermittency, as reported in Ref.\ \cite{reinoso2013extreme}. Intermittency route to extreme events is also reported when two neurons interact via excitatory chemical synapses in a coupled system of two Hindmarsh-Rose neurons \cite{mishra2018dragon}.

\subsubsection{Sliding bifurcation near a  discontinuous boundary}\label{SB}
\par Extreme events may also occur in a system for a suitably chosen parameter space if a discontinuous boundary is embedded in the phase space of that system. Recently, Kumarasamy et al.\ \cite{kumarasamy2018extreme} explored that extreme events are originated in a forced micro-electro-mechanical system (MEMS) due to sliding bifurcation near a  discontinuous boundary. The dimensionless model of MEMS is represented by,

\begin{equation}
\begin{array}{l}\label{eq.11}	
\dot{x} =y,\\
\dot{y} =-\gamma y-x+ \frac{\beta^2}{(1-x)^2}+ \alpha cos(\omega t),\\
\end{array}
\end{equation}
where $\alpha$ and $\omega$ are the amplitude and frequency of an external forcing, respectively. $\gamma$ is the damping term, and $\beta$ is the strength of the nonlinear electrostatic actuation force. The variables $x$ and $y$ delineate displacement and electrostatic voltage \cite{liu2004simulation, evans2014laser}.

\par
\begin{figure}[H]
	\centerline{
		\includegraphics[scale=0.75]{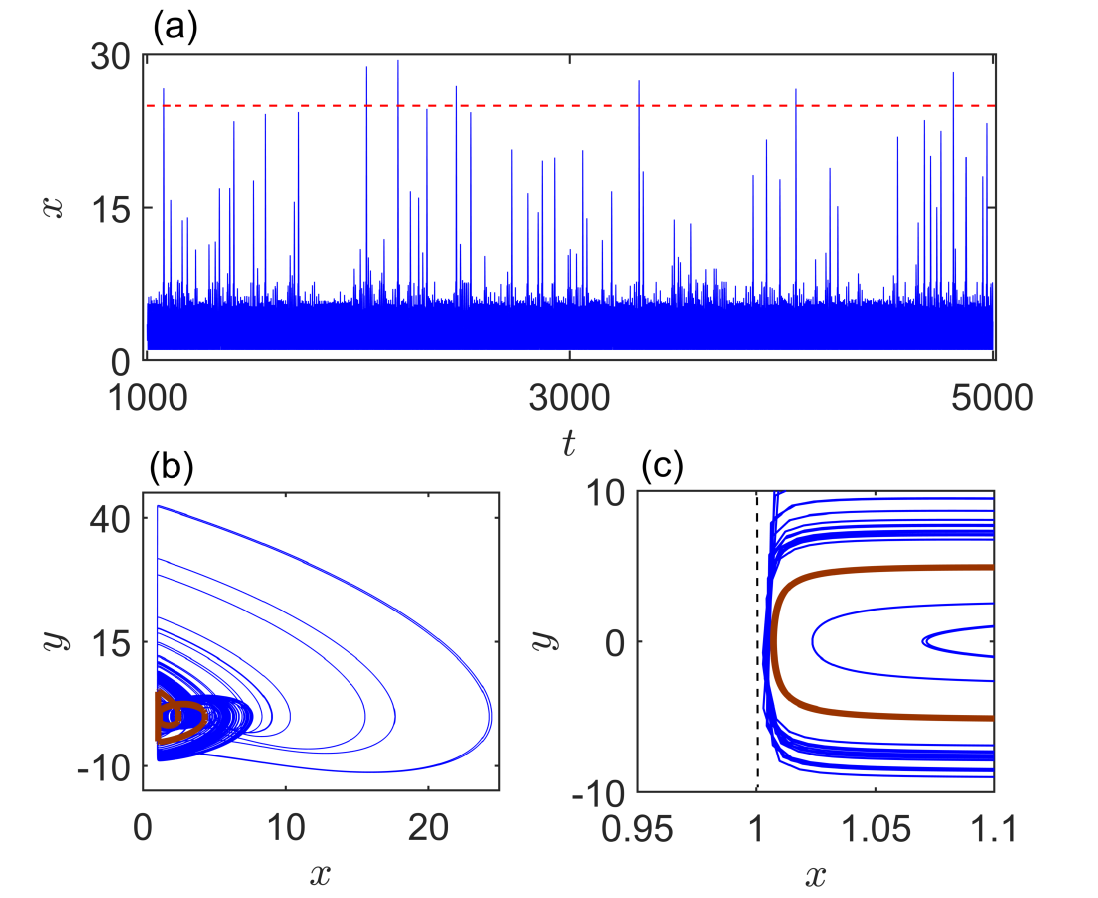}}
	\caption{(a) {\bf Time evolution of extreme events in MEMS:} Temporal evolution of $x$ for $\alpha=7.99$ and extreme event qualifier, $T=m+4\sigma$ (dashed horizontal red line) are plotted. The events (local maxima of $x$) exceeding $T$ are considered as extreme events. (b)-(c){\bf Phase portrait of the chaotic attractor:} (b) Two trajectories are drawn at $\alpha=7.99$ (blue) and $\alpha=5.0$ (brown). The trajectory for $\alpha=7.99$ tends to $x=1.0$ and is grazing along the discontinuous basin boundary before deflecting for a long excursion. This trajectory is able to exhibit extreme events in time evolution (See the subfigure (a)). 
	In contrast, the trajectory for $\alpha=5.0$ confines within a small bounded region, and there is no chance to exhibit extreme events. (c) An enhanced version of two trajectories near the discontinuous boundary $x=1.0$ (vertical dashed black line) is depicted here. The trajectory for $\alpha=5.0$ (brown) remains sufficiently far away from the $x=1.0$ line and hence fails to deviate for a large excursion in the phase space. On the other hand, the trajectory is deflected for a large excursion at $\alpha=7.99$, when it comes closer to the discontinuous boundary $x=1.0$. 
		Other parameters: $\gamma = 0.709$, $\beta = 0.318$, $\omega = 1.28$ .} 
	\label{fig_7}
\end{figure}

 This piece-wise smooth system possesses a switching manifold along with a point of singularity at $x=1.0$. The vector field becomes tangent to the switching manifold \cite{fang1999switching} and exhibits a sliding bifurcation \cite{jeffrey2011sliding}. The sliding trajectories travel a distance tangentially to the line $x=1.0$ and, finally, are repelled for a large expedition originating extreme events. At $\alpha=7.99$, the temporal dynamics of the displacement variable $x$ is shown in Fig.\ \ref{fig_7}(a). The occasional excursions of temporal dynamics become high so that local maxima of $x$ exceed the threshold $T$ (red dashed horizontal line) as shown in Fig.\ \ref{fig_7}(a). Two attractors for two different values of $\alpha$ (bifurcation parameter) are plotted for a comparison in Fig.\ \ref{fig_7}(b). The attractor (brown line) for $\alpha=5.0$ is confined within a small region far from the line $x=1.0$. However, for $\alpha=7.99$, the chaotic trajectory (blue line) comes very close to the line $x=1.0$ and is repelled. It makes occasional large excursions for a short duration so that extreme events may originate, as shown in Fig.\ \ref{fig_7}(a). Figure\ \ref{fig_7}(c) shows a closer view of the attractor near the singularity line $x=1.0$ (vertical dashed black line). The attractor for $\alpha=5.0$ remains far away from the line $x=1.0$, whereas the attractor for $\alpha=7.99$ comes close to that line. This line $x=1.0$ forms a discontinuous boundary that plays a leading role in the generation of extreme events in this system. By this process, the trajectory is repelled away to travel along the $y$-axis for originating extreme events. The trajectory of the system remains bounded within a dense (blue color) region for most of the time, however, occasionally makes large excursions for a short duration. The amplitude of extreme events is related to the  sliding distance of the trajectory along the $y$-axis. The sliding distance is a length that is traveled by the trajectory parallel to the line $x=1.0$. The amplitude of extreme events increases when the sliding trace along $y$-axis increases. 
 
\par Suresh et al.\ \cite{suresh2020parametric} investigated that extreme events also occur due to sliding bifurcation in the micro-electro-mechanical system under the influence of parametric excitation. The sliding bifurcation is also observed in a $CO_2$ laser model\ \eqref{eq.4} with a discontinuous boundary at $I=0$ \cite{kumarasamy2018extreme}. Physically, in this system, $I$ (which is proportional to the radiation density) cannot be negative, and thus, the system has a closed discontinuity at boundary $I=0$. As the trajectories approach the discontinuous boundary $I=0$, the system experiences a stick-slip bifurcation. This causes extreme events in the system for suitable choices of parameters. The system goes through the interior crisis, and before the crisis, the sliding bifurcation takes place.

\subsubsection{Noise-induced Intermittency} \label{noise}
\par Another kind of intermittency is discussed here that also originates occasional large events via noise-induced attractor-hopping in a multistable laser system that possess more than one coexisting stable states (steady state, oscillatory state, or both). Depending upon initial conditions, the trajectory converges to one of the coexisting stable states \cite{hens2015extreme,feudel2018multistability}. In the presence of noise, the trajectory of the multistable system may start hopping between the coexisting attractors \cite{feudel2008complex,kraut1999preference}. Such noise-induced intermittent attractor-hopping may lead to infrequent immense events. Pisarchik et al.\ \cite{pisarchik2011rogue} verified this type of intermittency in the erbium-doped fiber laser (EDFL) driven by harmonic pump modulation. This multistable system exhibits the appearance of extreme rogue waves due to the presence of noise, verified experimentally and numerically
\cite{pisarchik2011rogue,huerta2008experimental, pisarchik2012multistate}. The governing equations of the EDFL \cite{pisarchik2005dynamics} are given by,         
\begin{equation}
\begin{array}{l}\label{eq.8}	
\dot{P}=\dfrac{2L}{T_{r}}P(r_{w}\alpha_{0}[N(\xi_{1}-\xi_{2})-1]-\alpha_{th})+N\dfrac{10^{-3}}{\tau T_{r}}(\dfrac{\lambda_{g}}{w_{0}})^2\dfrac{r_{0}^2\alpha_{0}L}{4\pi^2\sigma_{12}},
\\
\\
\dot{N}=-\dfrac{\sigma_{12}r_{w}P}{\pi r_{0}^2}(N\xi_{1}-1)-\dfrac{N}{\tau}+P_{p}\dfrac{1-\exp[-\alpha_{0}\beta L(1-N)]}{N_{0}\pi r_{0}^2 L}.
\end{array}
\end{equation}
The intracavity laser power and averaged (over the active fiber length) population of the upper level are denoted by $P$ and $N$ $(0 \le N \le 1)$, respectively. The diode pump current at the fiber entrance is 
\begin{equation}
\begin{array}{l}\label{eq.10}	
P_{p}=p[1-m_{d}\sin(2\pi f_{d}t)+\eta G(\zeta, f_{n})].
\end{array}
\end{equation}
Here, $m_d$ is the external harmonic modulation, and $f_{d}$ is frequency of the external harmonic modulation. Additionally, noise amplitude is $\eta$, and random fluctuation $\zeta \in$ [-1, 1] with noise cutoff frequency $f_{n}$. Interpretations of all parameters are provided in Ref.\ \cite{pisarchik2011rogue}.

\begin{figure}[H]
	\centerline{
		\includegraphics[scale=0.23]{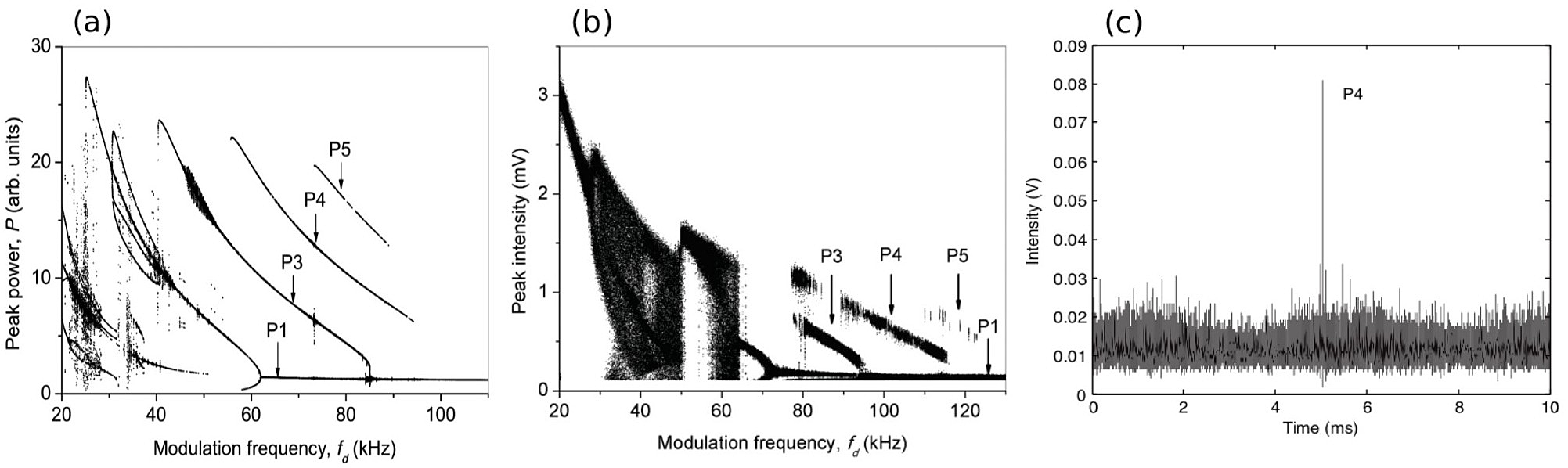}}
	\caption{(a) {\bf Numerical bifurcation diagram of EDFL without noise:} Laser peak power is displayed varying with external harmonic modulation $(f_{d})$ where $m_d=1$. The coexistence of period-1 (P1), period-3 (P3), period-4 (P4), and period-5 (P5) attractors are observed in a range of bifurcation parameter $f_{d}$. (b) {\bf Experimental bifurcation diagram in absence of external noise:} Experimental bifurcation diagram also exhibits similar dynamical features compatible to the  numerical bifurcation diagram where $m_d=0.8$ V. The coexisting periodic attractors (P1, P3, P4, P5) are
		found by switching on and off the signal generator. The attractor branches in this bifurcation diagram are shifted possibly due to presence of experimental noise. (c) {\bf Experimental time series:} A large intensity of the laser is observed. 
		This large event is considered as an extreme event. $m_{d}= 0.8$ V, $f_d=90$ kHz, $\zeta=0.5$ V are fixed. Other parameters are $L=88$ cm, $\alpha_{0}=0.4\rm{cm}^{-1}$, $\xi_{1}=2$, $\xi_{2}=0.4$, $\alpha_{th}=3.92 \times 10^2$, $\tau=10^{-2}$, $r_{0}=1.35\times10^{-4}$ cm, $\lambda_{g}=1.56\times10^{-4}$ cm, $f_n=7$kHz and $\beta=0.5$. 
		Reprinted figure with
		permission from Refs.\ \cite{pisarchik2011rogue,pisarchik2012multistate}.
	}
	\label{fig_6}
\end{figure}

\par A numerical bifurcation diagram and experimental bifurcation diagram are displayed in Figs.\ \ref{fig_6}(a) and \ref{fig_6}(b), respectively, for showing the multistability feature of the system\ \eqref{eq.8}. Here the peaks of laser intensity are plotted against the variation of frequency of the external harmonic modulation ($f_{d}$). Clearly, for different parameter values of $f_{d}$, coexisting periodic orbits of period-n $(n=1,3,4,5)$ are observed for different initial conditions. Now, if noise with an optimal intensity ($\zeta$) is induced in the system \eqref{eq.8}, a new kind of attractor appears in the system due to the loss of stability of the coexisting attractors. The trajectory starts infrequently switching between two coexisting attractors. As a result, the system exhibits intermittent behavior with respect to time, leading to occasional high intensity pulses. This kind of attractor hopping is noticed in the experimental time series, drawn for $f_{d}=90$~kHz as shown in Fig.\ \ref{fig_6} (c). In the presence of noise, the trajectory of this laser system occasionally transits to period-4 (P4) orbit for short time duration and then back to the period-1 orbit. This occasional transition to P4 orbit creates extreme intensity pulses(extreme events).

\subsection{Two coupled dynamical systems}\label{two}
\par Besides isolated dynamical systems, the researchers are equally interested in interpreting the underlying mechanism of the emergence of extreme events in the interacting oscillators. From this point of view, few studies are performed both numerically and experimentally. In the next section, we discuss the results on two coupled dynamical systems.

\subsubsection{On-off intermittency}\label{onoff}
\par
\begin{figure}[H]
	\centerline{
		\includegraphics[scale=0.5]{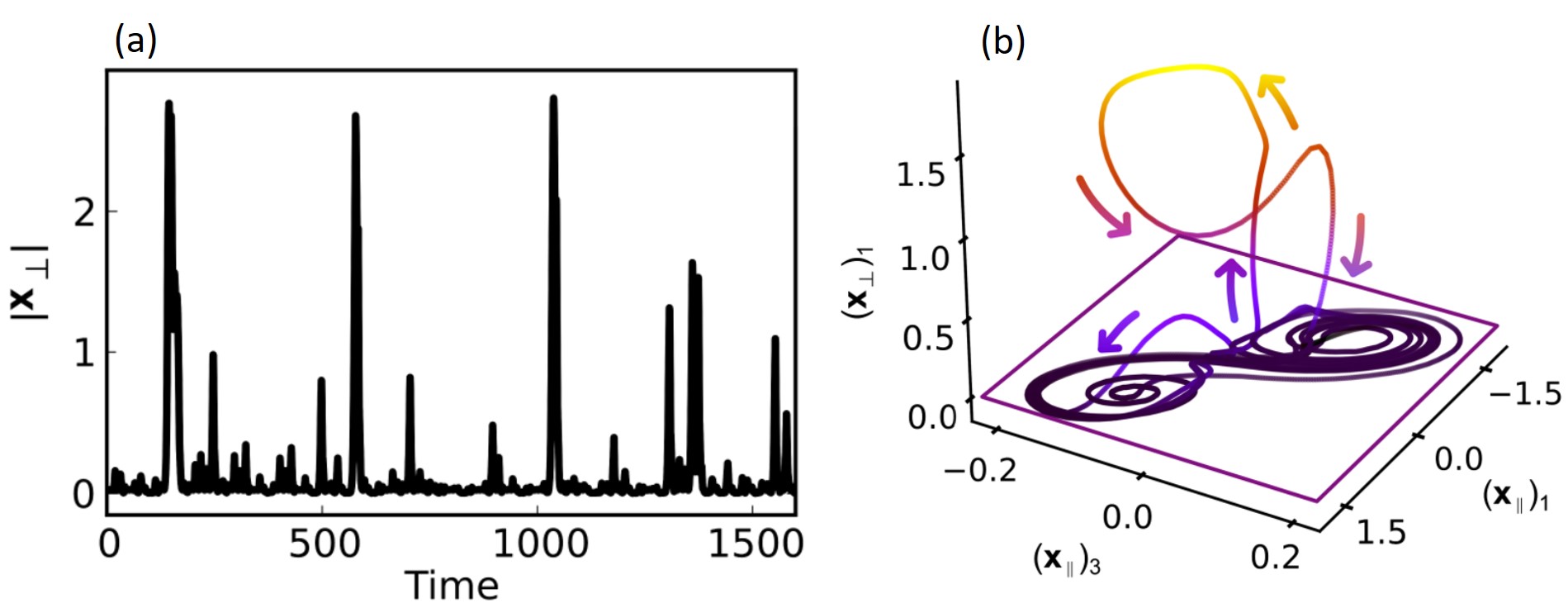}}
	\caption{(a) {\bf Loss of synchrony through On-off intermittency in coupled system:} Temporal evolution of $|\bf{x_{\perp}}|$ is the observable that exhibits occasional large events. The large events are extreme events. (b) {\bf $3$-D projection of $6$-D phase space:}  The system trajectory in the vicinity of a bubbling event is presented. Chaotic attractor lies on the invariant manifold  $(x_{\parallel})_1 (x_{\parallel})_3$-plane most of the time, and a bubbling event along the transverse direction of the manifold ($(x_{\perp})_1$) is shown here. The subscript indicates the $i$-th component of the corresponding vector. 
		Reprinted figure with permission from Ref.\ \cite{cavalcante2013predictability}.} 
	\label{fig_8}
\end{figure}

 In general, complete synchronization is the most desired behavior in coupled systems. This desired state is disrupted in the presence of noise, heterogeneity in the system, or both. A common scenario is noticed that the trajectory of the coupled system occasionally departs from an invariant manifold (the synchronization manifold) and jumps along the transverse direction of that manifold. As a result, occasional bursts with varying amplitudes are observed in the time evolution of the synchronization error. The trajectory, being repelled along the transverse direction of the synchronization manifold, ultimately comes back to the invariant manifold after a brief time interval. These short-lived, and intermittent excursions away from the synchronization manifold is known as \textit{attractor bubbling} \cite{boccaletti2002synchronization,ashwin1994bubbling,heagy1994characterization,heagy1995desynchronization, ashwin1996attractor, platt1993off, pradas2011noise, venkataramani1996bubbling, venkataramani1996transitions,gilson2013effect}. In such a situation, the synchronization error dynamics switch from zero to non-zero values intermittently. This phenomenon is also known as on-off intermittency of the error dynamics. This attractor bubbling is manifested as \textit{dragon king} (DK) events \cite{sornette2012dragon, wheatley2015multiple, premraj2021dragon}, which are highly informative outliers to a power-law distribution. This observation is delineated as extreme events in Refs.\ \cite{cavalcante2013predictability, motter2013control}. For demonstration, two unidirectionally coupled dynamical systems are taken in the form as,
	\begin{equation}
	\begin{array}{l}\label{eq.12}	
	\bf{\dot{x}_{M}}={\bf{F[x_{M}]}},
	\\
	{\bf{\dot{x}_{S}}}={\bf{F[x_{S}]}}+ c \bf{K}(\bf{x_{M}}-\bf{x_{S}}),
	\end{array}
	\end{equation}
	where  $\bf{x_M}$ and $\bf{x_S}$ are the state variables of the master and slave systems, respectively, and $c$ is the coupling strength, $\bf{K}$ is the coupling matrix. 	
	 In a state of synchrony, both the subsystems evolve in unison on an invariant synchronization manifold $\bf{x_{M}}$ =$\bf{x_{S}}$. Two new variables are introduced as $\bf{x_{\parallel}}  =(\frac{\bf{x_{M}}+\bf{x_{S}}}{2})$ and $\bf{x_{\perp}} =(\frac{\bf{x_{M}}-\bf{x_{S}}}{2})$ that describe the evolution within and transverse to the synchronized manifold. In a stable synchronized state, $\bf{x_{\parallel}}=\bf{x_{M}}=\bf{x_{S}}$ and
	$\bf{x_{\perp}}=0$. Cavalcante et al.\ \cite{cavalcante2013predictability} considered a simple, yet non-trivial pair of three dimensional electronic systems. For a certain choice of parameters, $|\bf{x_{\perp}}|$ remains zero for most of the time, unless it is interrupted by few aperiodic chaotic bursts. The expedition of the state of the system in the phase space from the invariant manifold $\bf{x_{\parallel}}$ is reported as extreme events. The temporal evolution of $|\bf{x_{\perp}}|$ is represented in Fig.\ \ref{fig_8}(a).  The trajectory eventually returns back to the invariant synchronization manifold $\bf{x_{\parallel}}$, due to the nonlinear folding of the flow. The trajectory is illustrated through the projection of the $6$D phase space onto a $3$D subspace containing components $(\bf{x_{\parallel}})_1$ and $(\bf{x_{\parallel}})_3$ of the invariant manifold and $(\bf{x_{\perp}})_1$ of the transverse manifold in the Fig.\ \ref{fig_8}(b). There is a probable scenario that a bubbling  may take place, when $\bf{x_{\parallel}}$ visits a neighborhood of the origin, as ${\bf{x_{\parallel}}}=0$ is an unstable saddle-like fixed point of the coupled oscillators. Oliveira et al.\ \cite{de2016local} also noticed later a similar mechanism of the formation of DKs in coupled chaotic electronic circuits\ \cite{de2014tunable,de2015trajectory}.

\subsubsection{Imperfect phase synchronization}\label{imperfect}
\par
\begin{figure}[H]
	\centerline{
		\includegraphics[scale=0.65]{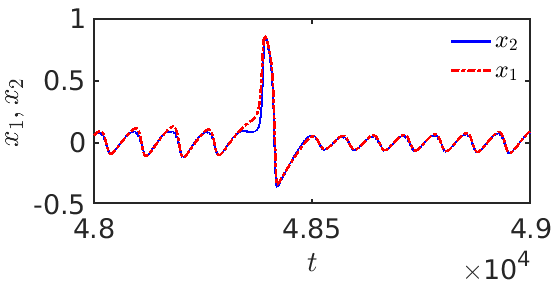}}
	\caption{{\bf Chaotic time series of $x_1$ and $x_2$:} Two time evolution of $x_1$ (blue) and $x_2$ (red) of two oscillators are shown. The trajectories of each unit are phase-synchronized for most of the time. But, the trajectories become desynchronized when those units become excited and hence a phase slip occurs with unit $1$ leading unit $2$ by $2\pi$. Parameters: $a = -0.025794, c =0.02, b_1 = 0.0135, b_2 = 0.0065,$ and $k=0.128$.} 
	\label{fig_9}
\end{figure}

 Besides the process mentioned above, extreme events may also be generated due to instability of phase synchrony in dynamical systems. Such type of mechanism has been reported by Ansmann et al.\ \cite{ansmann2013extreme} for two diffusively coupled heterogeneous FitzHugh–Nagumo (FHN) systems (Bonhoeffer–van der Pol model) \cite{bonhoeffer1948activation, fitzhugh1961impulses, nagumo1962active}. 
This coupled system is described by 
\begin{equation}
\begin{array}{lcl}\label{eq.5}
\dot{x_{i}} & =& x_{i}(a - x_{i})(x_{i}-1) -y_{i}+\dfrac{k}{N-1}\sum_{j=1}^{N}A_{ij}(x_{j}-x_{i}),\\
\dot{y_{i}} &=& b_{i} x_{i} - c y_{i},
\end{array}
\end{equation}
where $A=[A_{ij}]_{N\times N}$ is the adjacency matrix defining the topology, $k$ is the coupling strength, and $N=2$. Here, heterogeneity is introduced in the parameter $b$. $\bar{x}=\frac{x_{1}+x_{2}}{2}$ is considered as an observable. The coupled units are phase synchronized most of the time, when the oscillations are small in amplitude. But occasional desynchronization between two oscillators occurs, when both become excited. The temporal evolution of $x_1$ (dashed red line) and $x_2$ (solid blue line) are plotted in Fig.\ \ref{fig_9} which reveals the above-mentioned scenario. Basically, an imperfect phase synchronization \cite{zaks1999alternating, park1999phase} happens between the FHN oscillators. As a result, extreme events (as large spikes) arise in the temporal dynamics of $\bar{x}$ due to cohesion of large amplitude oscillation. A phase slip occurs just before an extreme event during desynchrony. Besides, a small channel-like structure is found in the state-space, which exists due to the alignment of the manifolds of the saddle focus at the origin. This alignment creates a gap in the phase space through which the bounded chaotic trajectories escape recurrently, but rarely. The trajectories enter aperiodically through this channel-like structure, causing high amplitude extreme events, as per their definition. The route to the emergence of extreme events is also reported as the interior crisis that occurs due to collision of period-doubling and period-adding cascades against a variation of a system parameter ($k$). We have discussed this collision of two bifurcation processes in Sec.\ \eqref{papd}.

\subsubsection{Instability of out-of-phase synchronization}\label{Reflection of extreme events through occasional in-phase synchronization} 
\par Now, we discuss about the generation of extreme events that occur due to occasional in-phase synchrony in two coupled neuronal models. This kind of formation of extreme events is quite different from others. For demonstration, Mishra et al.\ \cite{mishra2018dragon} considered two identical Hindmarsh-Rose systems \cite{hindmarsh1984model} coupled by bidirectional chemical synaptic interactions. The coupled system
is represented as
\begin{equation}
\begin{array}{lcl}\label{eq.7}
\dot{x}_j = -ax_{j}^3 + bx_{j}^2 + y_{j} - z_{j} + I -k_{j}(x_{j} - v_{s})\Gamma(x_{j}), \\
\dot{y}_j = c - dx_{j}^2 - y_{j},\\
\dot{z}_j = r[s(x_{j} - x_R) - z_j],
\end{array}
\end{equation}
where $j$=1,2. The chemical synaptic coupling function for the $j$-th neuron is defined by $\Gamma(x_j)=\dfrac{1}{\rm{e}^{-\lambda (x_{j}-\Theta)}}$ \cite{belykh2005synchronization}. Before coupling, two identical neurons show periodic bursting. The type of mutual interaction is contemplated in such a way that $k_{1,2} < 0$ and it creates a repulsive or inhibitory interaction between the coupled neurons. As a result, recurrent large amplitude events are observed in the temporal evolution of an observable $x_{\parallel}=x_1+x_2$. In this case study, $k_1=k_2=k$.

\par
\begin{figure}[H]
	\centerline{
		\includegraphics[scale=0.7]{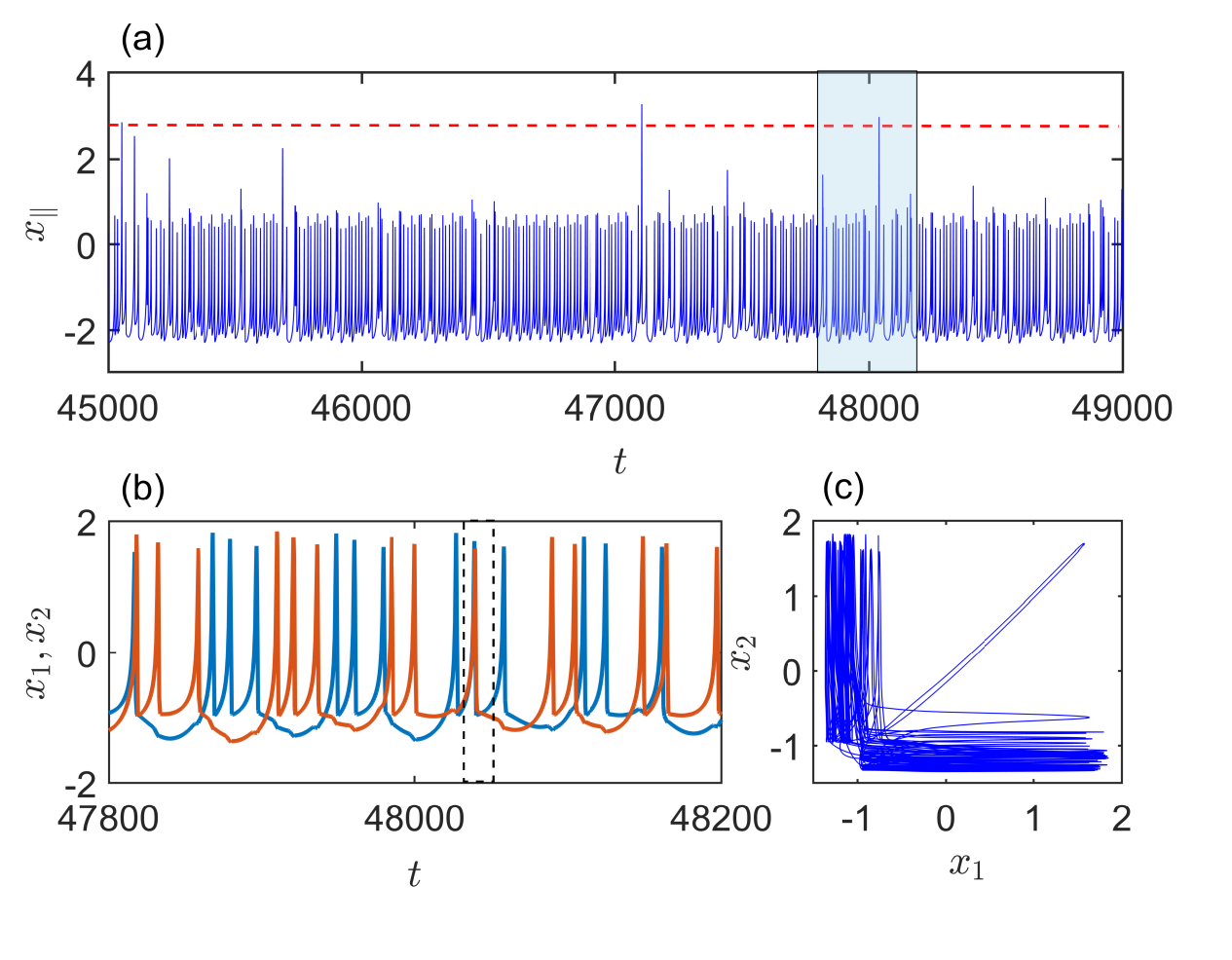}}
	\caption{(a) {\bf Temporal dynamics of $x_{\parallel}$}: A time series of $x_{\parallel}=x_1+x_2$ is plotted for $k=-0.1$. Few large spikes are observed, and these cross the threshold line $T$ (red dashed horizontal line). Now, a fragment of this time series (shaded box region) is selected consisting of an extreme event. Now, we focus on that shaded region in (b). (b) {\bf Temporal dynamics of $x_1$ and $x_2$}: Here, the time evolution of both variables exhibit bursting state as well as quiescent state. Occasionally two spikes of those time series almost overlap, and as a result, large spikes appear in the observable $x_{\parallel}$ (dashed box corresponding to the shaded box in (a)). (c) {\bf Phase portrait of $x_1$ vs.\ $x_2$:} Here, an occasional in-phase synchrony is captured here, when two trajectories move out along the in-phase direction. As a result, extreme event is perceived in the box region of (a). For numerical simulations, $a=1, b=3, c=1, d=5, x_{R}=-1.6, r=0.01, s=5, I=4, v_{s}=2, \lambda=10,$ and $ \Theta=-0.25$ is taken.} 
	\label{fig_10}
\end{figure}

 The origin of extreme events is regarded as the instability in the phase space due to a saddle point. A time evolution of $x_{\parallel}$ consisting extreme events and $T=\langle x_{\parallel} \rangle+6\sigma$ are plotted in Fig.\ \ref{fig_10}(a) for $k=-0.1$. A particular time interval (shaded box region) of this figure is highlighted to understand the origin of extreme events better. The temporal dynamics of two interacting neurons $x_1$ (blue line) and $x_2$ (red line) are plotted in Fig.\ \ref{fig_10}(b). These two trajectories remain out-of-phase most of the time. 
However, individual spikes of $x_1$ and $x_2$ occasionally come close to in-phase synchronization manifold (marked within a box in Fig.\ \ref{fig_10}(b)). This is further confirmed from a phase portrait $x_1$ vs.\ $x_2$ plot in Fig.\ \ref{fig_10}(c). It shows a large deflection of the trajectory along the in-phase direction. This rare and occasional transition to in-phase synchrony higher than the threshold $T$ manifests as an extreme event that repeats in the long run. Interestingly, such occasional excursions of the observable along the in-phase direction also form the dragon-king distribution. This kind of distribution is also reported earlier in Ref.\ \cite{cavalcante2013predictability}.


\subsubsection{In-out intermittency}\label{inout} 
\par {\it In-out intermittency} \cite{ashwin1999transverse,blackbeard2014synchronisation,covas2001out,saha2018characteristics} is a generalization of on-off intermittency \cite{ashwin1996attractor,platt1993off,pikovsky1984interaction}. In case of in-out intermittency, the trajectory of the coupled system gets attracted along the transversally stable direction of the  invariant sets, and spends in the neighborhood of the transversally attracting part of the strange attractor for a long time before coming closer to the neighborhood of a different invariant set, which is transversally unstable. Thus, ultimately it is repelling away along the unstable direction of the invariant manifold. In the case of on-off intermittency, the same invariant set plays the crucial role in both getting attracted towards and getting ejected away from the invariant manifold. Note that, in-out intermittency reduces to on-off intermittency if the system has a skew-product structure \cite{saha2018characteristics}. In-out intermittency plays a role to generate extreme events in a delay-coupled slow-fast system \cite{saha2017extreme}. Two identical FitzHugh-Nagumo oscillators coupled with two time delays can produce extreme events under a suitable choice of initial conditions \cite{saha2018riddled}, and appropriate delay coupling strength \cite{saha2017extreme}. The coupled system with time-delays is written as 
\begin{equation}
\begin{array}{lcl}\label{eq.30}
\dot{x}_{i} & =& x_{i}(a - x_{i})(x_{i}-1) -y_{i}+M_1 (x_{j}^{(\tau_1)}-x_{i})+M_2 (x_{j}^{(\tau_2)}-x_{i}),\\
\dot{y}_{i} &=& b x_{i} - c y_{i}+M_1 (y_{j}^{(\tau_1)}-y_{i})+M_2 (y_{j}^{(\tau_2)}-y_{i}),
\end{array}
\end{equation}

where $i \neq j$ and $i,j=1,2$. The parameter values are taken as $a=-0.025$, $b=0.00652$ and $c=0.02$. If $M_{2}=0$, then the only effective coupling strength is $M_1$ with single delay $\tau_1$. Here, $x_{j}^{(\tau_k)}=x_{j} (t-\tau_k)$ and $y_{j}^{(\tau_k)}=y_{j} (t-\tau_k)$, where $k=1,2$. The synchronization manifold $x_{1}^{(\tau)}=x_{2}^{(\tau)}$ and $y_{1}^{(\tau)}=y_{2}^{(\tau)}$ for all $\tau \in [0,max\{\tau_{k}\}]$ is the only stable attractor for small $M=M_1+M_2$. The period-adding cascade of mixed-mode oscillations situated at the synchronization manifold meets the period-doubling cascade of the limit cycle, causing extreme events. During the collision of both the cascades, extreme events occur for several suitable choices of parameters. The extreme events are almost equal in size, but appear irregularly in the time domain. For $M_1=0.005$ and $M_2 < 0.0048$, the invariant synchronization manifold is transversally stable. At $M_2 \approx 0.0048$, one or many periodic orbits in the synchronization manifold lose their transverse stability due to bubbling transition \cite{ashwin1994bubbling,ashwin1996attractor,venkataramani1996bubbling,venkataramani1996transitions}. These transversally unstable periodic orbits are responsible for repelling the trajectories away from the invariant synchronization manifold. However, a set of measure zero in the form of a saddle point resides in the synchronization manifold at the origin. Along the stable manifold of this unstable origin, the trajectories approach towards the synchronization manifold. Thus, the system possesses at least two distinct invariant sets out of which the origin is responsible for the ``in" dynamics, and the unstable periodic orbits correspond to the ``outward" repulsion from the manifold. Note that in-out intermittency is a transient phenomenon involving chaotic dynamics for a suitable choice of parameters because the trajectory finally converges to the chaotic attractor in the long-time dynamics. After the long transients, the trajectories execute chaotic synchronous large amplitude oscillations manifested as extreme events. The regime of in-out intermittency occurs between a bubbling transition and a blowout bifurcation \cite{ott1994blowout,ashwin1998unfolding}. At $M_2 \approx 0.0058$, the synchronization manifold loses its transverse stability due to blowout bifurcation, and the chaotic saddle outside of this manifold becomes the attractor. Only chaotic trajectories are observed after the blowout bifurcation consisting of out-of-phase large amplitude oscillations separated by small amplitude oscillations. Besides, the phase space of the system is quite complex, exhibiting a riddled basin of attraction \cite{saha2018riddled}. The state-space can be divided into two regions: (i) the ``pure" region, which is unable to produce extreme events, and (ii) the ``mixed" region, where extreme events may occur. This mixed region is very sensitive, since a small perturbation in the initial conditions can change the dynamics from the one that exhibits extreme events to the one that does not produce extreme events.

\subsection{Dynamical networks}\label{network}
Now, in this section, we discuss the progress of research on extreme events in networks \cite{barabasi2016network, newman2011networks} of coupled oscillators. A dynamical network consists of dynamical systems as nodes, which are connected by links or edges.
Based on network connectivity, two types of networks are observed, namely, (i) static networks, and (ii) time-varying networks. The network connectivity or the adjacency matrix remains invariant in time for static networks. In many real situations, the links that form a network's topology are time-varying. Recently, extreme events are observed in these two types of networks, and we will discuss them one by one.

\subsubsection{Static networks}
\par Few case studies have been performed regarding the generation of extreme events in static networks. Generally, heterogeneity in the nodal dynamics plays an important role in generating extreme events in such networks. In the following, we present two examples to demonstrate the formation of extreme events due to an interplay between the parameter mismatch and the nature of coupling, namely, repulsive and attractive interactions. 

\paragraph{Repulsive interaction}
\par  Extreme events are observed in a globally coupled network of Josephson junctions. It has been reported in Ref.\ \cite{ray2020extreme} that if any system possesses different kinds of oscillatory dynamics (like libration and rotation or pre-crisis and post-crisis), then under repulsive coupling, globally coupled oscillators can generate intermittent behavior that signifies extreme events during the transition between the two types of oscillations.

\begin{figure}[H]
	\centerline{
		\includegraphics[scale=0.55]{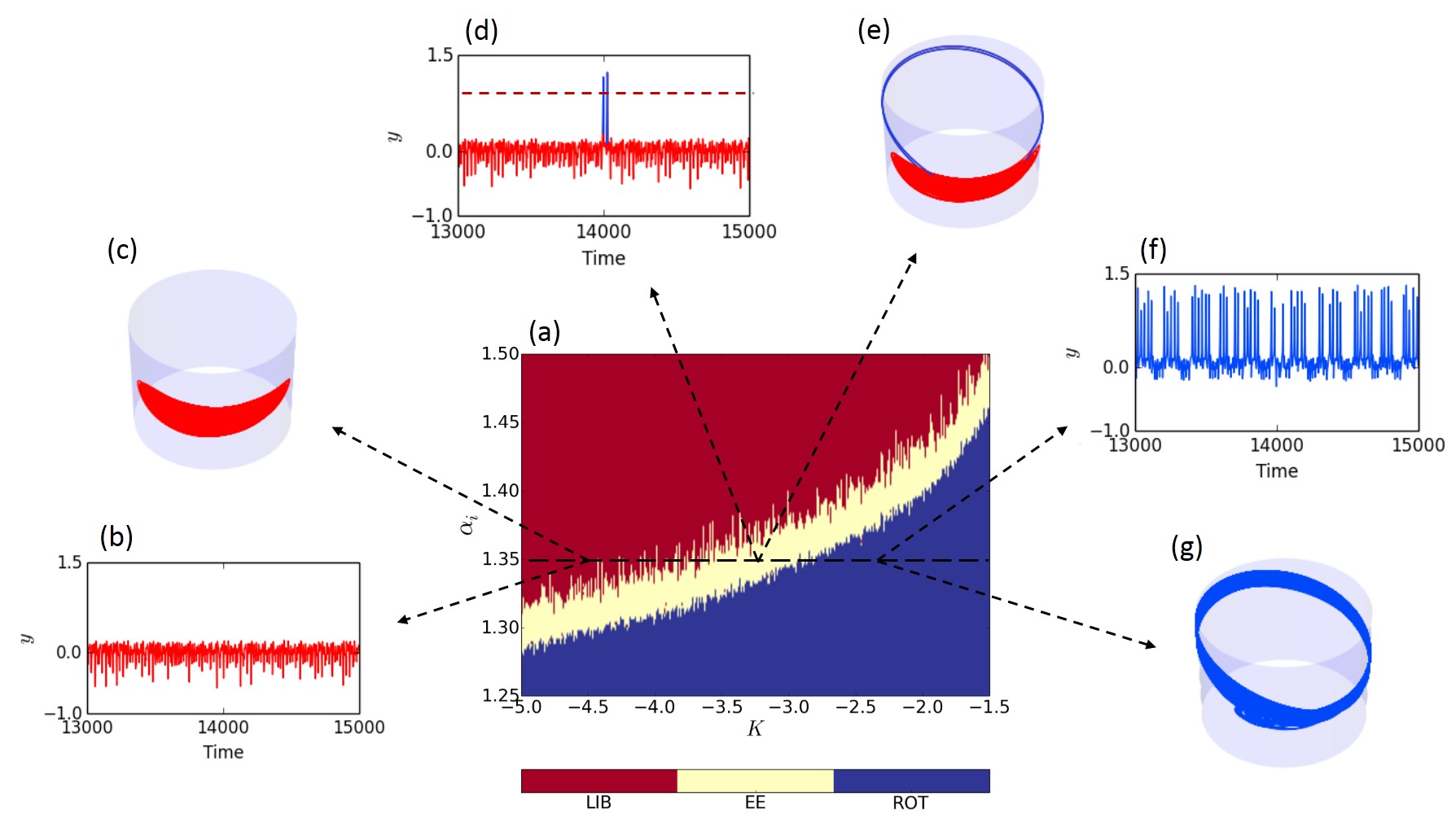}}
	\caption{(a) {\bf Transition of oscillation against $K$ in a complete  graph of dynamical units:} The vertical axis depicts $\alpha_i$ of each $i$-th node of the network, which are arranged in ascending order of values of $\alpha_i ~(\in [1.25,1.5])$ from bottom to top. Three subpopulations emerge with distinct collective dynamics, {\it i.e.,} (i) libration (red region), (ii) extreme events (yellow region) and (ii) rotation (blue region). A transition from rotational to librational motion with an intermediate range of extreme events is observed.
		(b, d, f) {\bf Temporal evolution of $y$ for $\alpha=1.35$:}
		Three regions (red, yellow and blue) show distinct types of oscillations, (b) libration (red line), (d) extreme events, small amplitude oscillation (red line) with occasional large spikes (blue line). Exemplary large events exceed a threshold ($T=\langle y_{max} \rangle+8\sigma$) (horizontal red dashed line) and, (f) rotation (blue line). (c, e, g) {\bf Phase portrait of the dynamics on a cylindrical surface:} Phase spaces corresponding to the figures (b), (d), and (f) are displayed here.  The phase portrait is wrapped onto the surface of a cylinder by considering the unit radius ($r$) of it, {\it i.e.}, $r=1$. The position of the trajectory is located on a 3-dimensional plane $(X, Y, Z)$, where $X=rcos\phi$, $Y=rsin\phi$, $Z=y$. 
		A horizontal color bar at the bottom depicts three different states, (i) libration (LIB in red), 
		(ii) extreme events (EE in yellow), (iii) rotation (ROT in blue). Parameter description: (b, c) $K=-4.5$ (d, e) $K=-3.3$, and (f, g) $K=-2.4$. $\alpha=1.35$ for subfigure (b) to (g).  Other parameters: $N=200, i_{dc} = 1.2, i_{rf} = 0.26$ and $ \Omega_{rf} = 0.4$. Reprinted figure from Ref.\ \cite{ray2020extreme}} 
	\label{fig_11}
\end{figure}

\par  An isolated Josephson junction \cite{strogatz2016nonlinear} possesses two types of oscillations: (i) libration and (ii) rotation, depending on internal parameter values. A complete graph of $N$ nodes is considered for the study where each node is represented by the superconducting resistive-capacitive-shunted junction (RCSJ) \cite{levi1978dynamics, dana2001chaotic}. The dynamics of the $i$-th node of the heterogeneous network of RCSJ array is given by,
\begin{equation}
\begin{array}{lcl}\label{eq.15}
\dot{\phi_{i}} & =& y_{i},\\
\dot{y_{i}} &=& i_{dc} - \sin \phi_{i} - \alpha_{i} y_{i} + i_{rf}\sin (\Omega_{rf}t) + KY,
\end{array}
\end{equation}
where  $Y$= $\frac{1}{N} \sum\limits_{j=1}^N  {y_j}$, and $\alpha$=$[h/4 \pi e I_{C} R^2 C]^{1/2}$$= (\frac{1}{\beta})^{\frac{1}{2}}$ is the damping parameter. For each $i$-th node, damping parameter is denoted by $\alpha_i$, and taken as $\alpha_i=1.1+0.002(i-1)$, for $i=1,2,\cdots,N$. $\beta, R$ and $C$  denote McCumber parameter, intrinsic resistance, and capacitance of a junction, respectively. $i_{rf}$ and $\Omega_{rf}$  are the normalized amplitude and frequency of a radio-frequency ($rf$) forcing signal, respectively.  $i_{dc}$ is a constant bias current normalized by the critical junction current $I_{C}$. $K$ defines the coupling strength of the mean-field interaction between the junctions. Two conditions are imposed in the network, (i) heterogeneity in the parameter $\alpha$ of the oscillators, and (ii) a repulsive global mean-field interaction, {\it i.e.}, $K<0$.
\par For selected parameter values, an isolated junction $(K=0)$ exhibits rotational motion like an inverted pendulum \cite{strogatz2016nonlinear}, however, it may transit to libration like a  simple pendulum motion \cite{mishra2017coherent} under repulsive interaction  ($K<0$). In Fig.\ \ref{fig_11}(a), three distinct subpopulations of junctions with changing of repulsive interaction are indicated by colored regions, {\it i.e.}, (i) librational motion or small amplitude oscillation (red), (ii) extreme events (yellow), and (iii) rotational motion or large amplitude oscillation (blue).
For illustrations of three kinds of oscillatory behavior, one single node (for $\alpha= 1.35$) is picked up (denoted by
a dashed horizontal line) and its dynamics (time evolution of y and phase-space in a cylindrical surface)
are demonstrated. Three kinds of qualitatively different features are exhibited through the temporal dynamics as shown in Figs.\ \ref{fig_11}(b), (d), and (f), respectively. Corresponding trajectories of the dynamics are demonstrated in a cylindrical plane in Figs.\ \ref{fig_11}(c), (e), and (g), respectively. The time evolution of $y$ in Fig.\ \ref {fig_11}(d) shows small amplitude librational motion that persists for a long time. But it is interrupted by a large amplitude oscillation, and the extreme events appear when maxima of $y$ ($y_{max}$) exceed a predefined threshold $T$. Extreme events originate due to the interplay between heterogeneity of parameters of the junctions and the repulsive interaction. This feature is also verified in a complete graph of the heterogeneous Li{\'e}nard systems. This phenomenon is not restricted to a complete graph of Josephson junctions only, but also observed in a network of Li\'enard systems under the same conditions of heterogeneity in parameters and repulsive interaction between the nodes. 

\paragraph{Attractive interaction}
\par
\begin{figure}[H]
	\centerline{
		\includegraphics[scale=0.65]{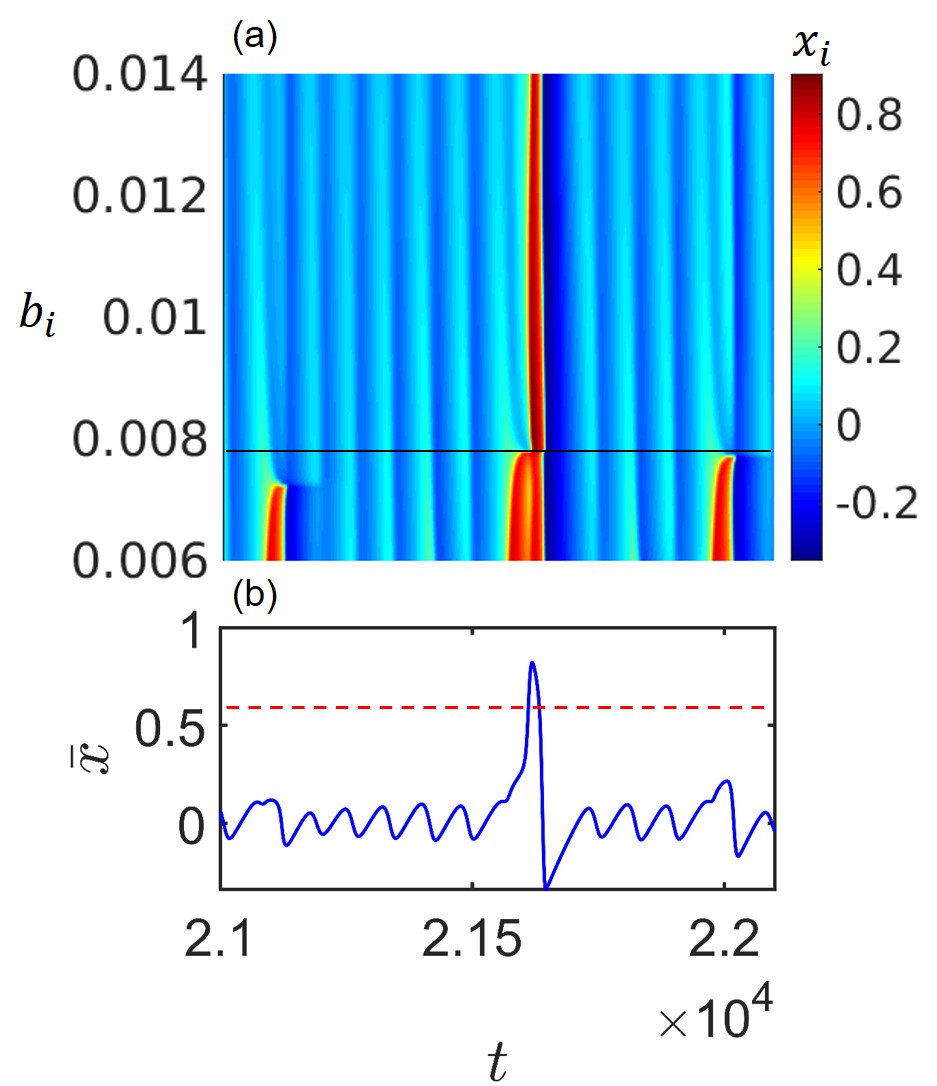}}
	\caption{ {\bf Temporal evolution of $x_i$ for each unit and $\bar{x}$:} In a particular time interval, the evolution of each $x_{i}$ (for $i$-th node) is displayed in the upper panel where $i=1,2,\cdots,101$. The color bar indicates the values of $x_i$. Nodes are kept in ascending order of values of $b_i$ from bottom to top. In (a), red stripes present the excited FHN units that $\bar{x}>0.6$. For the first red stripe, the number of proto-events cannot exceed a critical number (namely, critical mass) which is $23$. 
		But the situation changes for the second red stripe where the number of proto-events is equal to $23$. Immediately after that all the FHN units become excited and this fact is exhibited by the third red stripe. The fourth red stripe confirms again that it failed to cross the critical mass. As a result, an extreme event occurs in $\bar{x}$ as shown in the lower panel of the figure. Here, $k=0.00128$. 
		For detail description, see Ref. \cite{ansmann2013extreme}. }
	\label{fig_12}
\end{figure}

  Now, we discuss another example where extreme events occur in globally coupled oscillators under attractive coupling. In the Sec.\ \eqref{imperfect}, we have already discussed the origination of extreme events in two coupled excitable units of FitzHugh–Nagumo (FHN) system \cite{ansmann2013extreme, karnatak2014route}. 
Not only the two coupled system but Ansmann et al.\ \cite{ansmann2013extreme} also showed such extreme events in a network of $N (>2)$ non-identically coupled FHN units represented by Eq.\ \eqref{eq.5}. In this study, the observable is $\bar{x}=\frac{1}{N}\sum_{j=1}^{N}x_{j}$. The maxima of $\bar{x}$ become extreme events when they exceed a predefined threshold, $T=0.6$. The emergence of extreme events for the globally coupled FHN units is confirmed by taking $N=101$ oscillators\ \cite{ansmann2013extreme, karnatak2014route}. For this purpose, the system parameters are chosen as $a=-0.02651, c=0.02$, and $b_i\in [0.006,0.014]~\forall i$. The values of $b_i$ $(i=1,2,\cdots,N)$ are distributed in a equispaced manner, {\it i.e.}, $b_i=0.006+0.008\dfrac{i-1}{N-1}$. For the suitable coupling strength $k$, a portion of units becomes excited simultaneously.  As per Ref.\ \cite{ansmann2013extreme}, the values of the variable $x_i$ corresponding to that portion of units exceed $0.6$ during excitation. At this time, the variable ${x_i}$ exceeding $0.6$ is called proto-event. The number of excited units is denoted by $e=|\{i|x_{i}>0.6\}|$. The generation of extreme events depends on the appearance of proto events. Here, $|\{\cdot\}|$ denotes the cardinal number of a set. If the number of units exhibiting proto-events simultaneously exceeds a specific number, then the observable displays extreme events. This particular number is called “critical mass” (indicated by the horizontal black line in Fig.\ \ref{fig_12}(a)). So, if the number of exciting units is greater than or equal to the critical mass, then all the units in the network become excited after that. Thus, extreme event emerges in the globally coupled FHN units.

\par Time evolution of all the FHN units is exhibited in Fig.\ \ref{fig_12}(a). Here, around $t\approx21100$, some of the nodes whose values of $b_i$ are small exhibit proto-events, but $e<23$. So, the time evolution of ${\bar x}$ after the appearance proto-events fails to exhibit extreme event as shown in Fig.\ \ref{fig_12}(b). Ansmann et al.\ \cite{ansmann2013extreme} observed that the critical mass is $23$ for $N=101$. This fact is also clear from Fig.\ \ref{fig_12}. At $t\approx21600$, $e$ becomes $23$. After that, all units become excited (long red stripe as shown in  Fig.\ \ref{fig_12}(a)). As a result, an extreme event occurs in  Fig.\ \ref{fig_12}(b), where local maxima of ${\bar x}$ exceeds an extreme event qualifying threshold $0.6$ (indicated by red dashed line). Ansmann et al.\ \cite{ansmann2013extreme} reported that a system of $N=10000$ oscillators under a small-world network configuration\ \cite{watts1998collective} is also capable of exhibiting extreme events. Again for the suitable values of $k$, a group of FHN units fires simultaneously, and those exciting units form few localized clusters. Then, due to the spreading of excitation in the lattice structure underlying the small-world network, extreme events are observed from the time evolution of $\bar{x}$.

\par All these studies attest that parameter mismatch plays a crucial role in manifesting extreme events in global networks. In fact, the results contemplated in Ref.\ \cite{ray2020extreme} need additional repulsive mean-field interaction along with the parameter mismatch. Recently, an investigation on globally coupled maps demonstrates the occurrence of extreme events, where the attractive coupling through the mean-field can provide such fascinating behavior among an ensemble of identical coupled maps \cite{chowdhury2021extreme}. These coupled one-dimensional chaotic maps \cite{kaneko1990globally} form a two-cluster state before an analytically calculated critical coupling strength. The distance between these two clusters deviates abruptly beyond a properly justified threshold, and those states are characterized as extreme events. Its probability density function obeys the generalized extreme value distribution, and the Weibull distribution fits well with the distribution of recurrence time intervals between extreme events.  

\par In a recent work, Br{\"o}hl et al.\ \cite{brohl2020identifying} reported the generation of extreme events in complex networks of FitzHugh-Nagumo units represented by Eq.\ (\ref{eq.5}). Both the small-world and the scale-free network topologies are used in this context. The target is to locate the edges of the complex network, which are responsible for converting non-excited units into excited one and, consequently, leading to the emergence of extreme events. For this, the centrality of edges in a time-dependent interaction network and edge-based network decomposition technique are considered for addressing the problem.

\subsubsection{Time-varying networks}

\par So far, we have discussed about extreme events in isolated dynamical systems and under the framework of static network formalism. But in the real world, most of the existing interactions among physical, biological, and societal entities are time-varying. A variety of collective states has been explored earlier in time-varying dynamical networks of mobile agents \cite{majhi2019emergence,porfiri2006random,uriu2010random,kim2013emergence,majhi2017synchronization}, however, the studies of emergent extreme events in temporal networks are very few. Recently, extreme events in two distinct time-varying networks of mobile agents under the influence of attractive-repulsive interactions \cite{chowdhury2019extreme,9170822} are reported. For both of these arrangements, $N$ mobile agents move in any direction independently on a two-dimensional XY-plane $ \sum_{}^{}=\left[-g,g \right] \times \left[-g,g \right]$ with a velocity $ {\bf v}_i\left(t\right)= \left[v\cos\theta_i\left(t\right),v\sin\theta_i\left(t\right)\right], i=1,2,...,N$. Here, $v$ is the uniform modulus velocity of each agent and $\theta_i\left(t\right),$ $i=1,2,...,N,$ is chosen arbitrarily from an interval $\left[0,2\pi\right]$. Any kind of collision is forbidden among themselves. Thus, if the position of the $i$-th agent at any time $t$ is $(p_i(t),q_i(t))$, then the motion updating process can be represented by the following relations:

\begin{equation}
\begin{array}{lcl}\label{eq.100}
p_i(t+1)=p_i(t)+v\cos(\theta_i(t)),\\
q_i(t+1)=q_i(t)+v\sin(\theta_i(t)).
\end{array}
\end{equation}

\par To confine the agent's motion within the XY-plane $ \sum_{}^{}=\left[-g,g \right] \times \left[-g,g \right]$, whenever $p_i(t)$, $q_i(t)$ exceed $|g|$, a new $\theta_i(t)$ is re-generated so that $-g \leq p_i(t), q_i(t) \leq g$ remains for all the time.

\par
\begin{figure}[H]
	\centerline{
		\includegraphics[scale=0.44]{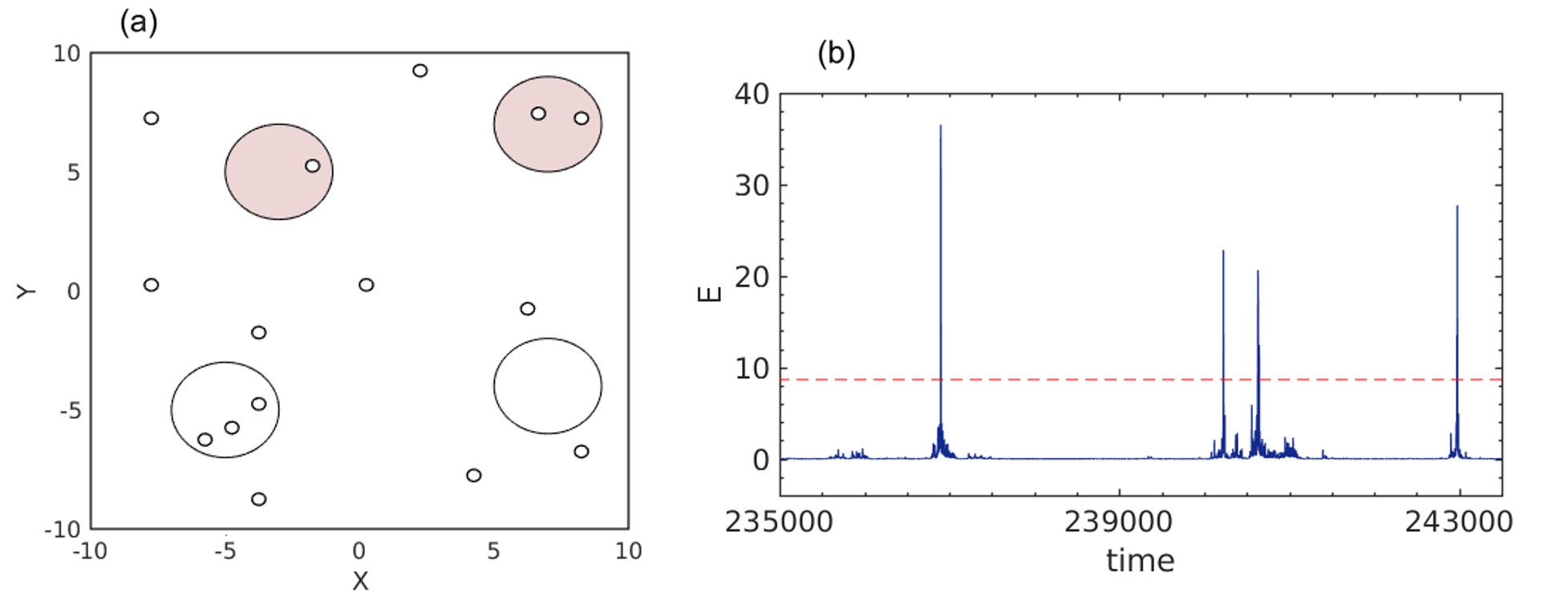}}
	\caption{(a) {\bf Schematic diagram of the time-varying network}: This picture shows the location of $N = 15$ moving agents (small circles) at any particular time instant. Big circles display interacting activation zones. Grey circles indicate attractive zones, and white circles represent repulsive zones. The oscillators sitting on top of those mobile agents are not able to communicate outside the coupling zones. Any interaction between agents only takes place when they are within the same coupling zone. For detailed discussion, please see Ref.\ \cite{chowdhury2019extreme}. (b) {\bf On-off intermittency}: Synchronization error $E$ is occasionally leaving the synchronization manifold ($E=0$) generating some extremely large deviated values. This irregular switching from zero to non-zero values of $E$ reflects the on-off intermittent behavior of the error trajectories. To distinguish among extreme events and other smaller intermittent events, red dashed line $T=m+8\sigma$ is drawn. One attracting coupling zone centered at $\left(g/2,g/2\right)$ and one repulsive coupling zone centered at $\left(g/2,-g/2\right)$ are considered with $g=10$. Each zone has a radius $r=4.0$ and identical area. $N=100$ R\"{o}ssler oscillators with modulus velocity $v=2.5$, display high amplitude fluctuation from the synchronous manifold for a combination of attractive coupling strength $K_a=1.0$ and repulsive coupling strength $K_r=-0.1$. Further discussions can be found in Ref.\ \cite{chowdhury2019extreme}.  }
	\label{timevarying}
\end{figure}

 In Ref.\ \cite{chowdhury2019extreme}, few interacting circular zones are predefined within the two-dimensional plane $ \sum_{}^{}$, and the interactions among those mobile agents take place only when they visit the same interacting zone. Figure \ref{timevarying}(a) shows a schematic illustration of the time-varying interaction. The smaller circles indicate the moving agents, the larger white circle represents the repulsive zone, and the gray circle denotes the attractive zone. These $m=4$ coupling zones are where $N=15$ agents can interact with each other. Interaction occurs only when the agents belong to the same fixed interaction zone. There is no interaction in the right-bottom coupling zone at that particular moment since it is empty. This argument also applies to the coupling zone on the left-top, containing a single agent. That agent in that left-top coupling zone can not interact with other agents at that time. The same argument can be made for other $9$ agents outside of coupling zones. Only for that time do the $2$ agents in the right-top coupling zone, and the $3$ agents in the left-below coupling zone interact among themselves.

 On top of each of these moving nodes, an oscillator is placed. And thus, the agents' motion affects the adjacency matrix at each time step and hence, influences the system's collective dynamics. However, one should note that the states of those oscillators situated on top of those agents do not influence the agents' mobility. If all the interaction zones are attractive, complete synchronization occurs in the system with a suitable uniform modulus velocity $v$ and for an appropriate coupling strength $K \ge K_c$. 
But, along with the attractive zones, if few repulsive zones with an appropriate repulsive coupling strength are introduced in the plane, the synchronization becomes intermittent and occasionally blows out from the synchronization manifold. The observed phenomenon is independent of the shape of the coupling zones and the number of zones. To verify this claim, two paradigmatic chaotic oscillators, namely Lorenz \cite{lorenz1963deterministic} and R\"{o}ssler \cite{rossler1976equation} oscillators  are assigned in each node to describe the agent dynamics in Ref.\ \cite{chowdhury2019extreme}.

\par Here, the synchronization error $E$ is chosen as the observable, and it is defined in terms of the standard Euclidean norm as
\begin{equation}\label{eq.32}
\begin{array}{lcl}
E = \Big\langle \frac{\sum_{i=2}^{N}\sqrt{\left(x^i_1-x^1_1\right)^2 + \left(x^i_2-x^1_2\right)^2 + \left(x^i_3-x^1_3\right)^2} }{\left(N-1\right)}\Big\rangle_t,
\end{array}
\end{equation}
where ${\langle \cdot \cdot \cdot \rangle}_t$ stands for time average. Figure \Ref{timevarying}(b) illustrates the temporal behavior of the error dynamics. Due to occasional interaction in the repulsive zone with appropriate coupling strength, few oscillators split from the coherent group and exhibit high amplitude deviation in the error $E$ compared to the regular behavior of the synchronized cluster. Hence, the temporal evolution of the error dynamics $E$ shows occasional large deflections (on state) from the synchronization manifold while it remains almost at zero value (off state) most of the time. Extreme events are characterized here as intermittent large deviations in the error function E and manifested by occasional excursions away from the synchroization manifold. They are larger than $T=m+8\sigma$ and they follow a non-Gaussian statistics. This threshold $T$ can also be analytically calculated using the method developed by Massel \cite{ryszard1996ocean}. 

\par It is possible in the above-described interaction policy that few mobile agents do not interact with any other agents and thus remain isolated at a particular instant. In Ref.\ \cite{9170822}, the authors treated the scenario in a different way, where the underlying network is always a global network. Nevertheless, the interaction among any two oscillators is either repulsive or attractive, depending on their mobile agents' relative distance. The relative distance between the $i$-th and $j$-th mobile agents is depicted by the standard Euclidean metric $d_{ij}=\sqrt{(p_i(t)-p_j(t))^2+(q_i(t)-q_j(t))^2}$. Suppose at any particular time $t$, $d_{ij}(t)$ is greater than a predefined distance $\beta$. In that case, two oscillators on top of those agents are attractively coupled, and if $d_{ij}(t) \leq \beta$, then the repulsive coupling is activated among those oscillators. Therefore, the coupling type may vary at each time step between the oscillators depending on the relative distance between those mobile oscillators. For numerical investigation, the Stuart-Landau oscillator is placed on top of each mobile agent. For suitable choices of coupling strength for two opposing type of interactions, the error trajectory becomes intermittent. This irregular away journey of the error trajectory from the synchronization manifold gives rise to infrequent large deviated events. These large excursions are characterized as extreme events. The chosen threshold $T$ shows one-to-one relation with the mean return interval $R_{T}$. This mean return interval is found to depend on the normalized distribution $f(x)$.

\section{Extreme events due to random walkers and Brownian motion} \label{Extreme events due to random walk in complex networks}

\par 
The efficient functioning of large networked infrastructures, such as the Internet, power grids, and socioeconomic and transportation networks, poses one of the greatest challenges among scientific communities. One needs to minimize the congestion and transit time emerging due to the continuous increment of traffic flow in most of the communication networks. These challenges are fairly commonplace experiences in the real world in the form of power blackouts due to tripping of power grids, information packet transmission via the Internet, traffic jams in transportation networks, to name but a few. Efficient interdisciplinary approaches to control such extreme events are recently addressed via modeling. Usually, routes to such congestion are investigated in different transport networks by considering the handling capacity of each node (or edge). The packets in the form of random walkers are sent from one node to another adjacent node within a static (time-independent) complex network. The congestion will take place in a particular node (edge), if and only if its capacity to provide a service to the incoming random walkers is exceeded. In the present section, we discuss the results of extreme events on complex networks due to random walkers. One should note that we demonstrate the onset of extreme events of random walkers in a network and not in the sense of arbitrary random walkers and Brownian motion. Here, random walk contemplates those dynamic processes, where a walker on a node of the finite undirected network hops to one of its nearest neighbors with equal transition probability at every time step. For a detailed understanding of random walk processes in complex networks, we refer interested readers to Refs.\ \cite{helbing2001traffic,barthelemy2011spatial}.

\subsection{Extreme events in a mono-layer network}

\par We carefully distinguish the results between monolayer and multilayer networks. Multilayer networks facilitate to study of several types of ties between the edges by accompanying multiple layers. The connections between different layers may produce various unexpected results, which can not be observed in monolayer networks. In the following sections, we first emphasize the results on monolayer networks and then focus on the rich playground of multilayer networks.

\subsubsection{Extreme events on nodes}
\begin{figure}[H]
	\centerline{
		\includegraphics[scale=0.42]{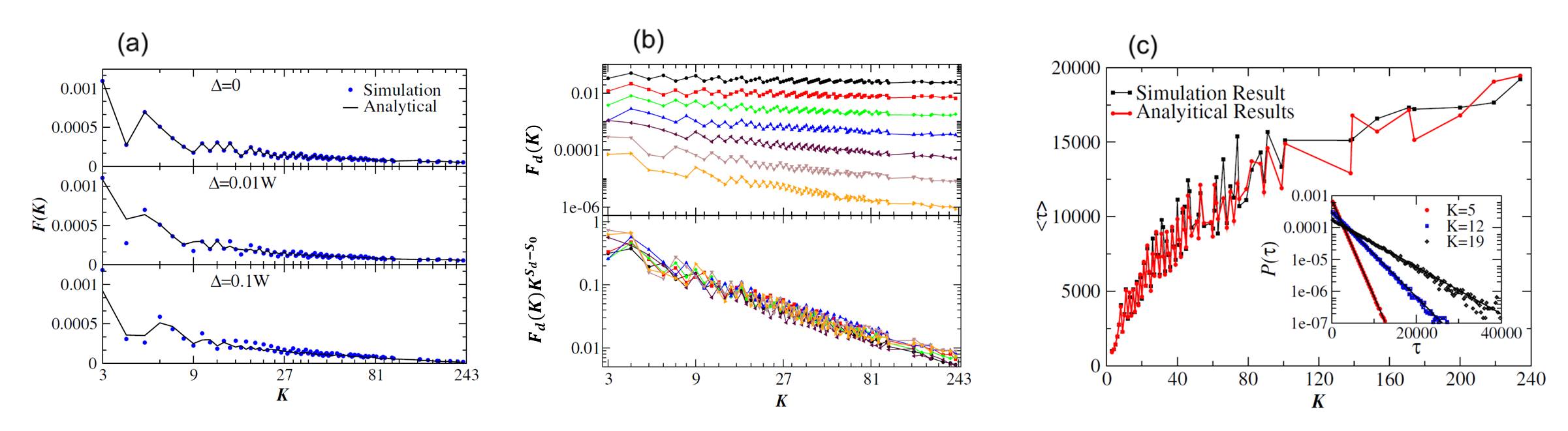}}
	\caption{(a) {\bf Probability for the occurrence of 
			extreme events in a semilog plot:} Here, the extreme event indicating threshold is $q=m+4\sigma$. The solid line is the analytical expression, given in Eq.\ (\ref{eq.44}). Each point in the figure is accumulated by averaging over all the vertices with the same degree. The oscillations are inherent in the analytical and numerical results and not due to insufficient ensemble averaging. The nodes with smaller degree $(K < 20)$, on an average, display a higher probability for the occurrence of extreme events as compared to the nodes with higher degree, say, $K > 100$. (b) {\bf Probability for occurrence of extreme events with different values of $q$:} The extreme event probabilities $F_{d}(K)$ in a log-log plot obtained from simulations with $\Delta=0$ (upper panel) show the scaling relation with respect to $d$. This scaling relation can not be determined analytically from the Eq.\ (\ref{eq.44}), but approximately falls into a curve using the relation (\ref{eq.46}) as shown in the lower panel. $S_0$ is the reference slope with $d=2$. The curves are drawn from top to bottom with $d=2.0,2.5,3.0,3.5,4.0,4.5,$ and $5.0$. (c) {\bf Mean recurrence time as a function of degree $K$:} The solid line is the analytical distribution, which agrees well with the observed numerical simulations. Recurrence time distribution obtained from simulations for three nodes (with different degrees $5$, $12$ and $19$ respectively) is represented in the inset. For numerical simulations, a scale-free network with degree exponent $\gamma=2.2$ is considered. The number of nodes is $N=5000$ and the number of edges is $E=19815$. All results are averaged over 100 realizations with randomly chosen initial conditions. $W=2E$ number of non-intercating random walkers are considered. Reprinted figure with permission from Ref.\ \cite{kishore2011extreme}.}
	\label{fig_21}
\end{figure}

A larger flux in a scalar time series may imply higher probabilities of occurrences of extreme events. In contrast, Kishore et al.\ \cite{kishore2011extreme} proved 
that a larger flux not necessarily gave rise to higher probabilities for extreme events (See Fig.\ \ref{fig_21}(a)) in the context of a node on a connected undirected network. For this purpose, a transport model based on random walk on complex networks \cite{noh2004random} is considered. 
As per this study \cite{kishore2011extreme}, even though hubs attract a huge amount of flux compared to smaller degree of nodes, but hubs are less susceptible to extreme events.

\par Let $A$ be a connected, undirected, finite networks with $N$ nodes and $E$ links, given by the adjacency matrix $A=[a_{ij}]_{N \times N}$, where $a_{ij}=1$, if there exists a connection between $i$-th and $j$-th nodes, otherwise, $a_{ij}=0$. The topology of this network may be taken as (i) random \cite{gilbert1959random,erdHos1960evolution}, (ii) small-world \cite{watts1998collective} and (iii) scale-free \cite{barabasi2009scale,broido2019scale}. Although, the dissimilitude in the occurrence probability of extreme events between hubs and smaller degree nodes is not so pronounced
in the case of random graphs. There are $W$ non-interacting walkers performing random walk on the considered network. A random walker sitting on the $i$-th node (say) at time $t$ can hop to any one of the neighboring nodes with
equal probability. Thus, $a_{ij}/K_{i}$ is the transition probability from the $i$-th to $j$-th node, where $K_{i}$ is the number of links connected with the $i$-th node. The corresponding master equation for the $n$-step transition probability of an independent walker starting from node $i$ at time $n=0$ to node $j$ at time $n$ is

\begin{equation}
\begin{array}{lcl}\label{eq.42}
P_{ij}(n+1)=\sum_{k} \dfrac{a_{kj}}{K_k} P_{ik}(n).
\end{array}
\end{equation}

The stationary state \cite{noh2004random} is

\begin{equation}
\begin{array}{lcl}\label{eq.43}
\lim\limits_{n \to \infty} P_{ij}(n)=p_j=\dfrac{K_j}{2E},
\end{array}
\end{equation}
which physically implies that more walkers will visit a given node if it has more links. An extreme event is defined in the spirit of extreme value statistics as that event which is typically associated with the tail of the probability distribution function of the events 
and has little probability of occurrence. 
A node exhibits extreme events if
the number of walkers’ arrival is larger than a threshold $q$. Clearly, $q$ should depend on the traffic flowing through the node. Otherwise, a uniform choice of $q$ leads to a situation, where few nodes always experience extreme events, while few of the rest nodes may never encounter any extreme events. To deal with this, $q=m+d\sigma$ has been considered in Ref.\ \cite{kishore2011extreme}, where $d \in \mathbb{R}$. Here, $m=\dfrac{WK}{2E}$ is the mean for a given node having degree $K$, and $\sigma=W\dfrac{K}{2E}\bigg(1-\dfrac{K}{2E}\bigg)$ represents the standard deviation. By considering the entire possible situations, the probability of finding more than $q$ independent random walkers traverse a given node with degree $K$ can be obtained as

\begin{equation}
\begin{array}{lcl}\label{eq.44}
F(K)=\sum_{j=0}^{2\Delta} \dfrac{1}{2\Delta+1}\sum_{k=\lfloor q \rfloor+1}^{\tilde{W}+j} \binom{\tilde{W}+j}{k} p^k (1-p)^{\tilde{W}+j-k},
\end{array}
\end{equation}
where $\tilde{W}=W-\Delta$ and $p$ is the probability of finding a walker at a given node of degree $K$. Here, it is assumed that the total number of walkers is a random variable uniformly distributed in the interval $ [ W-\Delta, W+\Delta]$.  

%
%

\par The probability for the occurrence of an extreme event with respect to the degree $K$ is shown in Fig.\ \ref{fig_21}(a). This figure clearly indicates that the probability of occurrence is higher for the nodes with smaller degree, on an average, compared to the hub. The solid lines in Fig.\ \ref{fig_21}(a) are the analytically derived result as given in Eq.\ \eqref{eq.44}, which agrees quite well with the numerical results. Note that, the entire distribution is usually plotted, and those events in the studies of dynamical systems, possessing a height greater than the predefined threshold, are considered as extreme events. Here, the probability for the occurrence of extreme events is only plotted instead of the whole distribution. The impact of different choices of $d$ is investigated in Fig.\ \ref{fig_21}(b). A smaller choice of $d$ leads to a higher probability of occurrence of extreme events, and thus, the uppermost curve of $F_d(K)$ in Fig.\ \ref{fig_21}(b) with $d=2.0$ is almost horizontal with $K$. This behavior is changed with an increment of $d$, and the curves of $F_d(K)$ exhibit a more negative slope. The following scaling relation

\begin{equation}
\begin{array}{lcl}\label{eq.46}
\dfrac{F_{d}(K)}{K^{1-S_{d}}}=\rm{Constant},
\end{array}
\end{equation}

is verified empirically. Here, $S_{d}$ is the slope of $F_{d}(K)$ representing extreme event probability for a threshold value $q$ with parameter $d \in \mathbb{R}$. Temporal correlations among the extreme events are inspected through the recurrence time distribution in Fig.\ \ref{fig_21}(c). The recurrence time distribution satisfies 

\begin{equation}
\begin{array}{lcl}\label{eq.45}
P(\tau)=\dfrac{e^{-\tau}}{\langle\tau\rangle},
\end{array}
\end{equation}

with mean recurrence time $\langle\tau\rangle=\dfrac{1}{F(K)}$, which fits well with the numerically observed simulations (See Fig.\ \ref{fig_21}(c)). The return interval clearly demonstrates that $\langle\tau\rangle$ increases with larger values of $K$. This again confirms that the small degree nodes are more prone to experience extreme events.

\par To implement their findings in a more realistic manner, an alternative intelligent routing algorithm \cite{medhi2017network} is employed instead of a random walk. But, the results obtained with the shortest path algorithm \cite{cormen2009introduction} do not change the perceived trend obtained with random walk qualitatively. Fei et al.\ \cite{wang2014extreme} further investigated the role of degree correlation \cite{newman2003mixing,newman2002assortative,johnson2010entropic,zhou2013link} of underlying networks on extreme events.

\par Kishore et al.\ \cite{kishore2012extreme} represented their obtained results in Ref.\ \cite{kishore2011extreme} for the unbiased standard random walk on networks. A measure of the ability of a node to attract walkers in the form of the generalized strength of the $j$-th node is defined by,

\begin{equation}
\begin{array}{lcl}\label{eq.48}
\phi_{j}=K_{j}^{\alpha} \sum_{i=1}^{K_j} K_{i}^{\alpha}.
\end{array}
\end{equation}

Here, $\alpha$ is the bias parameter, where $\alpha > 0$ implies that walkers preferentially hop to hubs and $\alpha < 0$ indicates biasedness towards smaller degree nodes. The master equation, in this case, can be written as

\begin{equation}
\begin{array}{lcl}\label{eq.49}
P_{ij}(n+1)=\sum_{l} A_{lj} \dfrac{K_{j}^{\alpha}}{\sum_{y=1}^{K_l}K_{y}^{\alpha}} p_{il}(n).
\end{array}
\end{equation}

\par Let $w$ be the number of random walkers out of $W$ noninteracting walkers passing through a node with generalized strength $\phi$. Then the probability of finding $w$ walkers is $p_{i}^w$, while the rest of $W-w$ walkers are distributed randomly on the rest of the nodes of the network. This leads to a binomial distribution while properly normalized.  Here, $p_i=\dfrac{\phi_i}{\sum_{j=1}^{N}\phi_j}$ is the stationary distribution for the number of walkers in the $i$-th node. So, the occurrence of more than $q_{i}$ walkers passing through the node is given by the probability 

\begin{equation}
\begin{array}{lcl}\label{eq.50}
F=\sum_{w=q_{i}}^{W} \binom{W}{w} p_{i}^w (1-p_{i})^{W-w}.
\end{array}
\end{equation}

\begin{figure}[H]
	\centerline{
		\includegraphics[scale=0.55]{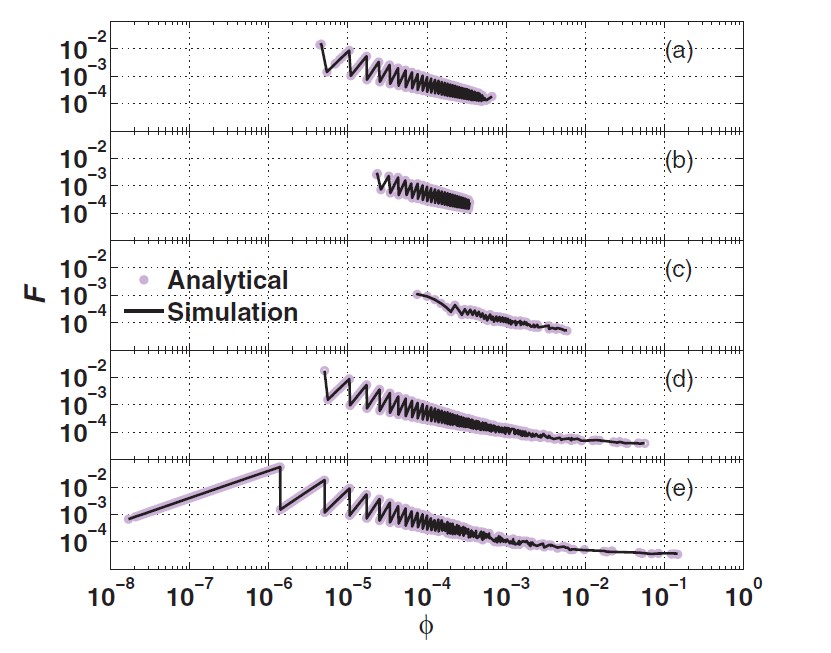}}
	\caption{ {\bf Probability of the occurrence of
			extreme events plotted as a function of the normalized generalized strength $\phi$:} The circles are obtained from simplified analytical expression (\ref{eq.50}) using standard incomplete beta function \cite{lopez1999asymptotic,ozcca2008extension}. The threshold for the extreme events are $q=m+4\sigma$. The numerical simulations are performed on a scale-free network ($N=5000$, $E=19915$) with $W = 2E$ random walkers averaged over 100 realizations with randomly chosen initial positions of walkers. The values of bias parameter $\alpha$ are taken as (a) $-2.0$, (b) $-1.0$, (c) $0.0$, (d) $1.0$, and (e) $2.0$. For details, please see the Ref.\ \cite{kishore2012extreme}. Reprinted figure with permission from Ref.\ \cite{kishore2012extreme}.}
	\label{fig_22}
\end{figure}

The threshold $q_i$ is taken as $q_i=m_i + d\sigma_{i}$, with $d \geq 0$. Here, $m_i=Wp_i$ is the mean flux and $\sigma_{i}=Wp_i(1-p_i)$ is the standard deviation. Figures \ref{fig_22}(a)-(e) depict that the nodes with smaller values of generalized strength $\phi$, on an average, raise to high probability of occurrence of extreme events, compared to the nodes with higher values of generalized strength $\phi$. 
The numerical findings match perfectly with the analytically derived relation \eqref{eq.50}. Also, the first-jump probability \cite{Tadic2001adaptive} is higher for hubs compared to smaller degree nodes for a standard random walk ({\it i.e.,} $\alpha=0$). The first-jump probability is defined as the probability of extreme events on a node at time $(n+1)$, when an extreme event takes place on a neighboring node at time $n$. In case of biased random walks ({\it i.e.,} $\alpha \ne 0$), $\alpha < 0$ are more prone to show first-jump probability, compared to $\alpha > 0$. Results \cite{kishore2011extreme,kishore2012extreme} suggest that extreme events on complex networks can be controlled by paying suitable attention in designing the capacity of the small degree nodes.

\subsubsection{Controlling strategies}

\par Extreme events occur in transportation networks due to excessive flux fluctuations in the flow dynamics. This overloading in the form of large fluctuations beyond the capacity disrupts the efficient functioning of the flow of transport \cite{holme2002vertex}. Thus, an optimal way of handling and delivery of the flow/information in the networks \cite{nicolaides2010anomalous,echenique2004improved} is one of the fundamental practical issues. The overarching challenge of such extreme events demands to be effectively controlled. These critical issues will also advance our understanding of the flow and transport properties of such large communication networks.

\par Now, we discuss a scheme for a partial reduction of the occurrence of extreme events in a scale-free network by tuning the nodal capacity $\alpha \ge 0$ \cite{kishore2013manipulation}. 
 An extreme event occurs on the $i$-th node, if the number of walkers at that instant crosses a predefined threshold 

\begin{equation}
\begin{array}{lcl}\label{eq.54}
q_i=m_i+d(\sigma_{i})^{\alpha} \approx Wp_{i}+d(Wp_i)^{\alpha},
\end{array}
\end{equation}
where $p_i=\dfrac{k_i}{\sum_{l=1}^{N} k_l}$ is the stationary probability to find a walker on $i$-th node with degree $k_i$. $W$ is the number of interacting walkers that perform random walks at each time step on a scale-free network of $N$ nodes and $E$ edges. Under the assumption of finite capacity of each node, the excess number of random walkers on $i$-th node can be treated as a queue of length 

\begin{equation}
\begin{array}{lcl}\label{eq.55}
Q_{i}(t)=[w_{i}(t)-q_i] \theta{[w_{i}(t)-q_i]},
\end{array}
\end{equation}
where $\theta{(\cdot)}$ is the Heaviside step function and $w_{i}(t)$ represents the number of walkers on the $i$-th node at time $t=1,2,\cdots,T$. The mean queue size $\langle S \rangle$ in excess of nodal capacity for a network with $N$ nodes at every time instant is given by,
\begin{equation}
\begin{array}{lcl}\label{eq.56}
\langle S \rangle=\lim\limits_{T \to \infty} \dfrac{\sum_{i=1}^{N} \sum_{t=1}^{T} Q_{i}(t)}{TN}
=\dfrac{\sum_{i=1}^{N} \binom{W}{w} p_{i}^{w} (1-p_{i})^{(W-w)} Q_{i}}{N}
=\dfrac{\sum_{i=1}^{N} f_{i}(w) Q_{i}}{N},

\end{array}
\end{equation}

\begin{figure}[H]
	\centerline{
		\includegraphics[scale=0.45]{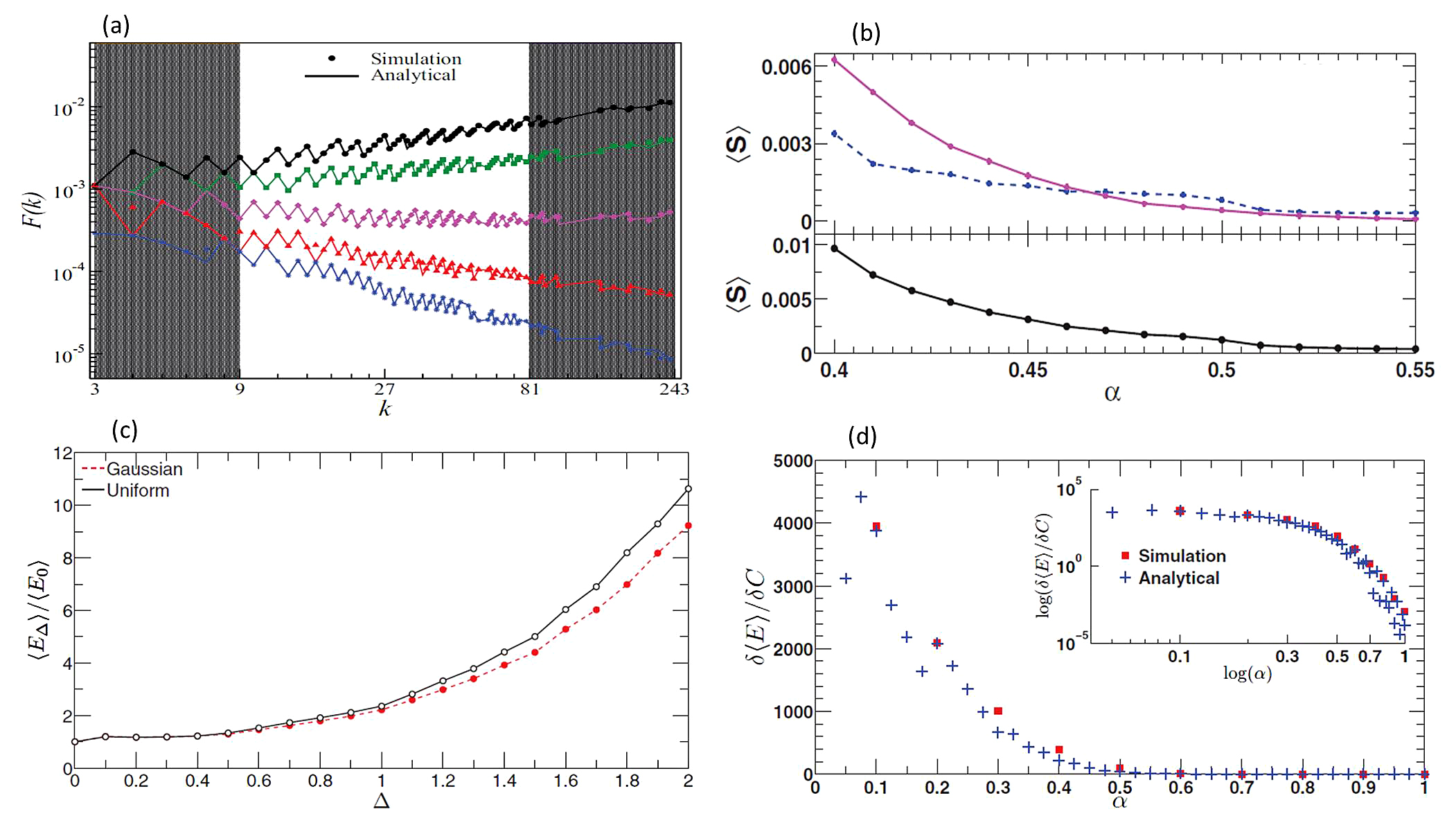}}
	\caption{(a) {\bf Probability for the occurrence of extreme events as a function of degree $k$:} The shaded regions represent the small degree nodes (left) and the high degree nodes (right). $\alpha$ is varied from bottom to top as follows $0.40, 0.43, 0.47, 0.50$ and $0.52$. The analytically simulated curve is obtained using the Eq.\ (\ref{eq.54}). (b) {\bf Mean queue size $\langle S \rangle$ as a function of $\alpha$:} The upper panel shows the functional dependence of $\langle S \rangle$ for the small degree nodes (dashed line) and large degree nodes (solid line). The curve in lower panel is for the whole network. (c) {\bf Scaled mean number of extreme events, $\dfrac{\langle E_{\Delta} \rangle}{\langle E_{0} \rangle}$ as a function of noise strength $\Delta$:} Here, $\alpha =0.5$. The analytical result (dashed and continuous), obtained from Eqs.\ (\ref{eq.59}) and (\ref{eq.54}), agrees well with the numerical simulations (solid and open). Also the results show the independence of the random numbers $\xi_i$, whether drawn from uniform or Gaussian distribution. (d) {\bf The dependence of $\alpha$ on the ratio of change in mean number of extreme events $\delta \langle E \rangle$ on the network by changing $\delta C$ in capacity of the network:} The inset shows the same figure, but in log-log plot. Beyond a certain nodal capacity, increment of nodal capacity does not contribute significantly in substantial reduction in the number of extreme events. All the simulations are performed on a scale-free network with degree exponent $\gamma =2.2$ of size $N=5000$ and $E=19915$. $100$ independent realizations are considered to obtain the results. Number of independent random walkers is $W=39380$. $m=4$ is taken without loss of any generality. 
	Reprinted figure with permission from Ref.\ \cite{kishore2013manipulation}.}
	\label{fig_24}
\end{figure}

where the distribution of the $w$ number of walkers (out of $W$ non-interacting number of random walkers) passing through the $i$-th node is $f_{i}(w)$. The probability for the occurrence of an extreme event on the $i$-th node is given by

\begin{equation}
\begin{array}{lcl}\label{eq.57}
F_{i}(q_{i})=\sum_{w=\lfloor q_{i} \rfloor+1}^{W} f_{i}(w)=I_{p}(\lfloor q_{i} \rfloor+1, W - \lfloor q_{i} \rfloor),
\end{array}
\end{equation}
where $I_{z}(a,b)$ is the incomplete beta function \cite{abramowitz1964handbook}. To reveal the dependence of $\alpha$ on manipulation of extreme events, Fig.\ \ref{fig_24}(a) is plotted for various values of $\alpha$. By the shaded regions of both sides, one can conclude that the probability of occurrence of extreme events has larger variation in the case of high degree nodes rather than the small degree nodes. This leads to an unequal dependence on the burden of extreme events on hubs and small degree nodes. Also, it is clear that the total number of extreme events on the entire network can be controlled by adjusting the nodal capacity parameter $\alpha$. In Fig.\ \ref{fig_24}(b), the effect of $\alpha$ on $\langle S \rangle$ is represented. For $\alpha > 0.5$, there is a significant reduction in the number of extreme events in the network, though $\langle S \rangle$ can never be exactly $0$, as the appearance of extreme events is due to the inherent fluctuations in the flux.  For $\alpha < 0$, the values of $\langle S \rangle$ are quite high indicating excessive mean number of walkers awaiting in the queue per node at every time instant. In the upper panel of the Fig.\ \ref{fig_24}(b), two separate curves, {\it i.e.,} the dashed line for small degree nodes and solid line for large degree nodes, intersect each other at about $\alpha \approx 0.47$. At the point of intersection, the nodes have an equal size of queue. For $\alpha > 0.5$, the small degree nodes have large queue sizes compared to the high degree nodes. This situation reverses for $\alpha < 0.5$, where large degree nodes display a larger burden of queue size. 

\par The time-independent capacity of the $i$-th node, assuming the actual capacity is a random variable \cite{tan2008empirical} assembled around the central tendency, is given with $\xi_{i}$ being a random number as

\begin{equation}
\begin{array}{lcl}\label{eq.58}
C_{i}(\Delta)=m_i +(d\sigma_{i} \pm \Delta \xi_{i} \sigma_{i})=C_{i}(0) \pm \Delta \xi_{i} \sigma_{i}.
\end{array}
\end{equation}

Clearly, this is equivalent to the threshold (\ref{eq.54}) for extreme events in absence of noise $\Delta$ and $\alpha=1$. The mean number of extreme events over the entire network $\langle E_{\Delta} \rangle$ scaled by  $\langle E_{0} \rangle$ is 

\begin{equation}
\begin{array}{lcl}\label{eq.59}
\dfrac{\langle E_{\Delta} \rangle}{\langle E_{0} \rangle}=\dfrac{\sum_{i} F_{i} (C_{i} (\Delta))}{\sum_{i} F_{i} (C_{i} (0))}.
\end{array}
\end{equation}

In Fig.\ \ref{fig_24}(c), the nonlinear increment of this quantity with respect to $\Delta$ is presented. As a consequence, one can come to the conclusion that a larger variability in the nodal capacity leads to an increase in the number of extreme events. But, there is a natural intuition that an increment of nodal capacity may lead to decrement of extreme events of the networks, on an average. To inspect this situation, the mean number of extreme events, when capacity changes by one unit, $\dfrac{\delta \langle E \rangle}{\delta C}$ is plotted as a function of $\alpha$ in Fig.\ \ref{fig_24}(d). $\dfrac{\delta \langle E \rangle}{\delta C}$ deteriorates rapidly for $\alpha < 0.5$. But for $\alpha > 0.5$, this quantity is very small. Even for $\alpha > 0.7$, extreme events are diminished by less than one event on an average. Thus increasing capacity beyond a certain level does not proportionately decrease the occurrence of extreme events.

\par
\begin{figure}[H]
	\centerline{
		\includegraphics[scale=0.55]{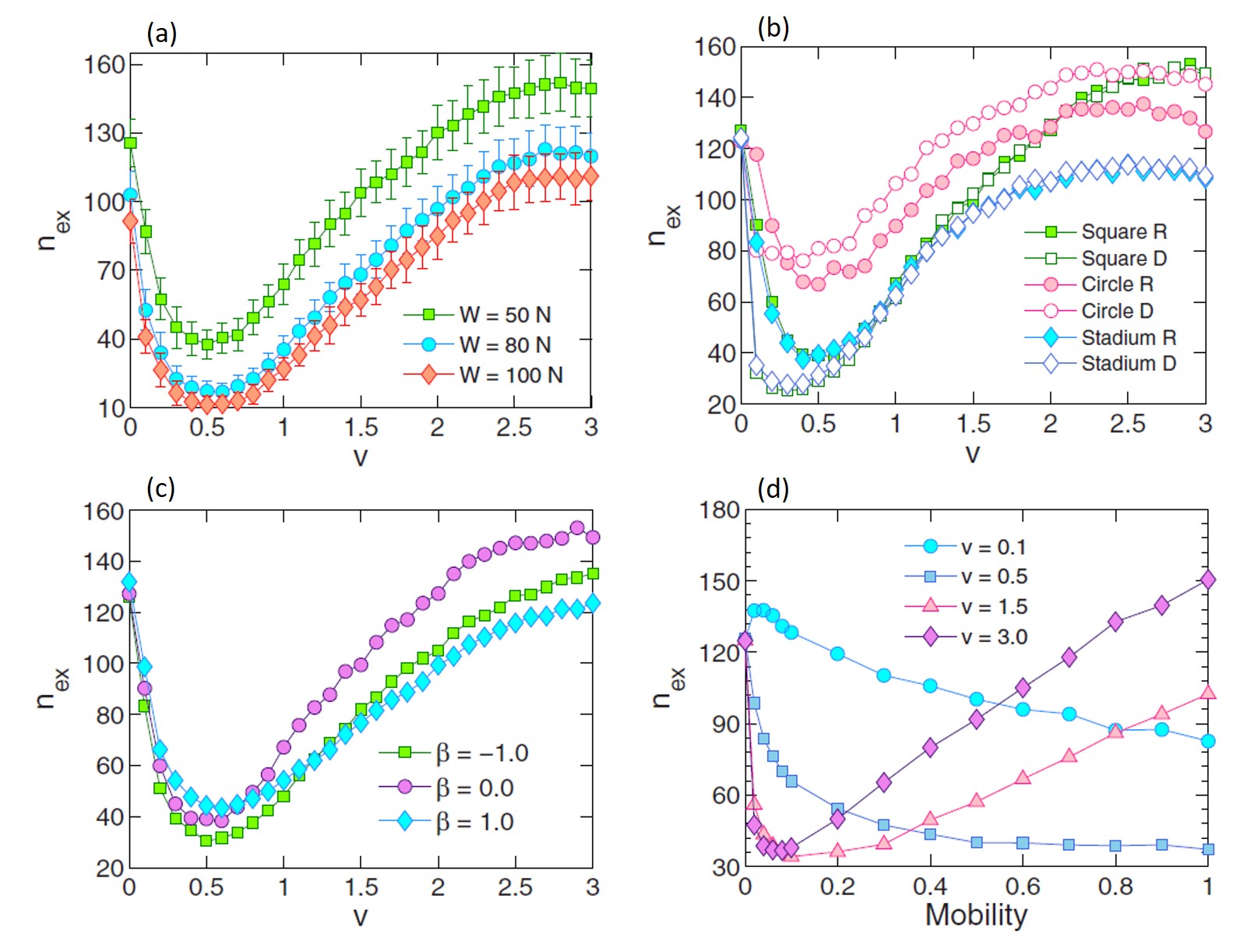}}
	\caption{(a) {\bf Dependence of agent velocity $v$ on the number of extreme events $n_{ex}$:} The non-monotonous correlation between $n_{ex}$ and $v$ is shown here from different numbers of packets $W$. Clearly, there exists an optimal moving velocity, irrespetive of number of packets present in the network, to suppress the number of extreme events in the network. (b) {\bf Effect of domain geometry:} Here, R stands for random walk and D stands for deterministic motion. Different shapes of domain is considered with same area $L \times L$. Clearly the boundary shapes do not affect the general observations of controlling extreme events using optimal moving velocity. (c) {\bf Effect of heterogeneous communication range:} $\beta$ is the parameter representing the nodal heterogeneity. $\beta = 0$ implies the uniform identical communication circles of each node, whereas positivity of $\beta$ signifies the presence of few nodes with significantly larger communication ranges. The opposite situation occurs for the negativity of $\beta$. (d) {\bf Effect of Mobility:} The term ``Mobility", here, represents the probability that any node moves with velocity $v$. The relatively higher values of $v$ possess an optimal velocity, close to zero, which is useful to control the number of extreme events
		in the network. All simulations are performed $100$ times independently with $T=1000$ time-steps.	The error bars are not shown in (b) and (c) for better visual representation. Further details with respect to the simulation
		set-up can also be found in Ref.\ \cite{chen2014controlling}. Reprinted figure with permission from Ref.\ \cite{chen2014controlling}.} 
	\label{fig_25}
\end{figure}

 A control scheme has been recently proposed by Chen et al.\ \cite{chen2014controlling}, in which mobility of the nodes helps to minimize the number of extreme events in any complex networks. There exists an optimal agent velocity, which diminishes the occurrence of extreme events in the network significantly (See Fig.\ \ref{fig_25}(a)). Figure \ref{fig_25}(a) portrays that the number of extreme events, $n_{ex}$ is initially decreasing up to a certain value of $v$, and beyond this critical velocity, $n_{ex}$ is increasing almost monotonically. Thus, there definitely exists a critical value of $v$, which will help to decrease the number of extreme events in the network. To demonstrate the findings, initially a domain of size $L \times L$ is considered. The shape of the domain does not effect their findings, which is shown in the Fig.\ \ref{fig_25}(b). In this figure, several shapes of the domain are considered with either random, or deterministic motion. In this physical domain, $N$ mobile agents with communication radius $a \ll L$ move in the domain. To confine the motion of these agents, a ``hard-wall" boundary condition is considered. The effect of a boundary is thoroughly investigated. The non-monotonous nature of extreme events due to the boundary effect is established. There is $W$ number of packets on the network at any given time. The number of packets in a node at time $t$ is given by $w(t)$ and each packet will be transmitted to any one of the neighboring nodes randomly at the next time step. In earlier works, people are concerned mainly regarding the extreme value distributions above the threshold, determined by the extreme value theory \cite{gumbel2012statistics,sabhapandit2007density}. Instead of the probability of occurrence of extreme events, the total number of extreme events is also an essential cornerstone of Ref.\ \cite{chen2014controlling}. To fulfill this motivation, an extreme event on a node is defined as the number of packets exceeding at least four standard deviations above the average. To represent their work \cite{chen2014controlling} more realistically, a heterogeneous nodal communication range is considered. The effect of heterogeneous vision range is contemplated in Fig.\ \ref{fig_25}(c), where an optimal value of $v$ is still observed. The detection of effective control scheme can be applied even in static networks by moving a few nodes, as the optimal mobility is close to zero in Fig.\ \ref{fig_25}(d).

\subsubsection{Extreme events on edges}

\par On the other hand, most of the works on extreme events based on random walkers are concerned about the appearance of extreme events on nodes. The probability of occurrence of extreme events is found highly dependent on the node and its degree. We have already discussed the novel analysis as discussed in Ref.\ \cite{kishore2011extreme}, in which they found the small degree nodes have comparatively higher extreme event probability rather than the hubs. In all these works, the probability of occurrence of extreme events on an edge is completely neglected. But, practical experiences suggest that jamming like situation can arise not only in nodes, but also on connecting edges. This study has been recently perceived by Kumar et al.\ \cite{kumar2020extreme}. The extreme event probability in the case of edges is solely dependent on the total number of walkers and the total number of edges. Thus, the probability of occurrence of extreme events on an edge is completely independent of graph topology. These numerical findings agree well with their perceived analytical results. The successive recurrence intervals of extreme event of flux and loads are uncorrelated. Here, the load is the sum of walkers on any edge and flux is the difference of the walkers traversing in opposite directions. Also, the maximum correlations between extreme events on an edge and on the two nodes connecting to it occur at a time lag of $0$ and $-1$, which implies that extreme events on an edge are preceded and also followed by those on a node. 

\subsection{Extreme events in multi-layer network}

\par
\begin{figure}[H]
	\centerline{
		\includegraphics[scale=0.45]{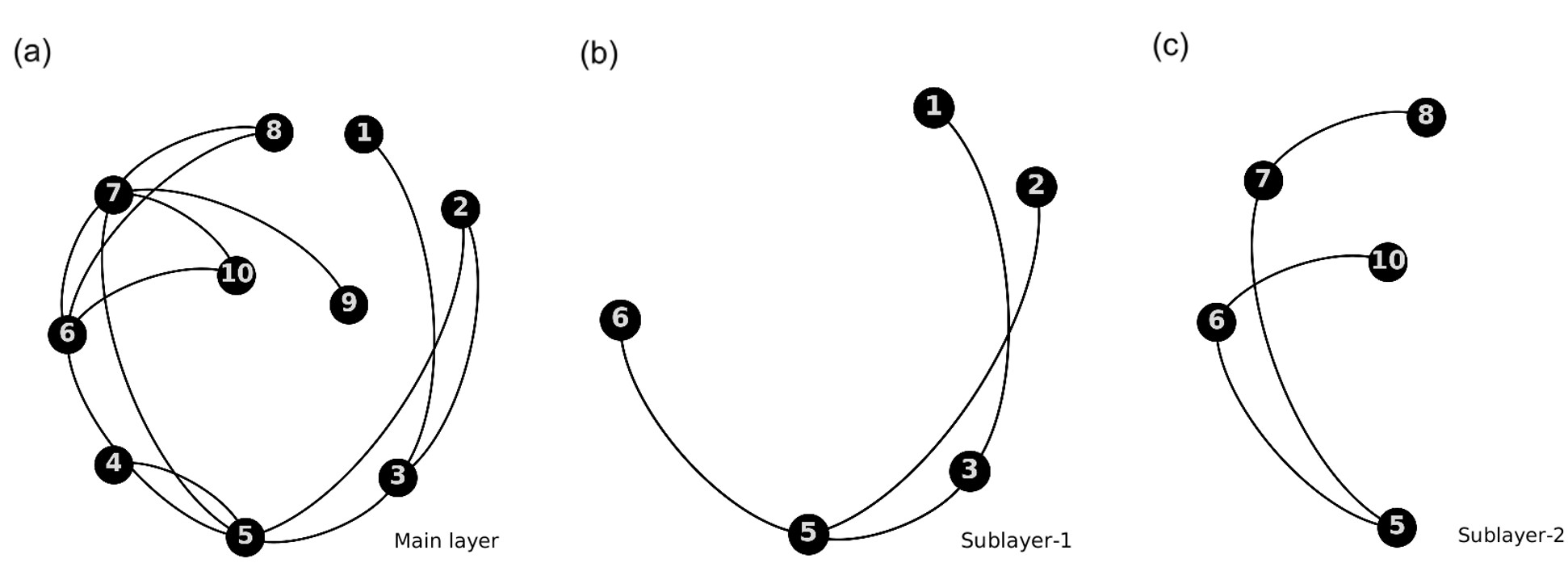}}
	\caption{{\bf Schematic illustration of an interdependent network subject to inter-layer resource competitions as proposed in Ref.\ \cite{chen2015extreme}:} In this particular example, the main layer in (a) has $N=10$ nodes. From this network, $M=2$ sublayers are constructed in subfigures (b) and (c) for illustrative purpose. First, we randomly select an arbitrary node from the network in subfigure (a). Randomly, then we select its $P_{m}.k$ neighbors. Here, $P_m=\dfrac{N_m}{N}=0.5$, where $N_m$ is the number of nodes in $m$-th layer for $m=1,2$. Until we reach $N.P_m=5$ number of total nodes in the sublayer, we again select $P_{m}.k$ neighbors for each selected neighbors. The combination of sublayer-1 and Sublayer-2 constitutes a multilayer network. Two sublayers compete  against each other for resources through the common nodes $5$ and $6$. Note that, few nodes like node $4$ and node $9$ may not present in any sublayers. All the figures are drawn in Gephi \cite{bastian2009gephi}.} 
	\label{fig_26}
\end{figure}

 The study of extreme events due to nodal flux fluctuations on single-layer transportation networks gains immense attention. However, the internal competition for common resources is one of the realistic thing, which uncovers a wide variety of phenomena in a society. Chen et al.\ \cite{chen2015extreme} considered a model that helps explore the extreme event dynamics using the dynamical features of the interdependent networks. Figure\ \ref{fig_26} is a schematic description of the model. There is a network consisting of $N=10$ nodes (See Fig.\ \ref{fig_26}(a)). This main network is divided into $M=2$ layers. Each of these sublayers provides packet transportation service within their respective sublayer through its nodes and links. Different layers can share few common nodes (node $5$ and node $6$ in this schematic figure given in Figs.\ \ref{fig_26}(b) and \ref{fig_26}(c)).

\begin{figure}[H]
	\centerline{
		\includegraphics[scale=0.335]{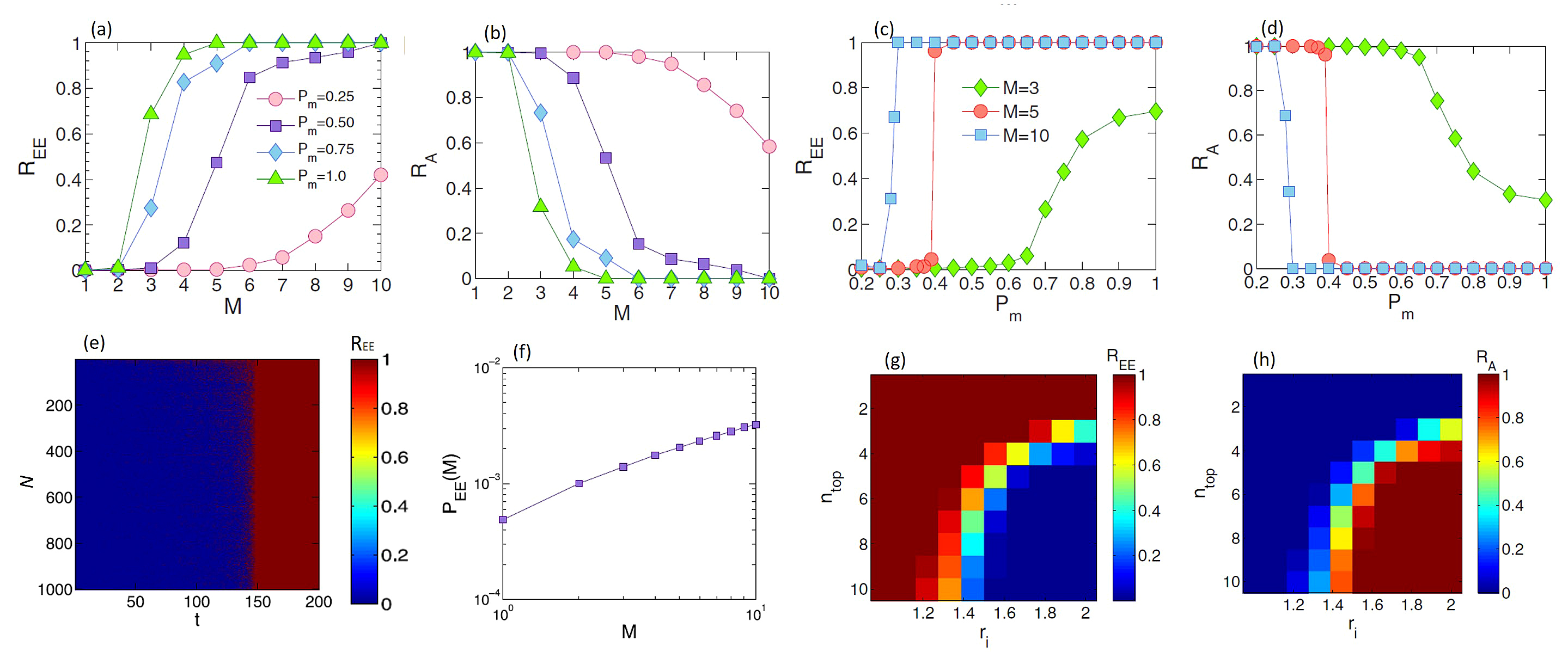}}
	\caption{(a) {\bf Extreme event occurrence rate $R_{EE}$ as a function of number of layers $M$:}  
		Clearly, $R_{EE}$ is almost close to $0$ for $M=1$. But the enhancement of number of layers leads to significant increment of $R_{EE}$. This implies excessive amount of load generation compared to a single layer network, due to internal resource competitions among the layers.  Although the capacity of a mono-layer network can handle all the generated load in the system, omitting the likelihood of the appearance of extreme events, the competitions in the interdependent network result in the occurrence of extreme events. (b) {\bf Packet arrival rate $R_A$ as a function $M$:} 
		If the network is free from extreme events, {\it i.e.,} $R_{EE} \approx 0$, then $R_A \approx 1$. In fact, the relation $R_A+R_{EE} \approx 1$ holds almost everywhere, after the system reaches to its equilibrium state. Even though, this relation does not hold in the transient state ({\it i.e.,} before the final equilibrium state) due to randomness and asynchronized updating. $R_A+R_{EE} = 1$ holds only when every single node of the entire network experiences extreme events. (c)-(d) {\bf Functional dependence of $R_{EE}$ and $R_A$ with the nodal coverage rate $P_m$, by keeping fixed the number of layers $M$:} With increasing $P_m$, there are more nodes will be shared among multiple layers. Thus, these common nodes enhance the interdependence of the entire system with severe internal competitions. Clearly, after a critical value of $P_m$, the curves of $R_A$ experience an abrupt decrease and simultaneously, the curves of $R_{EE}$ undergo a sudden increment. This shows a small change in the value $P_m$ can drive the entire system into a catastrophic state, where the entire system goes through extreme events globally. Since extreme events tend to occur on nodes shared by many layers, a straightforward control strategy is then to reduce the value of $P_m$ 	before it has passed the critical point. Also, note that the critical value of $P_m$ depends on the number of layers $M$. 
		(e) {\bf Emergence of two final equilibrium states in time for all nodes:} The system settles down between two final equilibrium states (i) (blue) free state ($R_{EE} \approxeq 0$), and (ii) (red) catastrophic state ($R_{EE} \approxeq 1$) in which every single node has an extreme event. There are no intermediate stable states because other intermediate states will eventually evolve into one of these final two states. The simulation is performed with $M=3$ layers and nodal coverage rate $P_m=1$ for a multilayer network of size $N=1000$. (f) {\bf Dependence of the probability for an extreme event to occur in a $M$-layer system, $P_{EE}(M)$ on $M$:} The monotonically increasing relation between $P_{EE}(M)$ and $M$ portray a qualitative explanation for the more recurrent manifestation of extreme events in systems with more layers. The positive correlation is shown here for $P_m=1$. (g)-(h) {\bf The extreme event occurrence rate $R_{EE}$ and the packet arrival rate $R_A$ in the parameter space:} All values of the parameters are same as in subfigure (a). $n_{top}$ is the number of top-degree hubs whose capacities are augmented. $r_i$ denotes ratio of the enhanced capacity to the original capacity. The system does not inhibit extreme events for small values of $n_{top}$ as well as for small values of $r_i$. Both figures contain a clear separating boundary, which distinguish between the free state ($R_A \approx 1$) and the catastrophic state ($R_{EE} \approx 1$). The extreme events occur at a small set of hubs, which is denoted by extreme event core. Reprinted figure with permission from Ref.\ \cite{chen2015extreme}.}
	\label{fig_27}
\end{figure}

\par Results \cite{chen2015extreme} indicate that extreme events have a propensity to arise on larger degree nodes, which oppose the earlier discussed results \cite{kishore2011extreme,kishore2012extreme} on single layer networks. In case of single layer networks, small degree nodes are more prone to exhibit extreme events. But, in case of multi-layer networks, the hub plays the decisive role in generating extreme events. There are two key parameters (i) number of layers, $M$ and (ii) the nodal coverage rate, $P_{m}=\dfrac{N_{m}}{N}$. Here, $N_{m}$ represents the number of nodes in a layer $G_{m}$, which is a subgraph of the original network $G$ consisting of $N$ nodes. Both these parameters possess a critical value beyond which the system undergoes a transition (See Figs.\ \ref{fig_27}(a)-(d)). Results are obtained through 20 independent simulation realizations for each of the $10$ network realizations of size $N=1000$ and average degree $\langle k \rangle  =  4$ with time-steps $500$. The quantities $R_{EE}$ and $R_A$ are averaged over last $300$ steps, as the system, in general, reaches its equilibrium state within $200$ time steps. The fraction of nodes at which extreme events occur at each time step is denoted here by $R_{EE}$. The ratio between the numbers of packets arriving and newly generated is denoted by $R_A$. The system does not have any intermediate state after the transient, except (i) a state free of extreme events, and (ii) a state completely dominated by extreme events (Fig.\ \ref{fig_27}(e)). The inclusion of layers leads to enhancement in the probability for an extreme event to occur in a $M$-layer system, $P_{EE}(M)$ as shown in Fig.\ \ref{fig_27}(f). To reduce the occurrence of extreme events significantly, an effective control strategy, based on locating the extreme events core formed by few hub nodes, is proposed. The extreme events core plays a vital role in the spreading of extreme events. By assigning larger capacities of the nodes in the core, one can enhance the network resilience. Instead of increasing the capacity of each node to prevent the occurrence of extreme events at a global scale, this idea is one of the low-cost control strategies as this is based on two simple outlines, viz., (i) finding the extreme event core, and (ii) increase the capacity of each node in the core. The effectiveness of this idea is shown in Figs.\ \ref{fig_27}(g) and \ref{fig_27}(h), respectively. The effective betweenness, and the empirical scaling law for betweenness centrality are developed to  understand the dynamics of extreme events.

\subsection{Spatial location dependent extreme events}
\par To explore the location dependence on the probability of extreme events, Amritkar et al.\ \cite{amritkar2018dependence} investigated the issue using the motion of a Brownian particle in a potential as members of an ensemble. The potential is treated as a location-dependent physical parameter in the dynamics. To affirm their claims, quadratic potential and periodic step function along with linear potential are considered in this study. The detected extreme events occur in the study due to inherent fluctuations in the model and thus, an integral part of the system. The probability of observing extreme events in a region is given by
\begin{equation}
	\begin{array}{lcl}\label{eq.51}
		\mathcal{F}= \sum_{k=\lfloor q \rfloor+1}^{N} \binom{N}{k} p^k (1-p)^{N-k},
	\end{array}
\end{equation}
where the total number of independent and noninteracting Brownian
particles is $N$ and $p$ is the probability of finding a Brownian particle in a region $R=\{x \in (c,d)\}$ is given by
\begin{equation}
	\begin{array}{lcl}\label{eq.52}
		p=p(R)=\int_{c}^{d} Q_{st}(x) dx.
	\end{array}
\end{equation}
Here, $Q_{st} (x)= Ae^{-\dfrac{ V(x) }{kT}}$ is the stationary solution of the Smoluchowski equation \cite{smoluchowski1916brownsche}, when the probability current is zero with $k$ being the Boltzmann constant, $A$ being the normalization constant, $V(x)$ is the potential and $T$ is the temperature of the heat bath. Instead of an uniform threshold $q$, depending on the location, $q$ is defined as
\begin{equation}
	\begin{array}{lcl}\label{eq.53}
		q=Np+d \sqrt{Np(1-p)},
	\end{array}
\end{equation}
where $d$ is the measure of rarity of extreme events. Hence, extreme events are defined as those events which occur in the tail of a probability distribution. Whenever, the number of particles in a region $R$ is greater than $q$, an extreme event occurs. The relation (\ref{eq.53}) clearly implies that any two among $p,N$ and $q$ are independent. Another consequence of the Eq.\ (\ref{eq.51}) that the probability distribution $\mathcal{F}$ is independent from $R$, though
depends on $(p,N,q)$.

\begin{figure}[H]
	\centerline{
		\includegraphics[scale=0.6]{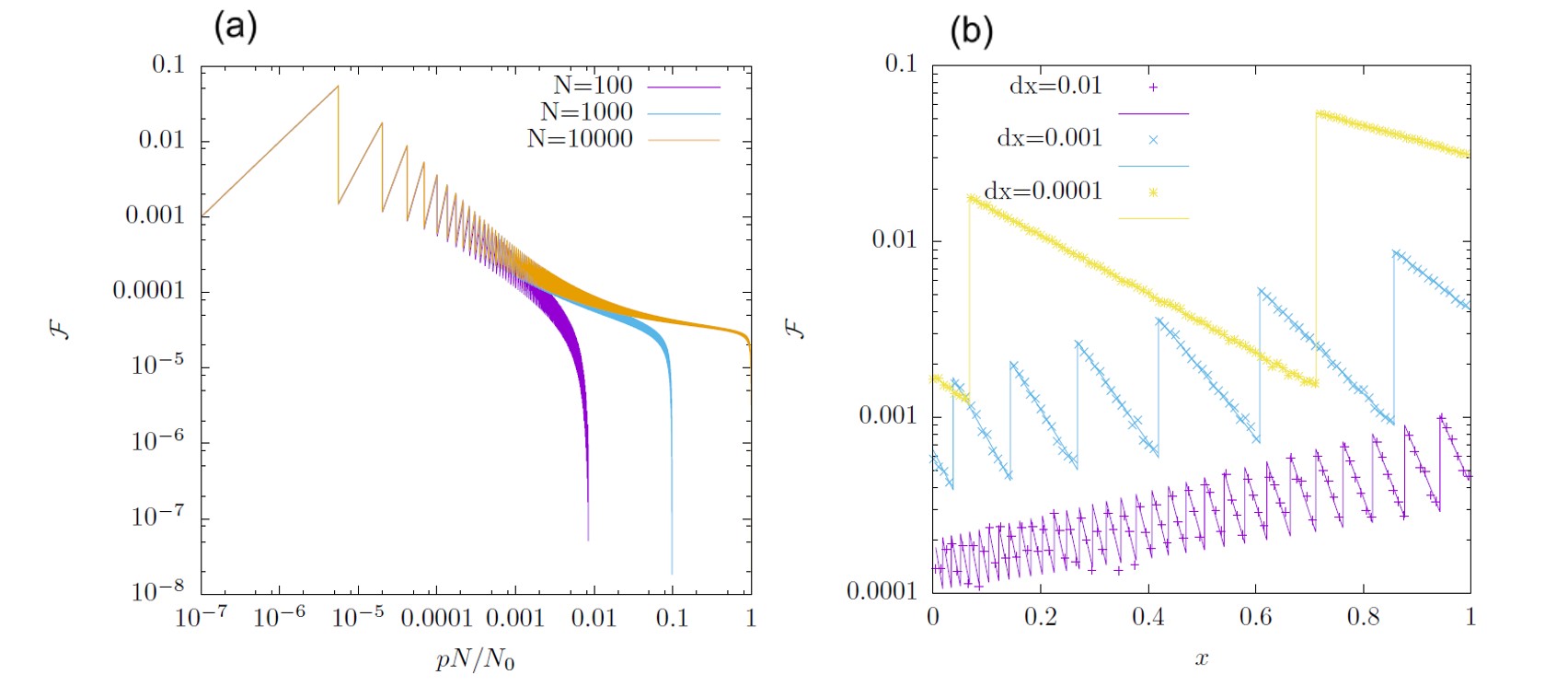}    }
	\caption{(a) {\bf $\mathcal{F}$ as a function of $\dfrac{pN}{N_0}$:} Here $N_0=10000$, the largest value of $N$ used in the article \cite{amritkar2018dependence}. Clearly, the probabilities of extreme events for different values of $N$ coincide for smaller values of $p$, although they reveal deviation for larger $p$. The oscillatory behavior of $\mathcal{F}$ undergoes an universal falling off, which divulges great consistency with the earlier perceived results in the letter \cite{kishore2011extreme}. (b) {\bf $\mathcal{F}$ as a function of $x$ for linear potential:} The values obtained by stochastic simulation of Brownian motion using Langevin equation are represented by symbols. The continuous curves are obtained analytically using the Eqs.\ (\ref{eq.53}), simplified form of (\ref{eq.51}) and $p=\dfrac{2e^c}{e^c-1} \sinh(\dfrac{cdx}{2}) e^{-cx}$, where $c$ is some temperature dependent constant. The results of the stochastic simulation agree well with the theoretical curves. Here $dx$ represents the width of the region $\bigg(x-\dfrac{dx}{2},x+\dfrac{dx}{2}\bigg)$. $\mathcal{F}$, on an average, increases as the potential increases. Same features of the probability, {\it i.e.,} increment of $\mathcal{F}$ is observed, when $dx$ decreases. Here, $c=2$ and $N=1000$. In both figures, $m=4$. 
		Reprinted figure with permission from Ref.\ \cite{amritkar2018dependence}.}
	\label{fig_23}
\end{figure} 
\par The above investigation suggests that this probability shows oscillations as well as exponential decay (See Fig.\ \ref{fig_23}(a)). Up to a certain value of $p$, the extreme event probability $\mathcal{F}$ coincides in Fig.\ \ref{fig_23}(a). Although beyond this critical value of $p$, large deviations occur. Figure \ref{fig_23}(b) reveals that if the width of region $dx$ decreases, then the probability of extreme events increases. However, for a fixed $dx$, the probability of extreme events increases with enhancement of $x$. The symbols in this Fig.\ \ref{fig_23}(b) represent the results of stochastic simulation of Brownian motion and the lines are the theoretical results. It is also notable that the probability of occurrence of extreme events, on an average, increases with the potential. In other words, $\mathcal{F}$ is smaller for smaller potentials, whereas larger for larger potential unless the average number of particles in an interval is less than one, which may be obtained for very large potentials and/or very small intervals.

\section{Prediction of extreme events in dynamical system}\label{prediction}
\par Prediction of extreme events is challenging, yet it is useful for mitigating such devastating events. Research efforts have been directed \cite{ghil2011extreme, farazmand2019extreme, alvarez2017predictability} toward the goal from two perspectives. One is the dynamical system approach \cite{kumarasamy2018extreme,cavalcante2013predictability, zamora2014suppression}, and another one is data-driven machine learning \cite{guth2019machine,qi2020using,pyragas2020using,lellep2020using}. Each of these approaches deals with questions of whether we can get an early warning indicator of a forthcoming extreme event before the trajectory of a system arrives in close vicinity of the region of instability. Addressing such questions with affirmative answers begin with recent research works \cite{farazmand2016dynamical}. Besides the dynamical system approach, recently, the machine learning approach has been used to predict extreme events. For this purpose, we need only recorded data of real events or simulated data of the observable from systems. The machine learning approach is helpful for prediction when no model is available but depends on data records of events. A few examples of both efforts are described below.

\subsection{Prediction of extreme events using dynamical instability}\label{instability}
\par In dynamical system approach, the challenging part is to locate the instability region embedded in the
phase space of a system. When the trajectory passes through this region before forming an extreme event,
then it may be possible to find an extreme event indicator. Here, we present a few explored system-dependent
studies, which help to predict extreme events.

\begin{figure}[H]
	\centerline{
		\includegraphics[scale=0.7]{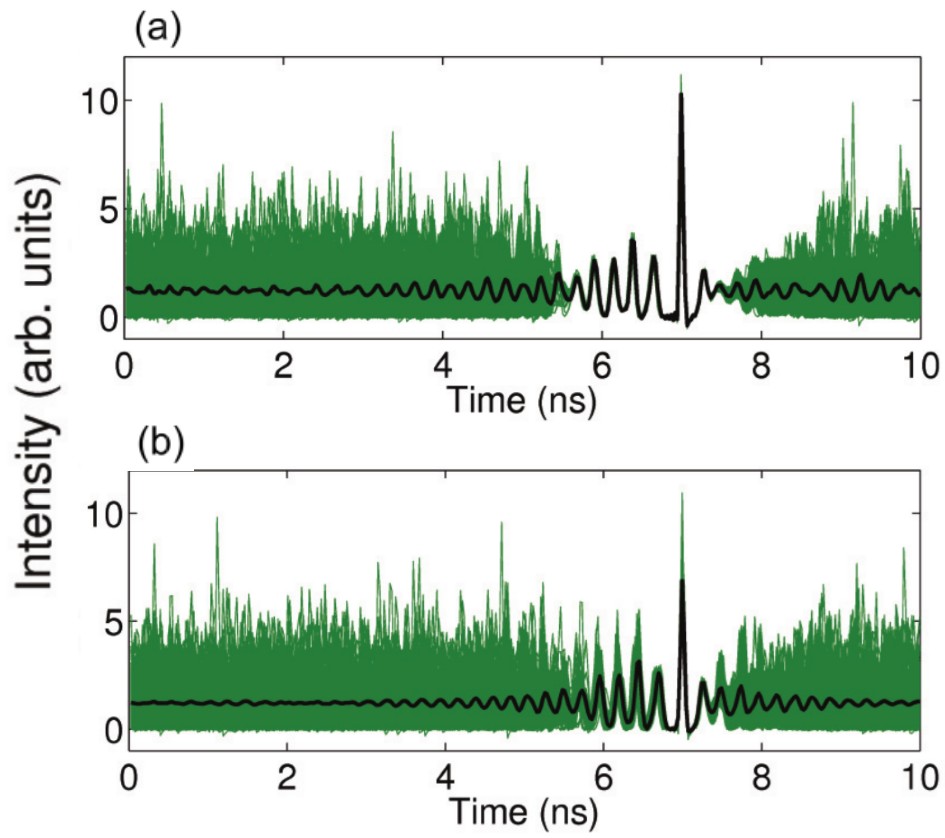}}
	\caption{{\bf Prediction of extreme events from the time evolution of the observable:} $459$ segments of equal time intervals around $459$ extreme events in a long time series are superimposed (green) in (a). All segments suddenly form a thin thread-like structure before the appearance of extreme events. This happens, due to a narrow channel embedded in the phase space, and a trajectory passes through it before the emergence of an extreme event. Extreme events are identified when maxima of laser intensity exceed a predefined threshold $T=m+8\sigma$. A similar process is applied for another case, where the extreme event qualifying threshold is chosen as $T=m+4\sigma$. For this case, the superposition of time segments is also plotted in (b). The average of the time segments is indicated by the thick black line. Compairing the two subfigures (a) and (b), a clear difference is observed that the first case (for $T=m+8\sigma$) is capable of a longer prediction time than the second case (for $T=m+4\sigma$). The band structure shrinks before a significant time, and then the rogue wave arises for the first case. But, for the second case, the significant time decreases. Parameters: $D=0.0$, $\alpha=3.0$, $P_{inj}=60$ n$s^{-2}$, $\nu =2.20$, $\gamma_n=1$ n$s^{-1}$, $\kappa=300$ n$s^{-1}$, $\Delta \nu=0.49$. Reprinted figure with permission from Ref.\ \cite{zamora2013rogue}.}
	\label{fig_28}
\end{figure}

\par An attempt is made in Ref.\ \cite{zamora2013rogue} to address the question of predictability of extreme events in an optically injected laser. For this purpose, a long time series of pulse intensity is captured from numerical simulation of the system described by Eq.\ \eqref{eq.2}. From the time series, $459$ segments are collected for equal time duration each around an extreme event and plotted all of them on top of each other as shown in Fig.\ \ref{fig_28}(a). Here, an extreme event is identified if a local maximum value of the observable (laser intensity) exceeds the extreme event qualifying threshold $T=m+8\sigma$. A special type of thread-like structure is made centered around extreme events as shown in Fig.\ \ref{fig_28}(a) and for the rest of the part, a band structure is noticed. As an extremely large intensity event approaches, the band structure dwindles, and all segments coincide in a thin curved line, which looks like a thread. It happens because the trajectories travel across a narrow channel or `rogue wave door' (region of instability) \cite{reinoso2013extreme} for each time before occurring extreme events. We have already explained the reason behind the generation of extreme events in optically injected laser system in Sec.\ \eqref{interior}. The external crisis-like process is responsible for generating extreme events in this system. The shrinking of the band starts quite ahead of time occurrence of the extreme event. This gives a positive impression that a knowledge of the laser intensity as a function of time can provide a clue on the prediction of large intensity events well before a time it appears. Notably, an increment of the threshold value leads to a longer prediction time for the appearance of extreme events.
For a comparison, a lower threshold $T=m+4\sigma$ is considered, and once again $459$ segments of data around the extreme events are superimposed in Fig.\ \ref{fig_28}(b), where narrowing of the band structure starts much later, thereby reducing the prediction time before the appearance of extreme events.    

\par
\begin{figure}[H]
	\centerline{
		\includegraphics[scale=0.65]{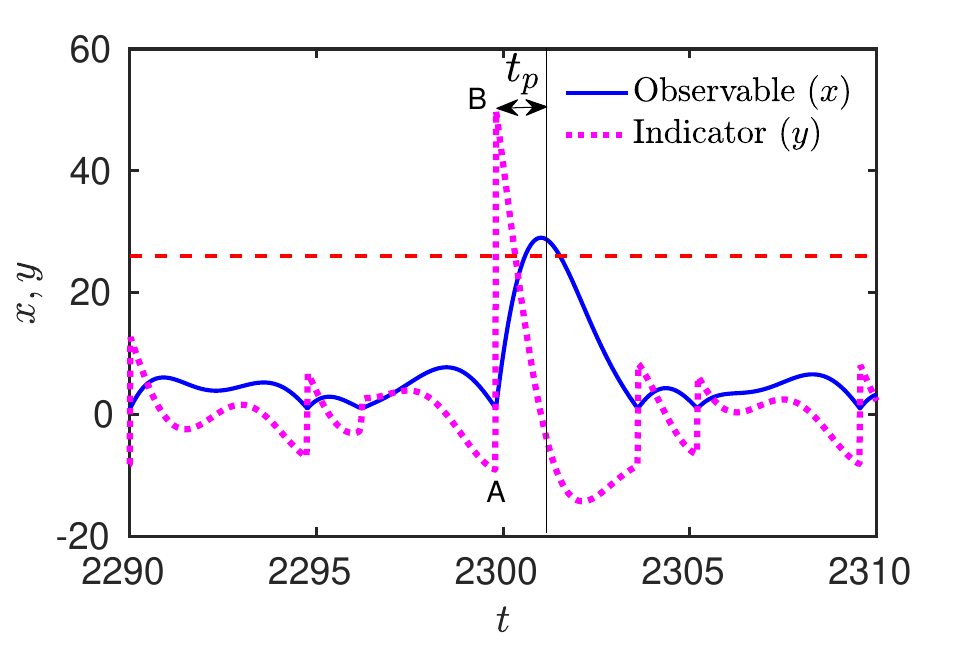}}
	\caption{{\bf Prediction of extreme events in MEMS:} Time evolution of $y$ (dotted magenta) and observable $x$ (solid blue) are portrayed here. When $y$ suddenly jumps to a large value from the point A to the point B, then $x$ attains its maximum value after a time $t_{p}$. Since sliding distance crosses a predefined threshold (which is not shown here), then an extreme event occurs as the maximum value of $x$ exceeds a threshold (red dashed horizontal line). It is reported in Ref.\ \cite{kumarasamy2018extreme} that if sliding distance crosses a predefined threshold, then the extreme events appear in the time evolution of $x$. Otherwise, extreme events do not appear. Parameters: $\gamma = 0.709$, $\beta = 0.318$, $\omega = 1.28$, and $\alpha=7.99$.} 
	\label{fig_29}
\end{figure}

 In Ref.\ \cite{kumarasamy2018extreme}, an early warning indicator is identified for the prediction of extreme events in the micro-electro-mechanical system (MEMS) described by Eq.\ \eqref{eq.11}. We have already discussed the mechanism of extreme events for this MEMS system in Sec.\ \eqref{SB}. The crucial observation is that, when the trajectory of this system comes sufficiently close to the discontinuous boundary $x=1$, then it is repelled away to produce a large excursion. The distance, a trajectory travels parallel to the discontinuous boundary at $x=1.0$ tangentially, is called the sliding distance. If the sliding distance crosses a predefined threshold, and the time evolution of $y$ reaches its maximum value, then the value of the $x$ variable gradually increases and gains its maximum value. Hence, extreme events occur in the system \eqref{eq.11} through the sliding bifurcation. In Fig.\ \ref{fig_29}, time evolution of $x$ (solid blue) and $y$ (dotted magenta) are shown for a short time interval. The variable $y$ suddenly jumps from point A to point B, when trajectory traverses parallel to $x=1.0$ line. Here, the difference between two points A and B is the sliding distance.
So, the maximum value of $y$ arises before the appearance of the maximum value of $x$ as shown in Fig.\ \ref{fig_29}. Since the maximum value of $x$ exceeds a predefined threshold (red dashed line), so an extreme event emerges. Here, the maximum value of $y$ plays as an extreme event indicator, whereas $x$ is the observable. $t_{p}$ is the significant time gap between extreme event indicator and occurrence of an extreme event in Fig.\ \ref{fig_29}. A similar approach has been made for the prediction of extreme events in the $CO_2$ laser model given in Eq.\ \eqref{eq.4}.

\begin{figure}[H]
	\centerline{
		\includegraphics[scale=0.4]{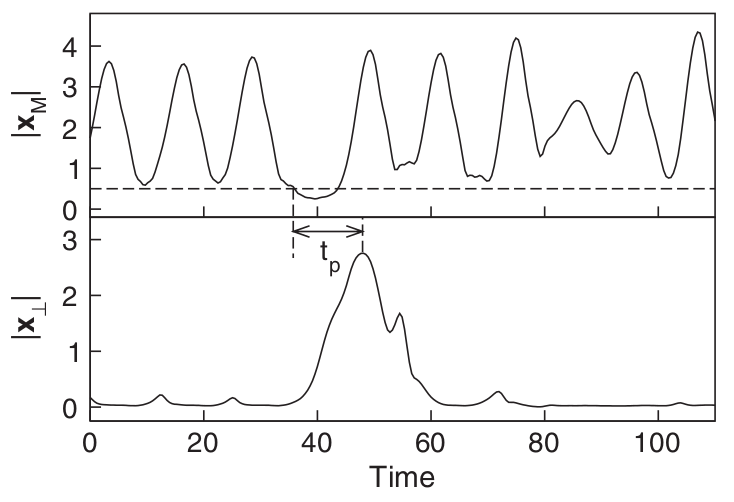}}
	\caption{ {\bf Time evolution of master subsystem and synchronization error from experimental observation:} As soon as, the time evolution of $|{\bf x_{M}}|$ (upper panel) drops below a predefined threshold $|{\bf x_{M}}|_{th}=0.5$ (horizontal dashed line), the time evolution  of $|{\bf x_{\perp}}|$ shows a rise and exhibits an extreme event. Crossing between $|{\bf x_{M}}|_{th}$ and time evolution of $|{\bf x_{M}}|$ plays as a precursor of a forthcoming extreme event. The time difference between that crossing and the appearance of an extreme event is denoted by $t_p$. 
		Reprinted figure with permission from Ref.\ \cite{cavalcante2013predictability}.}
	\label{fig_30}
\end{figure}

\par A prediction scheme is also suggested in Sec.\ \eqref{onoff} that uses a master-slave coupled electronic circuit to explore extreme events as described in Sec.\ \eqref{onoff}. When the trajectories of both the master and slave subsystems are confined in the synchronization manifold, two subsystems are collectively synchronized. But, when the trajectory of the master subsystem moves towards the saddle equilibrium point, located at the origin, then the master-slave system loses synchrony and it is reflected as a large signal in the error dynamics that is identified as an extreme event. Based on this study, a result is noticed that,
when the absolute value of ${\bf x_{M}}$ approaches zero, after some time, an extreme event may occur in the observable $|{\bf x_{\perp}}|$. This signature is used as an early warning signal of anticipated extreme events. In upper panel of Fig.\ \ref{fig_30}, when the value of $|{\bf x_{M}}|$ crosses below a predefined threshold  $\left|{\bf x_{M}}\right|_{th}$ (horizontal dashed black line), then after time $t_p$, an extreme event occurs in observable $|{\bf x_{\perp}}|$, as shown in lower panel of Fig.\ \ref{fig_30}. The predefined threshold values are $\left|{\bf x_{M}}\right|_{th}=0.32$ and $\left|{\bf x_{M}}\right|_{th}=0.5$ as per their consideration in case of numerical simulation and experiment, respectively. As the trajectory of the master subsystem reaches near the saddle origin, those trajectories will suffer a repulsion along the transverse direction of the invariant manifold for which synchronization error will become non-zero. Hence, the value of $|{\bf x_{\perp}}|$ becomes large enough, so that local maximum value of $|{\bf x_{\perp}}|$ is considered as an extreme event, which is reflected as dragon-king events in the probability density function (PDF). Here, the time difference between precursor and peak of an extreme event is denoted by $t_{p}$ as shown in Fig.\ \ref{fig_30}.

\begin{figure}
	\centerline{
		\includegraphics[scale=0.5]{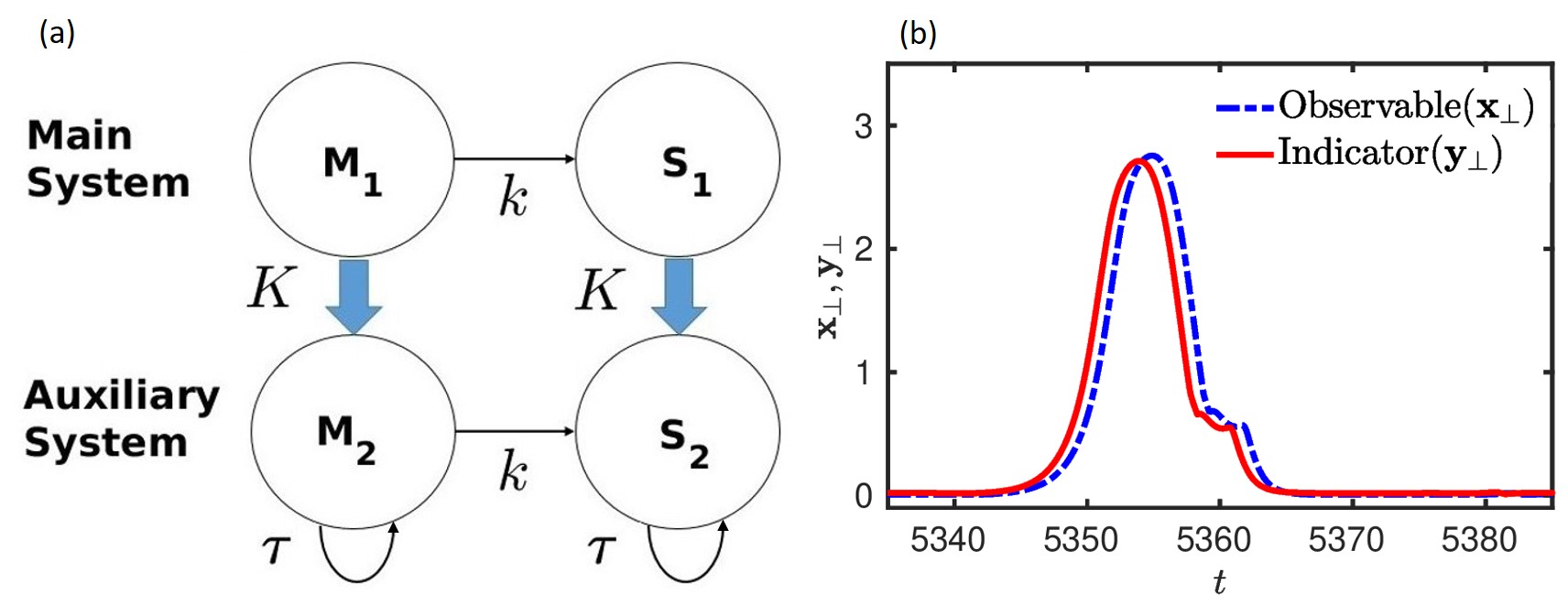}}
	\caption{(a) {\bf Schematic diagram of main and auxiliary systems:} Master subsystem $M_1$ interacts with the slave subsystem $S_1$ with a coupling strength $k$ and forms the main system. Similarly, a combination of master-slave subsystems ($M_2$ and $S_2$) represents the auxiliary system with the same coupling strength. In both cases, the master subsystem unidirectionally interacts with the slave subsystem. Again, $M_1$ unidirectionally interacts with $M_2$ with coupling strength $K$. Similarly, $S_1$ is unidirectionally coupled with $S_2$ by coupling strength $K$. Self-delayed feedback $\tau$ is applied on both the master and slave subsystems of the auxiliary system. (b) {\bf Time evolution of observable exhibiting extreme event and time evolution of extreme event predictor:} Time evolution of ${\bf x_{\perp}}$ (dashed blue) and ${\bf y_{\perp}}$ (solid red) are shown here. There is a time gap between two synchronization errors of main and auxiliary systems. Before occurring extreme events in the observable ${\bf x_{\perp}}$, the local maximum value of ${\bf y_{\perp}}$ takes place. So, the time evolution of ${\bf y_{\perp}}$ can able to predict the extreme events occurring in the time evolution of ${\bf x_{\perp}}$. 
	} 
	\label{fig_31}
\end{figure}

\par An alternative method has been proposed by Zamora-Munt et al.\ \cite{zamora2014suppression} for prediction of extreme events in unidirectionally coupled electronic circuits using anticipation synchronization \cite{voss2000anticipating, matias2017anticipated}. For this purpose, a set of two systems is considered. One system is unidirectionally coupled master-slave electronic circuits \cite{cavalcante2013predictability,gauthier1996intermittent} treated as the main system. 
Another system is unidirectionally coupled master-slave electronic circuits subject to a negative self-delayed feedback. Extreme events occur in the synchronization error of the system due to on-off intermittency. The second system with negative self-delayed feedback is treated as the auxiliary system, which helps to predict extreme events generated in the main system. 
  This auxiliary system is unidirectionally coupled with the main system in such a way that the dynamics of the main system is not changed. For a clear visualization of the scenario, a schematic diagram is portrayed in Fig.\ \ref{fig_31}(a). The set of state variables of the auxiliary system is given by $\{VA_{1,m}, VA_{2,m}, IA_{m},VA_{1,s},VA_{2,s},IA_{s}\}$, while for the main system, $\{VM_{1,m}, VM_{2,m}, IM_{m}, VM_{1,s},VM_{2,s},IM_{s}\}$ is the set of state variables. Here, $m$ stands for the master subsystem and $s$ signifies the slave subsystem. Now, synchronization errors of the main and auxiliary systems are defined as\\
\begin{equation}
	\begin{array}{l}\label{eq.8888}	
		{\bf x_{\perp}}=|VM_{1,m}-VM_{1,s}|+|VM_{2,m}-VM_{2,s}|+|IM_{m}-IM_{s}|,
		\\
		\\
		{\bf y_{\perp}}=|VA_{1,m}-VA_{1,s}|+|VA_{2,m}-VA_{2,s}|+|IA_{m}-IA_{s}|,
	\end{array}
\end{equation}
respectively. Here, ${\bf x_{\perp}}$ is the observable, whereas ${\bf y_{\perp}}$ plays a role of an indicator for the prediction of extreme events in the temporal evolution of the observable. As a result of attractor bubbling, trajectories of master and slave subsystems in the main system become desynchronized intermittently. So, synchronization error ${\bf x_{\perp}}$ becomes positive during this desynchrony. In Fig.\ \ref{fig_31}(b), the time evolution of ${\bf x_{\perp}}$ (blue dashed) and ${\bf y_{\perp}}$ (red solid) are plotted in short time interval. 
 The extreme event in the time evolution of ${\bf x_{\perp}}$ arises after the occurrence of the large bubbling event in the time evolution of ${\bf y_{\perp}}$ as
the main and auxiliary systems lead to anticipated synchronization. The result is shown in Fig.\ \ref{fig_31}(b). Here, a significant time difference between the time evolution of ${\bf x_{\perp}}$ (observable) and ${\bf y_{\perp}}$ (predictor) is observed.

\subsection{Prediction of extreme events using machine learning approach}


Reservoir computing is one of the powerful tools for model-free prediction of extreme events in dynamical systems. This approach employs a nonlinear input-output neural network with randomly generated values of the parameters, and uses linear regression to choose ``output weights" that fit the network output to a set of ``training data". This approach is computationally simpler compared to the other artificial neural network approaches, since only output weights are adjusted during the training process, while the network parameters are fixed. Recently, the reservoir computer has been successfully used to predict various low-dimensional and spatiotemporal chaotic systems which do not belong to the class of models generating extreme events. The underlying principle for reservoir computing is as follows:

\par Consider a dynamical system in the form of
\begin{equation}
	\begin{array}{l}\label{eq0}	
	\dot{{\bf x}}=f({\bf x}, \alpha).
	\end{array}
\end{equation}
We also assume that this model is capable of generating extreme events. We process the signal $x(t)=\{x_1(t), x_2(t),\cdots, x_M(t)\}$ to $u_i(t)=\frac{x_i(t)-\langle x_i(t) \rangle}{\sigma'}$   in such a way that the input signal has zero mean and unit variance, where $\sigma'$  is the standard deviation, $\langle x_i(t) \rangle$  is the mean of the given data and the angle bracket denotes time average. The reservoir computer has three components: an input layer, a nonlinear reservoir network with $N$ dynamical reservoir nodes, and a linear output layer.
\begin{figure}[H]
	\centerline{
		\includegraphics[scale=0.5]{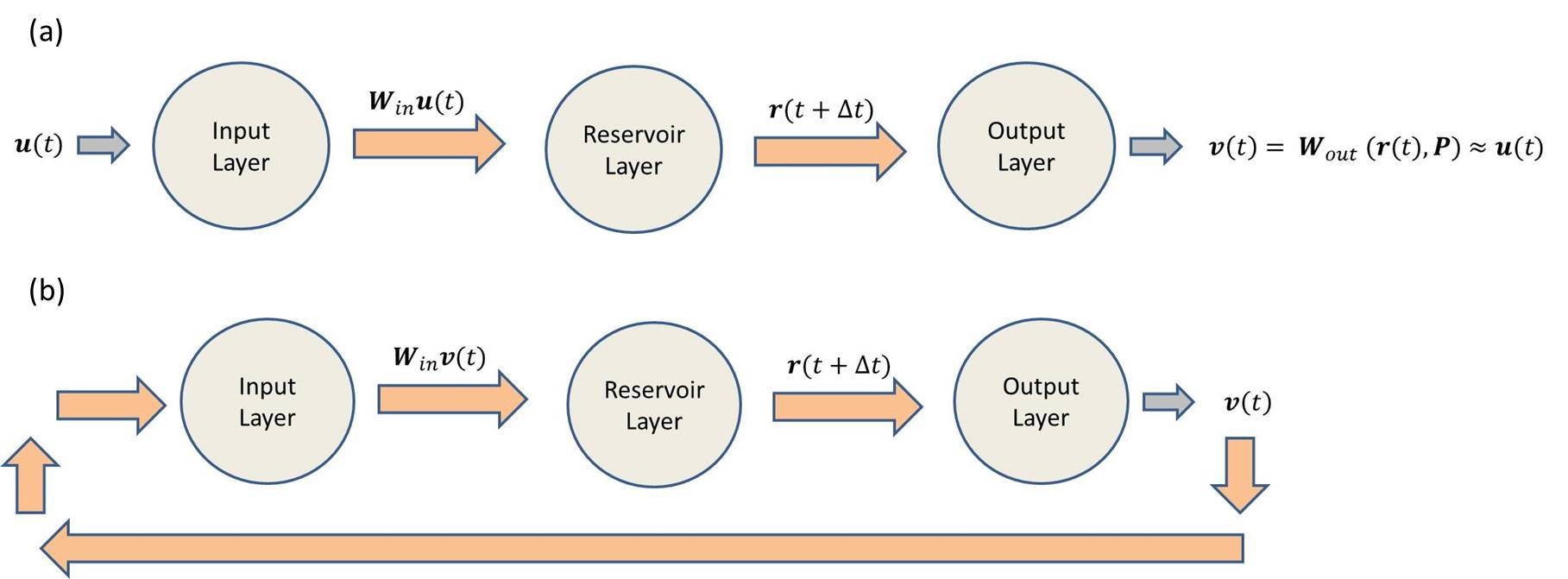}}
	\caption{ {\bf (a) The training phase corresponding to listening reservoir:} An input vector $\mathbf{u}(t) \in \mathbb{R}^{M \times 1}$ is transferred through the input layer at discrete time $t$. The input layer consists linear weight matrix $\mathbf{W}_{in} \in \mathbb{R}^{N \times M}$, chosen randomly from a uniform distribution $[-\chi, \chi]$. $\chi$ signifies the scalar input strength. The matrix $\mathbf{A}$ is a sparse random matrix, where the average degree of a reservoir node is $D$. The density $\frac{D}{N}$, {\it i.e.,} the non-zero  elements are randomly generated from a uniform distribution in the interval $[-1, 1]$. Both the matrices $\mathbf{A}$ and $\mathbf{W}_{in}$ are initially generated randomly, but then kept fixed for further iterations. The output layer contains $\mathbf{W}_{out} \in \mathbb{R}^{N \times M}$, which depends functionally on the matrix $P$. The elements of $P$ are the large number of parameters, which are trained to minimize the mean squared difference between the desired state and $\mathbf{v}(t)$ using Tikhonov regularized regression process.
		{\bf (b) The prediction phase corresponding to predicting reservoir:} After the initial training session, the future evolution of $\mathbf{u}(t)$ is predicted from earlier processed $\mathbf{v}(t)$, by replacing the input vector $\mathbf{u}(t)$. The evolution of this feedback loop is demonstrated in the bottom schematic figure (b). The parameters of the reservoir are chosen in such a way that all of the conditional Lyapunov exponents of the training reservoir dynamics conditioned on $\mathbf{u}(t)$ are negative for sufficiently large $t$ so that the reservoir state $\mathbf{r}(t)$ does not depend on initial conditions.}
	\label{fig_32}
\end{figure}

\par The state of the reservoir is determined by the $N$-dimensional state vector ${\bf r}(t)$  that satisfies a discrete time deterministic model
\begin{equation}
\begin{array}{l}\label{mla}	
{\bf r}(t+\Delta t)=f({\bf r}(t), {\bf u}(t)),
\end{array}
\end{equation}
where $\Delta t$  is the time-step and  ${\bf u}(t)$ is the input signal of the reservoir. There are many different ways to choose the nonlinear function $f$  of the reservoir. Using Tikhonov regularization, the output signal ${\bf v}(t)$  of the reservoir is minimized so that the output gives a good approximation of the input signal ${\bf u}(t)$. The generalized synchronization (GS) \cite{hramov2005generalized,hramov2005generalizedb,moskalenko2010generalized} between the system \eqref{eq0} and the reservoir (\refeq{mla}) means that the reservoir state ${\bf r}(t)$   becomes asymptotically a continuous function $\phi$  of the system state  ${\bf x}(t)$, {\it i.e.,} ${\bf r}(t)\sim \phi({\bf x}(t))$ as $t \to \infty$. The GS occurs if all conditional Lyapunov exponents are negative. The GS is a necessary condition for the reservoir to predict the input signal ${\bf u}(t)$. 
\par Prediction of extreme events by observing the dynamics of the output variable ${\bf v}(t)$, which can be treated as an early warning precursors (indicators) of the extreme events, is one of the active research topics nowadays. The prediction of extreme events is complicated by the fact that a local Lyapunov exponent \cite{abarbanel1992local,eckhardt1993local,wolff1992local} close to the  extreme events,  may be significantly larger than the global (average) Lyapunov exponent of a chaotic attractor.
Predictions of extreme events in deterministic systems can be done from the fact that the current state of the system uniquely determines its future state, but it is limited by a sensitive dependence on the initial conditions. But, yet extreme events exhibit a rich variety of statistical transition from symmetric near-Gaussian statistics to a highly skewed probability density function. The important questions to ask are whether the reservoir computer can be trained to learn the complex hidden structures in the highly nonlinear dynamics purely from data, and what are the essential structures required in the network to gain the ability to capture extreme events?

 
 \par Extreme events can be isolated low-probability events, or they can often be intermittent and even frequent in space and time. The curse of dimension forms one important obstacle for the accurate prediction of extreme events in large complex systems, where both novel models and efficient numerical algorithms are required. Another key aspect of extreme event is its exceptional amplitude from the average behavior. Prediction of the amplitude of extreme events is still now an incalculable puzzle. A dataset from a measured chaotic observable that consists of extreme events, can be separated into three distinct classes, 
 \begin{enumerate}
 	\item 	average behavior of the observable around the central tendency `$m$',
 	\item	the abrupt behavior of the observable in the regime $(m+d \sigma,\infty)$, where $m+d \sigma$ is taken as the extreme event qualifying threshold with  $d\in \mathbb{R}$, and finally,
 	\item 	the data lying within $(m, m+d \sigma)$ . 
 \end{enumerate}
 
\par Pyragas et al.\ \cite{pyragas2020using} used the reservoir computer for prediction of extreme events in systems studied in the articles \cite{ansmann2013extreme,cavalcante2013predictability,karnatak2014route}. Just like the pioneer work by Jaeger and Haas \cite{jaeger2004harnessing}, they introduced a reservoir computer consisting of three components. The input layer contains $M$ input nodes. Each of these $M$ components corresponds to each component of $\mathbf{u}(t)$. This $\mathbf{u}(t)$ is the normalized input vector so that it has zero mean and unit variance. The input matrix $\mathbf{W}_{in}$ helps to map the input vector $\mathbf{u}(t)$ to the reservoir state space $\mathbf{r}(t) \in \mathbb{R}^{N}$. $\mathbf{W}_{in}$ is drawn randomly from a uniform distribution $[-\chi, \chi]$, where $\chi$ is the scalar input strength. The reservoir dynamics is given by

\begin{equation}
\begin{array}{lcl}\label{eq.60}
\mathbf{r}(t+\Delta t)= (1-\alpha)\mathbf{r}(t)+\alpha \tanh(\mathbf{A}\mathbf{r}(t)+\mathbf{W}_{in} \mathbf{u}(t)+\xi \mathbf{1}),
\end{array}
\end{equation}

\par where $\Delta t$ is the time step, $\xi$ is a scalar, $\alpha \in (0,1]$ is the leaking rate and $\mathbf{1}$ denotes a column vector of ones. The reservoir adjacency matrix $\mathbf{A}$ is a sparse $N \times N$ random matrix, where the non-zero entries are drawn independently from a uniform distribution $[-1,1]$. Thus, $\mathbf{A} \in \mathbb{R}^{N \times N}$ is a function from the input state space to the output state space. $\mathbf{A}$ is multiplied by a positive factor to rescale the largest value of the magnitudes of its eigenvalues, commonly known as spectral radius, to the desired predefined value $\rho$. Here, $\tanh(\overrightarrow{b})$ is a vector, whose components are hyperbolic tangents ${[\tanh{b_1}, \tanh{b_2}, \cdots, \tanh{b_n}]}^{Tr}$ of the corresponding components of the argument vector $\overrightarrow{b}={[b_1, b_2, \cdots, b_n]}^{Tr}$ and ${Tr}$ denotes the transpose of the matrix. 

\par The transient time $t_{0}$ should be large enough so that the state of the reservoir is essentially independent of its initial state at time $t=0$. Starting from a random initial state $\mathbf{r}(–t_0)$, the reservoir evolves following Eq.\ (\ref{eq.60}) with input $\mathbf{u}(t)$. The data is recorded for distinct $\Upsilon$ reservoir states $\{\mathbf{r} (\Delta t), \mathbf{r} (2 \Delta t), \cdots, \mathbf{r} (\tilde{T}) \}$ for the training $0 < t \le \tilde{T}=\Upsilon \Delta t$. The output of the listening reservoir is given by
\begin{equation}
\begin{array}{lcl}\label{eq.61}
\mathbf{v}(t)= \mathbf{W}_{out}^{Tr} \mathbf{r}(t).
\end{array}
\end{equation}
The elements of the matrix $\mathbf{W}_{out} \in \mathbb{R}^{N \times M}$ is adjusted by minimizing the following quadratic form with respect to $\mathbf{W}_{out}$

\begin{equation}
\begin{array}{lcl}\label{eq.62}
\sum_{j=1}^{K} {|| \mathbf{W}_{out}^{Tr} \mathbf{r} (j \Delta t) - \mathbf{u} (j \Delta t) ||}^{2} + \beta \text{tr} (\mathbf{W}_{out}^{Tr} \mathbf{W}_{out}),
\end{array}
\end{equation}
where $||\mathbf{b}||^{2}= \mathbf{b}^{Tr} \mathbf{b}$ denotes the sum of the squares of elements of $\mathbf{b}$ and $tr$ is the trace of a square matrix. The ridge regression parameter $\beta > 0$ is chosen such that regularization term $\beta \text{tr} (\mathbf{W}_{out}^{Tr} \mathbf{W}_{out})$ helps to avoid overfitting of $\mathbf{W}_{out}$. If the training is successfully occurred based on the Tikhonov regularization, then the reservoir output should yield a good approximation to the input, $\mathbf{v}(t) \approx \mathbf{u} (t)$ for $t > \tilde{T}$. Interested readers may consult Refs.\ \cite{jaeger2004harnessing,pathak2018model,lu2017reservoir} for further understanding of the reservoir computing.  

\par To verify the effectiveness of the proposed algorithm \cite{pyragas2020using}, two coupled Fitzhugh-Nagumo systems are considered in the form of Eq.\ \eqref{eq.5}. All the values of parameters are same as already discussed in the Sec.\ \eqref{imperfect}.
The distance between two successive extreme events in this two coupled system \eqref{eq.5} is approximately around $T_{EE} \approx 100$. The necessary condition for the generalized synchronization between the system (\ref{eq.5}) and the listening reservoir is verified using the negativity of all conditional Lyapunov exponents. The maximal conditional Lyapunov exponent is approximately equal to $\lambda \approx -0.036$. Using the reservoir computer, extreme events can be predicted and up to the prediction time $\tau \le 40$, the RMS (root mean square) error among the predicted output of the reservoir and the two coupled system is relatively small. Based on this predicted signal,  a control $p_1$ is applied only to the variable $x_1$, whenever the predicted signal crosses a predefined value $s^{*}=0.7$. The perturbation amplitude is $\epsilon=10^{-3}$. Due to this perturbation, it destroys the generalized synchronization and as a result of that the reservoir computer fails to provide accurate prediction of the input signal immediately after the control pulse. However, the characteristic time of the generalized synchronization $\frac{1}{|\lambda|} \approx 27.78$ is significantly less than $T_{EE}$, which helps to re-gain the generalized synchronization between them within two successive applied control pulses. Increment of system size reduces the prediction time $\tau$. For $101$ globally coupled Fitzhugh-Nagumo units, $\tau$ is approximately around $16$, which is almost half the width of the spikes of extreme events. Although the prediction time is comparatively lower, but still sufficient enough to suppress extreme events with small perturbations. The route of generation of extreme events for this system \eqref{eq.5} is already discussed in Sec.\ \eqref{imperfect}. For this system, the control is applied to all $x_i$'s. The perturbation strength $\epsilon=5\times10^{-3}$ is larger due to the choice of bigger system size. The system is not perturbed for the next $25 \Delta t$ time units after each control event. 

\par For further validation of the proposed strategy \cite{pyragas2020using}, the two nearly identical unidirectionally
coupled chaotic oscillators are contemplated in the master-slave configuration, originally proposed by Cavalcante at al.\ \cite{cavalcante2013predictability}. We have already discussed a prediction mechanism for this system in the Sec.\ \eqref{prediction}. Instead of predicting the extreme events for this system (\ref{eq.12}), the prediction of the proposed precursor, as discussed in Sec.\ \eqref{prediction}, is found to be advantageous by reservoir computer. Not only this strategy helps to reduce the dragon kings effectively, but also it requires less average control energy compared to the proposed strategy by Cavalcante et al.\ \cite{cavalcante2013predictability} almost by two orders of magnitude.

\par Recently, a comparative study \cite{amil2019machine} for predicting the amplitude of an upcoming chaotic pulse of an optically injected semiconductor laser consisting of rogue waves is discussed to understand the role of various machine learning algorithms in terms of data requirements and robustness. Using three Deep Learning frameworks, namely Multi-Layer Perceptron, Convolutional Neural Network, and Long Short-Term Memory, Meiyazhagan et al.\ \cite{meiyazhagan2021model} predicted the extreme events in a parametrically driven nonlinear dynamical system. Apart from
those works, the prediction of extreme events in the ensemble of identical coupled oscillators \cite{chowdhury2021extreme} has been made using deep learning architecture, viz.\ long short-term memory (LSTM). Another work\ \cite{ray2021optimized} reports that the convex combination of forecasts achieved from three frameworks, viz.\ feed-forward neural networks, reservoir computing, and long short-term memory, can predict the dynamics consisting of extreme events better than the forecasting of individual framework.

\par Prediction of extreme events can also be done using different neural dynamics and network topologies. Lellep et al.\ \cite{lellep2020using} considered a fully connected feed forward neural network to forecast upcoming extreme events. To establish the claims, two-dimensional H\'{e}non map 

\begin{equation}
\begin{array}{lcl}\label{eq.63}
x_{n+1}=1-a{x_{n}}^{2}+y_{n},\\
y_{n+1}=bx_{n},
\end{array}
\end{equation}
is considered. To set the system dynamics in the chaotic regime, values of the parameters are taken as $a=1.4$ and $b=0.3$. The prediction task is assigned here by means of a predefined threshold $\theta$. This approach is quite different from the early discussed article \cite{pyragas2020using}. Suppose, we have $k$ distinct data points $(x_i,y_i)$ for $i=1,2, \cdots, k$. We want to predict in advance whether $y_{k+\alpha} \ge \theta$, {\it i.e.,} whether $y$-component of the trajectories passes the threshold $\theta$ at exactly after $\alpha$ iterations or not. Of course, trajectories may pass the threshold at any intermediate iterations $i=1,2,\cdots,(\alpha-1)$. Here, $\alpha$ is the prediction time. But, prediction of the trajectories at those intermediate iterations is not subject of interest here. Without loss of generality, $\theta=0.3$ is chosen. The prediction task is actually equivalent to the classification task here, as the state space is divided into two regions, (i) $y_n \ge \theta$, and (ii) $y_n < \theta$.  

\begin{figure}[H]
	\centerline{
		\includegraphics[scale=0.45]{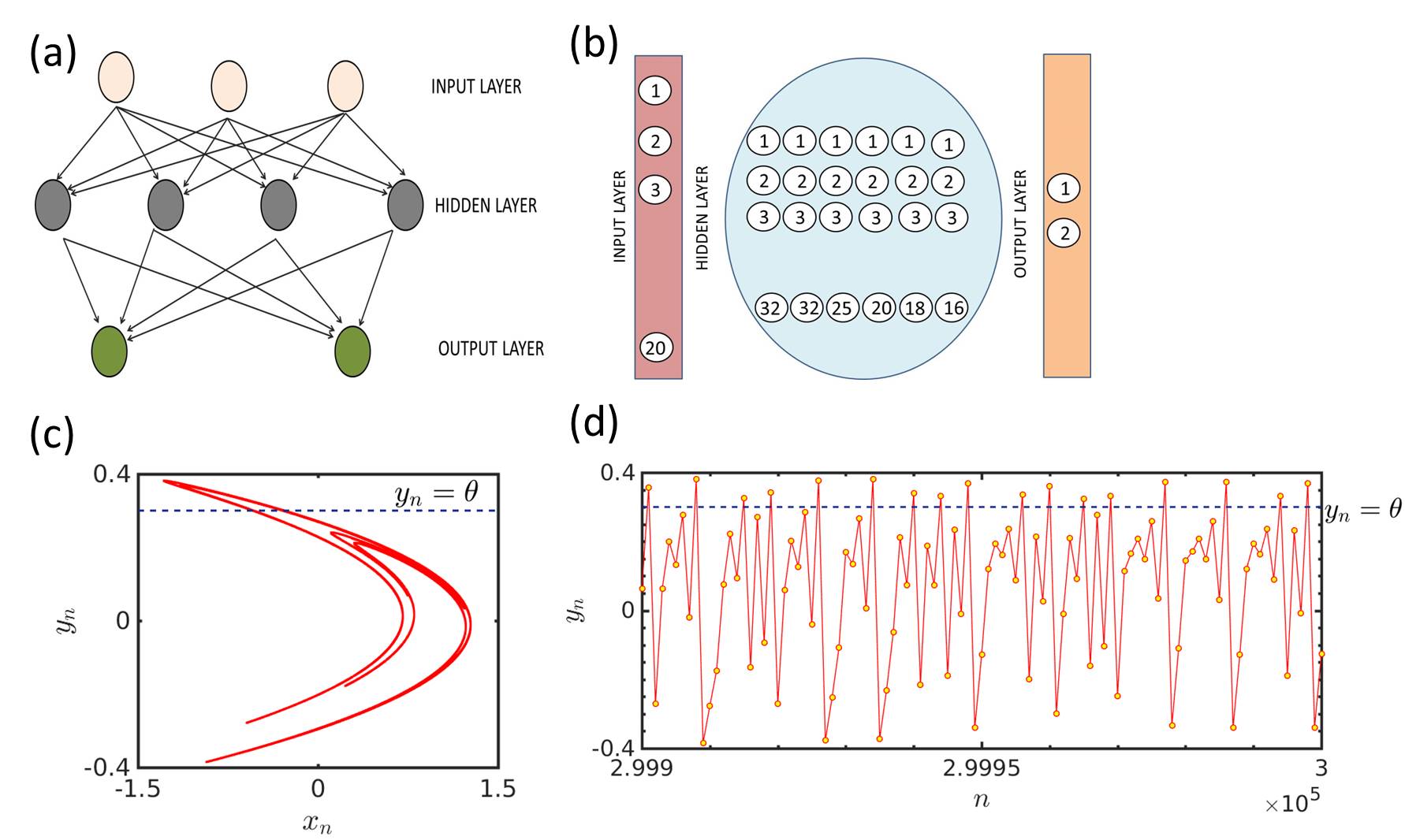}}
	\caption{(a) {\bf Schematic diagram of a feed forward neural network:} A schematic diagram is drawn to illustrate the features of the artificial neural network. It has three distinct layers, (i) input layer, (ii) hidden layer, and (iii) output layer. All intralayer links are absent, but all inter-links are present there. (b) {\bf The proposed feed forward neural network:} The neural network consists of three layers is proposed. The number of neurons in each layer is $20,32,32,25,20,18$ and $16$. All layers are globally connected as shown in the schematic diagram (a). (c) {\bf Classification task of machine learning:} The phase space is classified into two parts, (i) $y_n \ge \theta$, and (ii) $y_n < \theta$. The ultimate challenge is whether the machine can predict position of the trajectory, belonging to one of these two regions,  after exact $\alpha$ iterations, or not. (d) {\bf Time series of the chaotic map:} An exemplary trajectory along with the threshold, $\theta$ for prediction task is shown for $100$ iterations. For both the figures (c) and (d), the initial condition is chosen randomly from the interval $[0,1]$ and the threshold, $\theta=0.3$ is chosen in Ref.\ \cite{lellep2020using}.  }
	\label{fig_33}
\end{figure}

\par For the prediction purpose, an artificial multilayer neural network is considered, where the intralayer connections are completely absent, but fully connected between layers. Thus, the links between nodes do not form a cycle. The proposed feed forward neural network contains three distinct sections as follows: 

\begin{enumerate}
	\item Input layer: We have the information about the system's current position at $k$-th iteration as well as its past $(k-1)$ iterations. The dimension of the system (\ref{eq.63}) is $2$. Thus, the number of neurons required in the input layer is $2k$. As the chosen system is fully observed and is a deterministic chaotic map, smaller values of $k$ improve the prediction performances. Lellep et al.\ \cite{lellep2020using} considered $k=10$.
	
	\item Hidden layers: Initially, $6$ hidden layers are considered. By increasing the number of hidden layers of fixed size, one can observe an improved performance in the sense of expanded prediction horizon $\alpha$. The number of neurons per  hidden layer is chosen using the algorithm suggested by the Ref.\ \cite{heaton2008introduction}. The first hidden layer contains $32$ neurons. Other hidden layers contain $32, 25, 20, 18, 16$ following an approximately linear interpolation. ReLU activation function, ReLU($x$)=max($0,x$) is used for the hidden layers. The binary categorial cross-entropy \cite{goodfellow2016deep} is minimized by Adam optimizer \cite{kingma2014adam}. This helps the baseline topology to learn the task optimally during training.
	
	\item Output layer: The output layer consists $2$ nodes. Softmax activation function \cite{goodfellow2016deep} is used for this layer. For the two output classes, this function is equivalent to the sigmoid function, $\sigma(x)=\frac{1}{1-{e}^{-x}}$. The output layer returns two numbers representing the probabilities of the input being in the respective regions, (i) $y_n \ge \theta$, and (ii) $y_n < \theta$. The input is then classified according to the higher probability.
	
\end{enumerate}

\par Along with the prediction of extreme events using data-driven approach, the role of key parameters, viz.\ (i) the network size $N_P$, (ii) the prediction time, $\alpha$, and (iii) the number of training samples $N_\alpha$ is discussed \cite{lellep2020using}. In order to maintain a certain accuracy of $80\%$ or more than that, there exists a relation between $N_P$ and $\alpha$ as $N_P \propto \exp(h\alpha)$, where $h=0.465$ is the topological entropy \cite{artuso1990recycling}. However, this exponential behavior is not maintained for large $\alpha$, as once the parameters of
the network are exhausted, new features cannot be learned even by increasing the training data set. The scaling rule $N_\alpha \propto exp(2h\alpha)$ also holds up to a point. After that, the performance can not be improved by the introduction of more training samples. The decisive role of network parameters and structure is highlighted through the saturation of performance for larger $\alpha$.


\par Instead of predicting the exact trajectory of the system, final equilibrium statistics can be predicted using model free prediction.	Qi et al.\ \cite{qi2020using} considered a convolutional mixed-scale dense neural network for predicting different statistical regimes of truncated Korteweg-de Vries (tKdV) equation. This tKdV equation is capable of exhibiting several statistics including near-Gaussian to highly skewed
PDFs only by tuning the inverse temperature parameter. Since exact recovery of a single time series is not the basic motivation here, thus a small perturbation in the extreme value location does not affect significantly in the prediction of statistical features in the extreme events. To achieve desired closeness among the shape of the distributions between the target data and network's output data, Kullback–Leibler divergence is used as relative entropy loss function. The deep neural network is trained using data set only near-Gaussian statistics. Thus, the neural network cannot know about the large extreme events appearing in other statistical regimes. The input data are stored in the form of a tensor $\mathbf{x} \in \mathbb{R}^{J \times N \times C}$, whereas the output data is presented as $\mathbf{y} \in \mathbb{R}^{J \times N \times 1}$. Here, $C$ is the number of channels. The first layer contains a single channel. The other layers contain a combination of all the previous layer data in history. $J$ is the spatial grid points and $N$ reflects the time steps. In each single convolution layer, the input data are updated from the previous layer using the operator

\begin{equation}
\begin{array}{lcl}\label{eq.64}
\mathbf{y}=\sigma(g_{h}(\mathbf{x})+b),
\end{array}
\end{equation}
where $g_h= \sum_{i=1}^{C} h^i * x^i$ is the convolution operator. Here, $h^i$ is the convolutional filter kernel and it covers a small window with size $w_1 \times w_2$, where $w_1$ governs the correlation in the spatial direction and the temporal correction determines $w_2$. The scalar parameter $b$ is treated as bias term. For different layers and for different output channels, $b$ and $g_h$ change. The rectified linear unit (ReLU) function is
taken as the nonlinear operator $\sigma(x)=\max\{0,x\}$. The prediction time in the truncated KdV equation is found to be beyond the decorrelation time scale of the state, defined as integration of autocorrelation function. For the prediction purpose, a moderate number of layers $L=80$ is used. Further increment in the number of layers does not improve the results significantly. After the saturation of the relative error, the result can not be improved even after inclusion of larger number of training iterations.

\par Due to system's nonlinearity and its subsequent instabilities, the amplitude of an extreme event is larger compared to that of a regular event. So, a predictor of an extreme event should classify optimally between quiescent events and the extreme events. Guth et al.\ \cite{guth2019machine} proposed a prediction metric, which possesses superior optimization properties as compared to $F_1$-score. The constructed metric is

\begin{equation}
\begin{array}{lcl}\label{eq.65}
\alpha^{*}={\max}_{q \in [0,1]}  (\alpha(q)-q),
\end{array}
\end{equation}
provides a better prediction in a qualitative sense than total accuracy. Here, $q$ is the extreme event rate and $\alpha(q)$ is given by

\begin{equation}
\begin{array}{lcl}\label{eq.66}
\alpha(q)=\int_{-\infty}^{\infty} s(\hat{b})\abs{\frac{\partial r}{\partial \hat{b}}} d\hat{b} .
\end{array}
\end{equation}

Here, $\hat{a}$ and $\hat{b}$ indicate threshold values, which are defined for the binary classification problem. Any value that exceeds these thresholds is treated as an extreme event, otherwise called quiescent. The recall $r$ is defined by 

\begin{equation}
\begin{array}{lcl}\label{eq.67}
r(\hat{a},\hat{b})=\frac{P(TP)}{P(TP)+P(FN)},
\end{array}
\end{equation}
and the precision $s$ is defined by
\begin{equation}
\begin{array}{lcl}\label{eq.68}
s(\hat{a},\hat{b})=\frac{P(TP)}{P(TP)+P(FP)}.
\end{array}
\end{equation}

The interpretation of TP, FP and FN is represented in table \ref{binary}. 

\begin{table}[h]
	\begin{center} 
		\begin{tabular}{|c| c|}
			\hline
			True Positive (TP) & an extreme event which is predicted to be extreme \\
			\hline
			False Positive (FP) & an quiescent event which is predicted to be extreme \\
			\hline
			True Negative (TN) & an quiescent event which is predicted to be quiescent \\
			\hline
			False Negative (FN) & an extreme event which is predicted to be quiescent \\
			[1ex]
			\hline	
		\end{tabular}	
	\end{center} 
	\caption{Binary classification of predictor}\label{binary}
\end{table}

\par The predictor's performance is decided depending on the term $\alpha(q)-q$. If $\alpha(q)-q < 0$ or, it remains close to zero, then the predictor is poor at that extreme event rate $q$. To demonstrate their ideas, Majda–McLaughlin–Tabak model \cite{majda1997one} and the Kolmogorov flow \cite{platt1991investigation} are selected.


\section{Control of extreme events in dynamical systems}\label{control}
\par
\begin{figure}[H]
	\centerline{
		\includegraphics[scale=0.5]{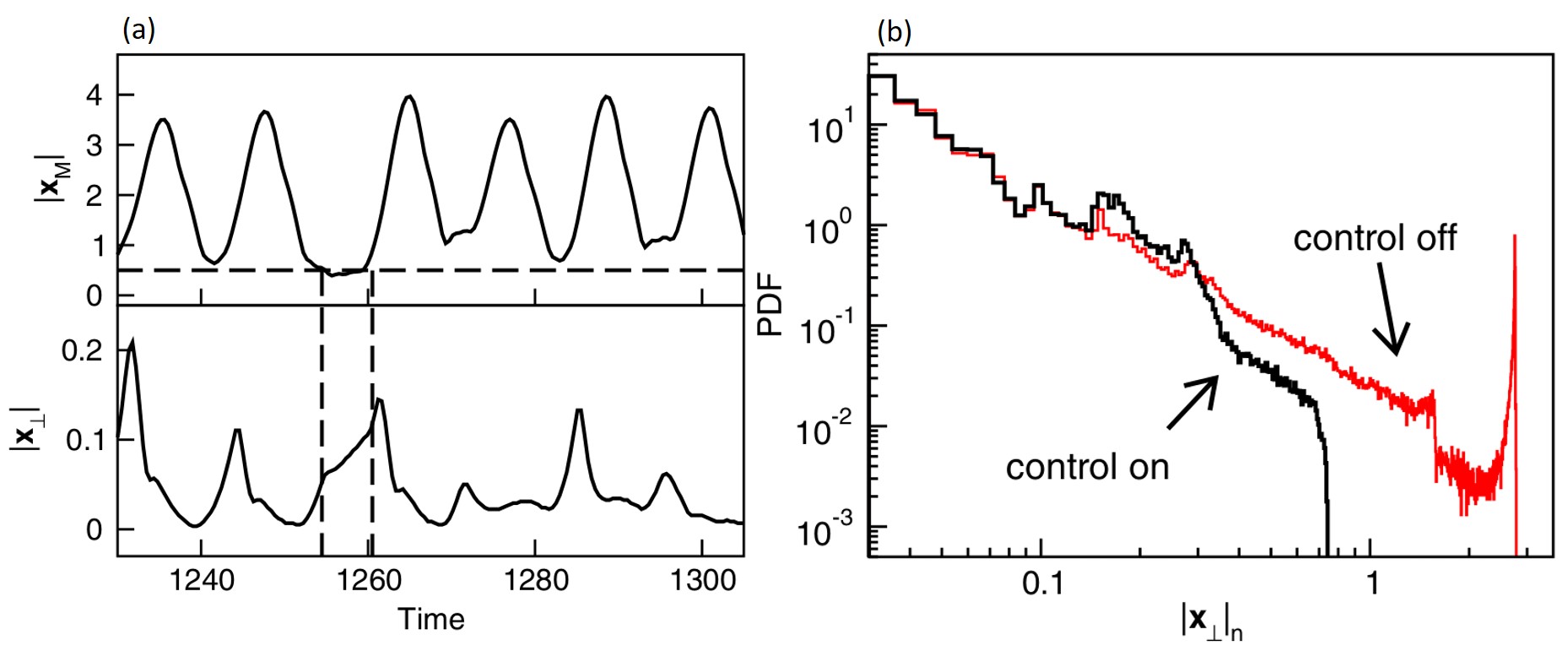}}
	\caption{(a) {\bf Suppression of extreme events by applying feedback control:} An upcoming extreme event can be successfully predicted using the temporal evolution of $|{\bf x_{M}}|$ as shown in the upper panel of (a). Whenever the values of $|{\bf x_{M}}|$ lie below the dashed horizontal line $|{\bf x_{M}}|_{th}=0.5$, occasional feedback is introduced in the slave oscillator only. In the lower panel of (a), two dashed vertical lines indicate one of such time intervals, where such control strategy is implemented. Clearly, the extreme event is completely suppressed due to the successful application of the control technique in the time evolution of the observable $|{\bf x_{\perp}}|$.
		(b) {\bf Probability density function of event sizes before and after suppression:} PDF of $|{\bf x}_{\perp}|_{n}$ is portrayed before applying control scheme, {\it i.e.,} $c_{DK}=0$ (red) and after applying control scheme, {\it i.e.,} $c_{DK}=0.5$ (black), respectively. Predefined threshold is taken as $|{\bf x_{M}}|_{th}=0.5$ for experiment. Dragon king events are mitigated successfully using the prediction through $|{\bf x_{M}}|$ variable and occasional activation of feedback. 
		Reprinted figure with permission from Ref.\ \cite{cavalcante2013predictability}.}
	\label{fig_37}
\end{figure}

 The control or suppression of extreme events in nature, such as Tsunami, 
floods, cyclones, droughts, etc., is difficult in principle, if not impossible. The main reason for the difficulty is the lack of any model, low- or high-dimensional. On the other hand, in humans-made systems, power grids, share market crashes, etc., a control strategy can be attempted if any well defined model or a network structure of a system is available. Any control strategy aims to apply additional feedback or temporary perturbation in a system to suppress any instability whenever any spurious event occurs. 
Then, one can perturb the system with an appropriate function to mitigate such events. Such a method is successfully applied in two nearly identical unidirectionally coupled chaotic electronic circuits in a master-slave configuration \cite{cavalcante2013predictability}. This approach is very cost-effective as only one needs to activate the perturbation based on the forecasting indicator. This occasional activation of perturbation to the slave subsystem effectively mitigates the dragon kings. For this purpose, the slave system \eqref{eq.12} is re-written with an addition of a feedback term,
\begin{equation}
\begin{array}{l}\label{eq.38}
\dot {\bf x}_{S}={\bf F}[{\bf x}_{S}]+c{\bf K}({\bf x_{M}}-{\bf x_{S}}) +[1-\theta(|{\bf x_{M}}|-|{\bf x_{M}}|_{th})]c_{DK}{\bf K_{DK}}({\bf x_{M}}-{\bf x_{S}}),
\end{array}
\end{equation}
where the coupling matrix of the feedback is denoted by ${\bf K}_{DK}$ with $({\bf K}_{DK})_{ij}=1$ for $i=j=1$ and $0$ otherwise. $\theta(\cdot)$ is the Heaviside step function. The subfigure in the upper panel of Fig.\ \ref{fig_37}(a) displays how the trajectory of $|{\bf x_{M}}|$ occasionally crosses the predefined threshold (horizontal dashed line), indicating the emergence of upcoming extreme events as shown in the temporal evolution of $|{\bf x_{\perp}}|$ (see the lower panel of Fig.\ \ref{fig_37}(a)). 
Based on this forecasting, an introduction of the perturbation in the form of feedback helps control the appearance of dragon kings, as illustrated in Fig.\ \ref{fig_37}(b). 
\par Using the prediction of the occurrence of extreme desynchronization
events, Zamora-Munt et al.\ \cite{zamora2014suppression} also discussed a mechanism to control extreme events in unidirectionally two coupled electronic circuits. A scheme for prediction of extreme events proposed in Ref.\ \cite{zamora2014suppression} for this system has already been discussed in Sec.\ \eqref{instability}. For prediction, an auxiliary system is introduced, and this auxiliary system is unidirectionally coupled with the main system subject to negative delayed feedback. Due to this interaction, anticipated synchronization occurs, and it helps in predicting extreme events in the main system. With the support of this prediction scheme, extreme events are also suppressed using a direct corrective resetting technique \cite{mayol2012anticipated}. Whenever we know in advance using anticipated synchronization that dragon kings may occur in the following few times, we reset the observable value under a safety amplitude by applying corrective reset to the main system. The effectiveness of this control scheme is tested successfully for the stochastic system also.

\par In Ref.\ \cite{ray2019intermittent}, a threshold-activated coupling scheme \cite{sinha1993adaptive, meena2017threshold} is found to be effective for terminating the extreme events in coupled Ikeda maps. This suppression technique is implemented by an exchange of information between the two maps only when the observable of any map crosses below a predefined threshold. This control scheme is also a kind of resetting approach \cite{ray2021mitigating, phogat2021phase}. This control scheme has some relevant applications in reality. For example, when a deficiency of food emerges in one patch of an ecological population, that undersupply is somehow controlled with the help of the neighboring patches. In fact, their implemented scheme does not need any prediction of such devastating events in advance too. The prediction of extreme events in most of the cases is out of our hands. So, scientific communities are trying to do something in systems such that the number of occurrences of extreme events is partially or entirely reduced. This is what exactly Varshney et al.\ \cite{varshneysuppression} did. 

\par The two coupled FHN neuronal models with environment is described as  
\begin{equation}
\begin{array}{lcl}\label{eq.41}  
\dot{x}_{i} & =& x_{i}(a - x_{i})(x_{i}-1) -y_{i}+k\sum_{j=1}^{2}A_{ij}(x_{j}-x_{i})+\epsilon E,\\
\dot{y}_{i} &=& b_{i} x_{i} - c y_{i},\\
\dot{E} &=& -dE-\dfrac{\epsilon}{M}\bigg(\sum_{i=1}^{M} x_{i}\bigg),
\end{array}
\end{equation}
where $d$ is the decay constant. Here, $M=2$ and $i,j=1,2$ with $i \neq j$.

\begin{figure}[H]
	\centerline{
		\includegraphics[scale=0.53]{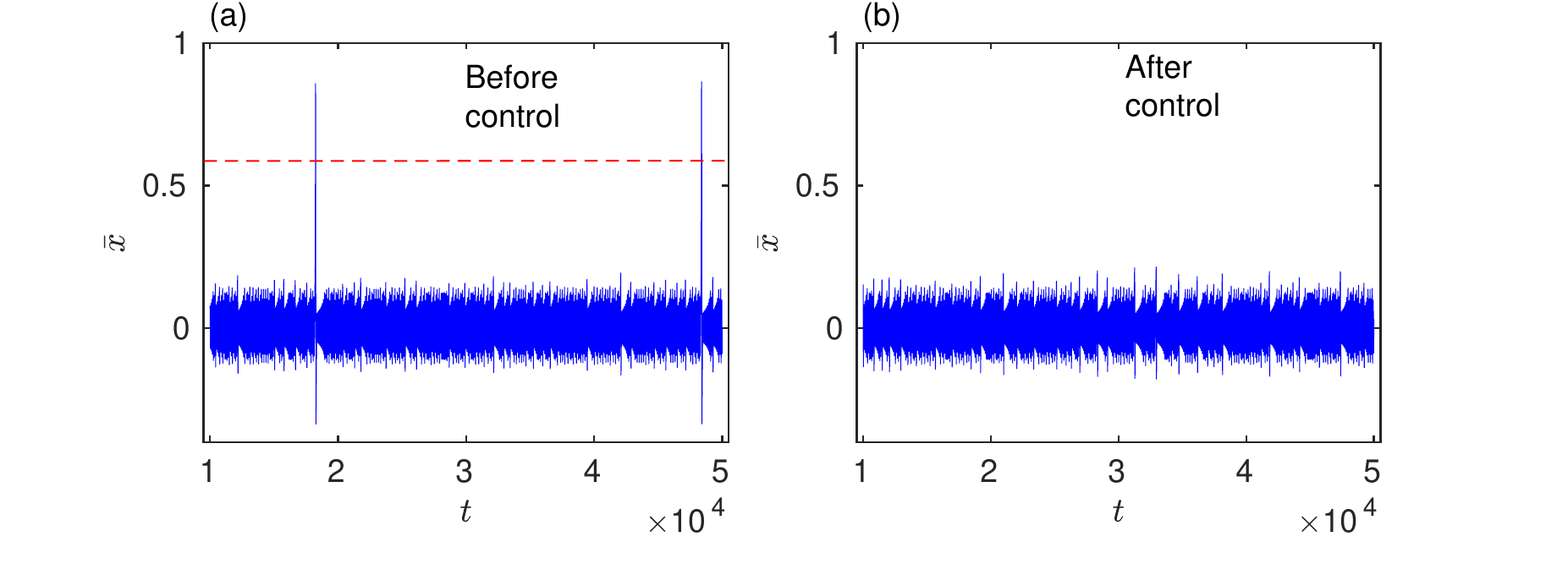}}
	\caption{ {\bf Temporal dynamics of $\bar{x}$ for two coupled FHN systems:} (a) Infrequent large amplitude oscillations are observed in the absence of interaction ({\it i.e.,} $\epsilon=0$) between FHN oscillators and the environment. Extreme events are considered in the time series of $\bar{x}$, when maxima of $\bar{x}$ exceeds a predefined threshold $0.6$ \cite{ansmann2013extreme}. (b) Due to interaction with environment, extreme events are completely vanished in the coupled FHN systems for suitable environmental coupling strength $\epsilon=0.025$. Other parameters: $a= -0.025794, b_1= 0.0065, b_2= 0.0135, c= 0.02, k=0.128, d=3,$ and $M=2$.}
	\label{fig_39}
\end{figure}

This system is interacted here with the environment \cite{sharma2012phase} with an environmental coupling strength $\epsilon$. The environmental variable $E$ is introduced so that the dynamics of this variable exhibit decaying nature. Occurrence of extreme events is exhibited in the time evolution of the observable $\bar{x}=\dfrac{x_1+x_2}{2}$, when maxima of $\bar{x}$ exceeds a predefined threshold $0.6$. In absence of any interaction strength ($\epsilon=0$), then extreme events occur in time series of $\bar{x}$ for a suitable choice of $k$ as shown in Fig.\ \ref{fig_39}(a). With increasing the value of $\epsilon$, the number of occurrences of extreme events in the two coupled FHN units is annihilated after a critical coupling strength.
Figure \ref{fig_39}(b) shows that extreme events completely disappear from the time evolution of $\bar{x}$ for a suitable coupling strength $\epsilon=0.025$. Besides two coupled FHN units, two coupled $CO_{2}$ laser models\ \eqref{eq.4} \cite{bonatto2017extreme} is also considered in Ref.\ \cite{varshneysuppression}, and interact with environment $E$. 
 The interplay between the decay constant of the environment $d$ and mean-field coupling strength $\epsilon$ is responsible for reducing the extreme events. 
 
\par The addition of a self-time-delay feedback \cite{ikhlef2008time, yamashita2015continuous} term in a system can also help for suppressing extreme events. The autonomous Li{\' e}nard system exhibits a dual character of conservative and dissipative dynamics in the appropriate range of parameters' values. The system's basin of attraction splits into two smooth regions of conservative and dissipative dynamics as depicted in Ref.\ \cite{suresh2018influence}. So, depending on the choice of initial conditions, this autonomous system either converges to stable focus (dissipative dynamics) or possesses neutrally stable periodic orbit (conservative dynamics). But in the presence of self-time-delay feedback, the autonomous system reveals dissipative dynamics in the whole phase space, and time delay feedback plays as a damping term. Here, it creates an obstacle for the large excursion of the chaotic trajectory for forced Li{\' e}nard system. Consequently, a suitable choice of the self-time-delay feedback strength helps suppress occasional large excursions of the chaotic trajectory of the system.

\par Also, it may be noted that machine learning algorithm has been successfully used \cite{pyragas2020using} for control of undesirable large events by applying a feedback signal. The machine first predict the extreme events, then a feedback control at the predicted time is applied to the system dynamics. 
This scheme has been successfully applied in two coupled oscillators 
and also in an ensemble large number of oscillators to suppress extreme events.

\par Finally, we want to conclude this section by mentioning the limitations of these proposed mitigating strategies. We discuss different existing methods, viz.\ i) self-time-delay feedback approach, ii) threshold-activated coupling, iii) feedback control, iv) corrective resetting approach, and v) controlling through environmental coupling. This summary suggests that mitigation of extreme events, although, is an essential helpful topic. But, the existing literature on this topic is very thin, and most of these control policies are dynamical systems-dependent. A generic systematic theoretical framework needs to be proposed soon to suppress the appearance of extreme events.

\section{Experiments on extreme events}\label{Experimental observation of extreme events}
\begin{figure}[H]
	\centerline{
		\includegraphics[scale=0.65]{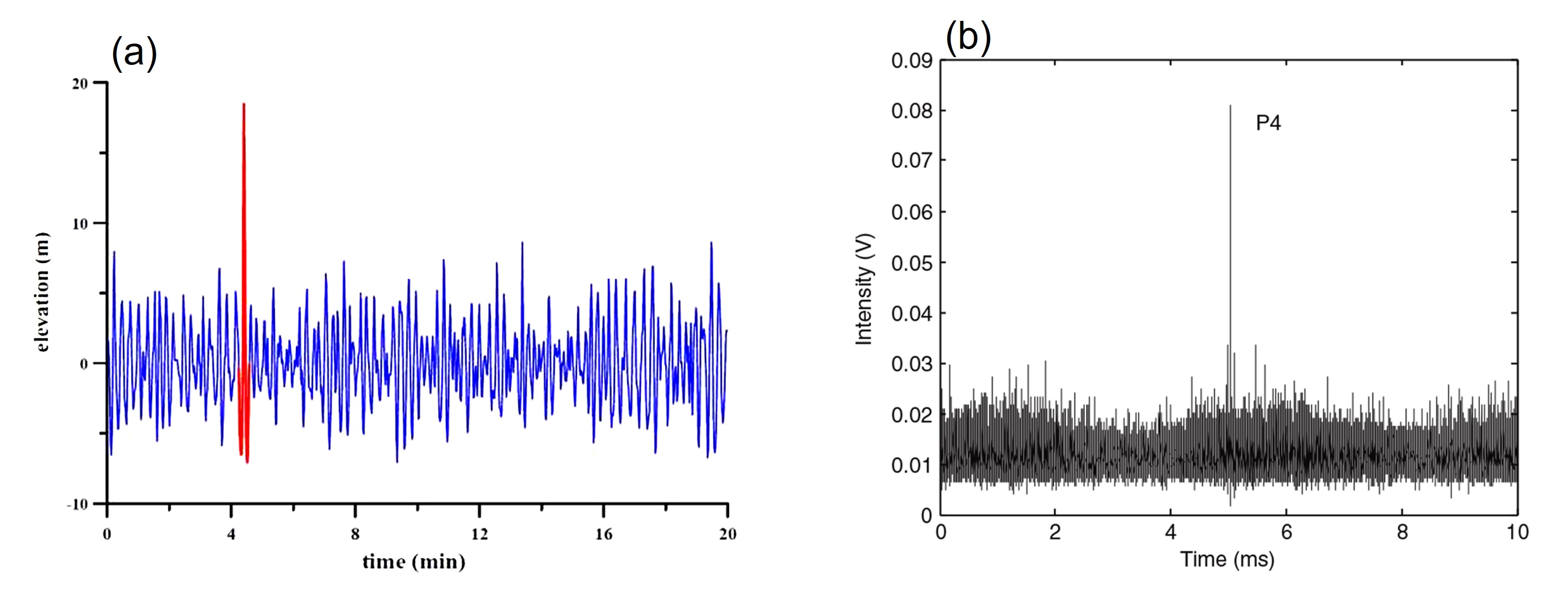}}
	\caption{ {\bf Temporal dynamics of rogue waves:} (a) Time record of "New year wave" at  ``draupner'' in the North see in 1995\ \cite{kharif2003physical}, (b) Optical rogue wave in an multistable laser experiment\ \cite{pisarchik2011rogue}. Reprinted figure from Refs.\ \cite{pisarchik2011rogue,kharif2003physical}.}
	\label{fig_40}
\end{figure}

Planners and researchers mostly talk about extreme events as natural calamities due to their huge impact. For a long, they have been analyzed statistically from recorded data such as locational variation of rainfall and temperature, which has been used for studies of flood \cite{gumbel2012statistics}. However, those events are difficult to be reproduced in laboratory experiments due to the lack of models and high dimensional character. On the other hand, devastating giant waves in the high sea, as reported by the seamen, are almost mythical until their existence has been found in the mid-nineties \cite{hopkin2004sea}. These are recorded as a large size wave in the North Sea from oil platforms\ \cite{kharif2003physical} in 1995, as shown in Fig.\ \ref{fig_40}(a). It immediately attracted the attention of oceanographers and physicists, who called those events as ocean rogue waves. 
The nonlinear processes involved in the origin of rogue waves have been explored theoretically using the nonlinear Schr\"odinger equation \cite{akhmediev2009rogue}.
The first laboratory experiment on extreme events called as optical rogue waves by analogy with the ocean rogue wave as solitary large amplitude events reported \cite{solli2007optical} in an optical fiber in 2007. Experimental results are confirmed in numerical studies of the nonlinear Schr\"odinger equation. 
A series of simple experiments on extreme events have started in semiconductor laser \cite{bonatto2011deterministic, zamora2013rogue}, extended microcavity laser \cite{selmi2016spatiotemporal}, liquid crystals \cite{clerc2016extreme} and electronic circuits \cite{kingston2017extreme, cavalcante2013predictability, mishra2018dragon} in the laboratory. The results are also simulated by deterministic nonlinear dynamical models. These simple experiments mainly recreate the temporal dynamics of extreme events. Occasional large intensity pulses arrive in a long run and lie in the tail of non-Gaussian distribution of all events. 

\par
\begin{figure}[H]
	\centerline{
		\includegraphics[scale=0.9]{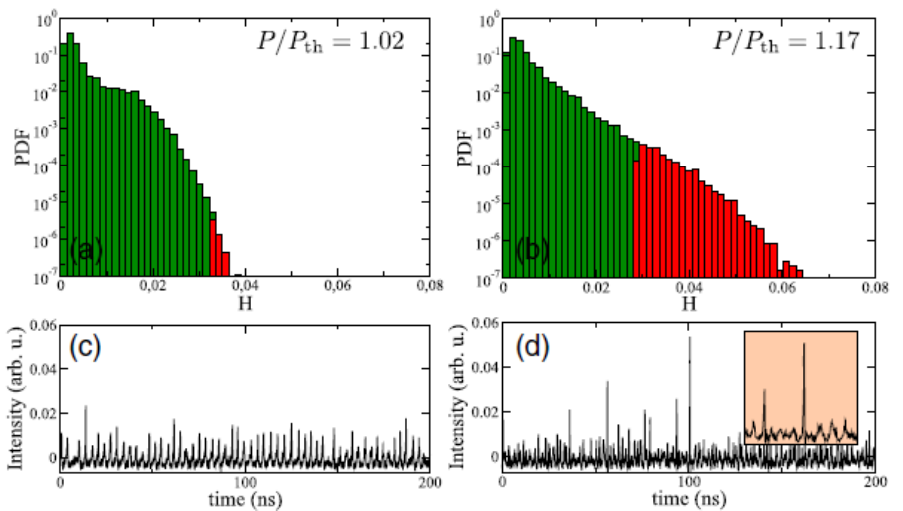}}
	\caption{ {\bf Experiment on spatiotemporal chaos and extreme events in a microcavity semiconductor laser:} Probability distribution of (a) spatiotemporal chaos and (b) extreme events. Time evolution of laser intensity for (c) chaos, and (d) extreme events. Reprinted figure from Ref.\ \cite{selmi2016spatiotemporal}.}
	\label{fig_41}
\end{figure}

 
The most important question has been addressed what is the mechanism of the origin of large intensity events in the laser system. Two important routes, interior-crisis and intermittency, have been identified as responsible for the formation of extreme events that are basically confirmed by the numerical studies obtained from deterministic nonlinear dynamical systems \cite{zamora2013rogue,reinoso2013extreme}. The question of predictability and control have also been addressed. 
Noise is suitably included in the semiconductor laser for diminishing the probability of occurrence of extreme events\ \cite{zamora2013rogue}. A control of the occurrence of extreme events has also been suggested in another experiment with a diode laser under phase-conjugate
feedback \cite{dal2013extreme}. Here, the number of extreme events can be increased by enhancing the mirror reflectivity. The attractor-hopping in multistable laser system has been confirmed \cite{pisarchik2011rogue}  that leads to the origin of rogue waves in the presence of noise, as shown in Fig.\ \ref{fig_40}(b)  and discussed in detail in Sec.\ \eqref{noise}. Spatiotemporal extreme events have also been explored in micro-cavity laser experiments \cite{selmi2016spatiotemporal} and liquid crystal devices \cite{clerc2016extreme} that form due to the collision of coherent structures. The upper panel consisting of subfigures (a)-(b) in Fig.\ \ref{fig_41} portrays the histograms of the intensity heights with respect to the pump parameter. The red portions of those histograms represent the extreme events whose heights are higher than the significant height. The lower panel of Figs.\ \ref{fig_41}(c)-(d) demonstrates the emergence of complex dynamics with irregular occurrences of large amplitude pulses due to the presence of spatial coupling. With an increment of higher pump intensities, the mean pulse period increases, as depicted in Fig.\ \ref{fig_41}(c)-(d).

\par Besides optical systems and lasers,  electronic experiments have also been performed, where the manifestation of extreme events is concerned. Numerical observations are  validated with real-time experiments \cite{kingston2017extreme} in an analog circuit of the  forced Li{\' e}nard system. This experiment also confirms the origin of  extreme events via interior-crisis and PM intermittency, which is discussed elaborately in Sec.\ \eqref{interior}. Extreme events may emerge in coupled systems due to attractor bubbling. A master-slave coupled chaotic circuit is designed for verification of on-off intermittency-driven extreme events \cite{cavalcante2013predictability}. On the other hand, Mishra et al.\ \cite{mishra2018dragon} set up two analog circuits of Hindmarsh-Rose neuron model. A single neuron circuit reproduces the periodic bursting dynamics as observed in numerical simulations. The coupled electronic circuit reproduces the temporal dynamics under the influence of repulsive coupling. The probability density function of peak values of an observable confirms the manifestation of dragon-king like behavior.  

\par However, the elegant investigation on absence epilepsy by Frolov et al.\ \cite{frolov2019statistical} attests to the fact that the statistical distribution of extreme epileptic events satisfies classical extreme value distribution instead of displaying such bumpy dragon-king like distribution. They study the epileptic brain activity of five male WAG/Rij rats. These rats are capable of emanating frequent spike-wave discharges (SWD) per day. This SWD is treated there as a single event of extreme hypersynchronization of neuronal activity in cortical layers and thalamic nuclei. The EEG recordings of these rats open new avenues for early prediction of such SWDs by capitalizing real time tracing of the variance of the wavelet energy PDF. Also, Pisarchik et al. demonstrated similar properties of extreme behavior in epileptic EEGs of rodents after artificially induced ischemic stroke \cite{pisarchik2018extreme}. Recently, the time series of EEG spectral power of ten human participants with generalized epilepsy in a frequency band of $1–5$ Hz is also found to obey a heavy-tailed Weibull distribution \cite{karpov2021noise}. Pre-bifurcation signal (noise) amplification on EEG signals gradually increases before the onset of an epileptic seizure, just like the behavior of many dynamical systems near the critical bifurcation point. 
\section{Summary and future perspectives} \label{Conclusion and Future directions}
\par The studies on extreme events have become increasingly promising for their huge impact. Tsunami, floods,
droughts, power blackouts, share market crashes, traffic jamming are a few devastating familiar examples which
attest to the necessity of this evolving topic. A long history of human endeavor exists to save the life from
the aggression of devastating natural events caused by unknown processes. Prediction of extreme events
is thus a vital task to mitigate its’ harmful effect. An early warning indicator may help different disaster
management groups to take pre-emptive measures for saving life. Efforts to understand extreme events from
available data records motivate several scientific communities. By studying this topic from the perspective of
statistics, it is possible to extract information about the probability of its occurrence and its return interval.
However, the availability of limited data, particularly, in the case of natural events, makes it really difficult to
reach statistical convergence. In fact, most of the models representing such high dimensional phenomena do
not necessarily contain the whole ingredients. The challenges lie in the discussion of such devastating events
that demand to be looked at from a different perspective in order to provide a better understanding of the topic. Driven by this motivation, we have revisited the studies of extreme events based on two different
approaches, viz. i) dynamical systems and ii) random walkers in this review.

\par The studies of extreme events in dynamical systems have been started using mathematical models
and dynamical networks. This initiates a plethora of research activities from a dynamical system perspective
using physical, climate, biological models, and so on. The results of these studies are very encouraging with quite a few possibilities. Although no common unique definition of extreme events can be established, still some effective measures exist in the literature. These measures can identify
extreme events from a long time series and thereby scrutinize their existing statistical properties. The
prediction of extreme events remains challenging, however, attempts are being made both from the dynamical
system perspective and machine learning. Few control methods for the suppression of extreme events have also
been explored.

\par On the other hand, researchers are equally interested in the problems of extreme events in traffic dynamics. The increasing volume of flux challenges the efficient functioning of the system. The theory of random walkers in networks can explain uniquely many real-world paradigms that arise in public infrastructures such as road networks, communication networks, power grids, to name a few. The important effect of network parameters like hubs, vertices, edges has been inspected recently on the study of extreme events in models of random walkers. These studies enrich our understanding of extreme events. We have also incorporated some relevant studies on extreme events in models of random walkers, that deal with some controlling strategies for mitigating extreme events. These control strategies have significant applications, including cybersecurity.

\par In this review, we have presented the recent progress of interdisciplinary research done to advance our understanding of extreme events that have become of increased interest among various disciplines in recent years. Usually, such events occur with low probability compared to the regular behavior of the system. But, a sudden occurrence of extreme events draws the attention of several scientific communities due to its destructive impact.  We organize our review into separate sections as follows.


\par In Sec.\ \eqref{Dynamical origin of extreme events}, we have focused on the issue of how extreme events originate in nonlinear dynamical systems and try to cover up some known mechanisms with examples in isolated systems, two coupled systems, and networks of dynamical systems. We have discussed few important routes to the chaos that may lead to the origin of extreme events in isolated and two coupled dynamical systems. Besides these routes, instability of the synchronization manifold may generate extreme events in the coupled dynamical systems. We have also observed that extreme events may also appear in static as well as time-varying dynamical networks due to various possibilities like heterogeneity of parameters, coupling topology, network architecture, or others.  

\par In Sec.\ \eqref{Extreme events due to random walk in complex networks}, we have discussed the emergence of extreme events on random walkers. Models of self-driven many-particle systems reflect many aspects of cooperative congestion phenomena. These models resemble many real-life instances ranging from the diffusion of e-mail viruses to traffic jams in road networks, and disease spreading on spatial networks. Congestion on such large infrastructures arises unexpectedly and gives rise to a breakdown of the system’s normal functioning. 
We have contemplated the effect of several attributes like the degree of nodes, enhancement of nodal capacity, biased strategies of random walkers, the velocity of random walkers, to name but a few, on the extreme events due to transport on complex networks modeled via random walkers.  

\par In Secs.\ \eqref{prediction} and \eqref{control}, we have emphasized the key challenges of prediction and mitigation of extreme events in dynamical systems.  These topics undoubtedly need more effort to propose any reliable control strategy and prediction scheme. From the dynamical system perspective, in general, the prediction schemes have been proposed based on observation of the instability region in the phase space of a system. In most cases, the prediction schemes and control strategies for mitigating extreme events are model dependent. In some cases, control of extreme events has been proposed with a simultaneous prediction procedure. Sometimes, the prediction scheme is unknown although control of extreme events is possible by choosing suitable control strategies. Both prediction and control strategies have been reported here based on numerical performance from the dynamical system perspective. We have also discussed data-driven approaches for forecasting upcoming extreme events. Without any prior knowledge of the explicit model dynamics, these machine learning algorithms allow data-driven predictions of extreme events. In spite of its youth and the benefits of simpler model implementation, very little has been done on the literature on extreme events from the perspective of machine learning algorithms, and further attention needs to be paid to the problem of prediction of extreme events. We have briefly discussed some techniques that capture signatures as an early warning before an extreme event really appears.
\par We provide a concise discussion on the advancement of experimental studies on extreme events in optical systems and electronic circuits in Sec.\ \eqref{Experimental observation of extreme events}. A few experiments on extreme events related to epileptic seizures have also been reviewed here.

\par Finally, the open problems discussed below promise interesting discoveries and tremendous progress in our understanding of extreme events.

\begin{enumerate}

	\item In case of extreme events, the trajectory of the dynamical systems occasionally visits a region of instability resulting in a faraway excursion to locations in the phase space, and ultimately it returns to that region where trajectory stays most of the time after a short duration. Whenever this large excursion in a dynamical state variable (observable) exceeds a predefined threshold level, it qualifies as an extreme event. However, the choice of this extreme event qualifier threshold is somewhat arbitrary and system-dependent. A wide variety of difficulties arises in the case of the determination of a clear-cut threshold. Thus, a promising future prospect will be the discovery of a non-arbitrary unique threshold.
	
	\item The most common routes of such devastating events for an isolated dynamical system are generally associated with the onset of chaos. As a parameter of the system is varied continuously, the dynamics of those systems become chaotic. This review suggests that most of the emergence of extreme events in an isolated dynamical system is dependent significantly on the route to chaos.
	Generally, thus a trivial query arises here does there exist any correlation between the route of formation of extreme events in an isolated dynamical system and the route to chaos? Does a question arise if a more generic process is involved in the origin of extreme events in dynamical systems besides the route to chaos, in general?

	\item We have tried our best to provide a rather complete overview of the current results regarding the different possible types of mechanisms responsible for extreme events in dynamical systems. A very relevant subject of investigation along this aspect, that will certainly attract attention, is which type of isolated and coupled dynamical systems are capable of generating extreme events.
	
	\item Our extensive review discloses several ways of formation of extreme events in dynamical systems. Most of them are due to instabilities caused by various factors in phase space. It demands an integrated mathematical procedure that will help to locate these instabilities.
	
	\item Undoubtedly, other routes to extreme events may exist and mostly remain undiscovered. The task becomes more daunting in high-dimensional systems, where many interacting degrees of freedom contribute to extreme event formation. Very few works have been done on the occurrence of extreme events in the dynamical network. But it is still elusive to understand the exact mechanism behind the occurrence of extreme events from the perspective of network science. In fact, it would be really significant to investigate the possible types of dynamical networks that are capable of generating extreme events.

	\item Results presented here related to the extreme events due to random walkers in a network are found to occur in all circumstances due to inherent fluctuations in the model. The absence of any external driving force makes extreme events an integral part of the systems. Thus, one can use the capacity of each node \cite{kishore2012extreme, chen2015extreme}, and the velocity of the walkers \cite{chen2014controlling} as tuning parameters. These parameters can help to diminish the number of extreme events in the system if chosen appropriately. But, this capacity enhancement of the node is a very costly approach. Besides, the choice of velocity for all the walkers is nonuniform generally in real life for most of the cases. Under these circumstances, can anyone suggest a design approach for networks that are resilient to extreme events?
	
	\item In the occurrence of extreme events due to random walk, all the works basically consider the noninteracting independent walkers in the network. But in reality, the interactions among those walkers play an important role, and this correlation among the walkers can play a decisive role in the understanding and the probability of occurrences of extreme events.
	
	\item We have observed that some regions of instability embed in the phase space through this review. This instability region is responsible for the origination of extreme events. This region of instability may arise due to the presence of the saddle \cite{ray2019intermittent,cavalcante2013predictability,saha2017extreme}. An analytical approach \cite{babaee2016variational} is already found to be successful for locating the region of instability. Using this approach, a prediction scheme for extreme events is proposed in Ref.\ \cite{farazmand2016dynamical}.  But still, it is a challenging issue to find out a general framework to identify the region of instability, which is responsible for generating extreme events. As a benefit of capturing the region of instability, can we form a general set-up for getting a prediction scheme of extreme events from isolated dynamical systems to dynamical networks?
	
	\item We discuss the dynamical system approaches as well as the data-driven approaches for predicting upcoming extreme events. The extreme event indicator predicts an extreme event before a significant prediction time. One important question is whether we can enhance the prediction time of extreme events. Besides, how large is the prediction time against the time scale of a system?
	
	\item Machine learning algorithms have been found to be very beneficial for forecasting extreme events under certain accuracy. But, the performance of such machine learning depends crucially on several factors. For instance, the number of available data records of past extreme events needs to be passed to the machine for training. How much data is sufficient for tracking such events accurately? This is an important question. In fact, how many extreme events should the data contain for successful prediction of upcoming extreme events?
	
	Moreover, the choice of a threshold is very significant to broaden the horizon of prediction time as revealed in Ref.\ \cite{zamora2013rogue}. One needs to investigate systematically the reason behind such influence on the choice of thresholds in the future.

	\item The control schemes so far reviewed in this report are applicable for low dimensional dynamical systems only.  A  question may arise if these schemes can be applied to high dimensional systems? In particular, how such control strategies are applicable to complex networks with a variety of coupling topologies?
	
\end{enumerate}

\par We hope this review will succeed in demonstrating  
different ways to detect the underlying mechanisms that trigger extreme events and to suppress the formation of such events based on current researches. We believe that the research reviewed above has the potential to broaden the scope of prediction of such events using the instability of phase space, and machine learning algorithms. It is noteworthy to mention that there are lots of other relevant important questions related to extreme events, and directions for future research are many. In fact, the domain of extreme events is vast spreading over many disciplines, and it is a subject of numerous theoretical and experimental investigations. Presenting an exhaustive account of them is really not an easy task.  Any unintentional omission of relevant references has been apologized for. We aim to bring together a substantial body of literature published over the last few decades to provide a comprehensive picture of extreme events. We conclude with the hope that this review will provide a guiding path for future researchers interested in extreme events' studies.



\section*{Declaration of competing interest}
The authors declare that they have no known competing financial interests or personal relationships that
could have appeared to influence the work reported in this paper.

\section*{Acknowledgments}
This work is supported by Science and Engineering Research Board (SERB), Government of India (Project no. CRG/2021/005894). SNC and DG were supported by the Department of Science and Technology, Government of India (Project No. EMR/2016/001039). DG was supported by the Department of Science and Technology, Government of India (Project no. INT/RUS/RFBR/360). SNC would also like to thank Physics and Applied Mathematics Unit of Indian Statistical Institute, Kolkata, for their support during the pandemic COVID-19. SNC would
also like to acknowledge the financial support from the CSIR (Project No. 09/093(0194)/2020-EMR-I) for funding him during a later stage of our work. We would like to thank our colleagues and collaborators
Tomasz Kapitaniak, Gopal K. Basak, Matjaz Perc, Mahmut Özer, Chittaranjan Hens, T. Chakraborty, Arindam Mishra,
Soumen Majhi, and Sarbendu Rakshit. We are indebted to T. P. Sapsis, Hugo L. D. de S. Cavalcante, A. E. Hramov, C. Masoller, V. Lucarini, H. Kantz, A. N. Pisarchik, N. Frolov, M. S. Santhanam, U. Feudel, K. Lehnertz, R. E. Amritkar, C. Bonatto, C. Nicolis, Y. C. Lai, C. Grebogi, M. Senthilvelan, D. Sornette, A. Prasad, E. Ott, Pinaki Pal, S. Sinha, S. Havlin, J. Kurths, M. Farazmand, T. Kathamuthu, M. Paulsamy, R. I. Sujith, V. K. Chandrasekar, V. Kishore, S. Kumarasamy, S. Boccaletti, B. Barzel, I. Belykh, V. Belykh, E. M. Bollt, J. M. Buldú, J. Burguete, A. Buscarino, A. Cardillo, T. Carroll, M. Clerc, R. Criado, P. De Lellis, M. Di Bernardo, A. Diaz Guilera, R. D’Souza, E. Estrada, S. Focardi, L. Fortuna, L.V. Gambuzza, J. Garcia-Ojalvo, A. Garcimartin, G. Giacomelli, J. Gömez-Gardeñes, J. Hizanidis, P. Hövel, S. Jafari, S. Jalan, M. Jusup, S. Kurkin, M. Lakshmanan, S. Lepri, V. Latora, D.Maza, R. Meucci, L. Minati, G. Mindlin, Y. Moreno, D. Musatov, A. Raigorovskii, G. V. Osipov, P. Parmananda, L. M. Pecora, N. Perra, A. Politi, M. Porfiri, R. Ramaswamy, P.L. Ramazza, M. Romance, E. Schöll, M. D. Shrimali
for their contribution of knowledge, which encouraged and
benefited us to provide the current account on this rapidly growing field of research. SNC thanks M. S. Santhanam for valuable discussions during the SERB School on Nonlinear Dynamics held at the Department of Physics, Guru Nanak Dev University, Amritsar.

\bibliographystyle{elsarticle-num}
\bibliography{sample_20_02_2021}

\end{document}